\documentclass[aps,prx,reprint,superscriptaddress,longbibliography]{revtex4-1}

\usepackage{amsthm,amsmath,amssymb}
\usepackage{tikz}
\usepackage{bm}

\usepackage{mathrsfs}
\usepackage{graphicx}
\usepackage{amsmath,amssymb,amsfonts,bm}
\usepackage[colorlinks, linkcolor=black, anchorcolor=black, citecolor=blue, runcolor=black, urlcolor=black, CJKbookmarks=true]{hyperref}
\usepackage{lipsum}
\usepackage{enumitem}

\def\be{\begin{equation}} \def\ee{\end{equation}}
\def\bea{\begin{eqnarray}} \def\eea{\end{eqnarray}}

\def\Tr{\text{Tr}}

\newcommand{\Z}{\mathbb{Z}} 
\renewcommand\[{\begin{equation}}
\renewcommand\]{\end{equation}}

\begin{document}

\title{Anomaly in open quantum systems and its implications on mixed-state quantum phases}
\author{Zijian Wang}
\affiliation{Institute for Advanced Study, Tsinghua University, Beijing 100084,
People's Republic of China}
\author{Linhao Li}
\email{linhaoli601@163.com}
\affiliation{Department of Physics and Astronomy, Ghent University, Krijgslaan 281, S9, B-9000 Ghent, Belgium}

\begin{abstract}  
In this paper, we develop a systematic approach to characterize the 't Hooft anomaly in open quantum systems. Owing to nontrivial couplings to the environment, symmetries in such systems manifest as either strong or weak type. By representing their symmetry transformation through superoperators, we incorporate them in a unified framework that enables a direct calculation of their anomalies. In the case where the full symmetry group is $K\times G$, with $K$ the strong symmetry and $G$ the weak symmetry, we find that anomalies of bosonic systems are classified by $H^{d+2}(K\times G,U(1))/H^{d+2}(G,U(1))$ in $d$ spatial dimensions. To illustrate the power of anomalies in open quantum systems, we generally prove that anomaly must lead to nontrivial mixed-state quantum phases as long as the weak symmetry is imposed.  
 Analogous to the ``anomaly matching" condition ensuring nontrivial low-energy physics in closed systems, anomaly also guarantees nontrivial steady states and long-time dynamics for open quantum systems governed by Lindbladians. Notably, we identify a novel $(1+1)$-D mixed-state quantum phase that has no counterpart in closed systems, where the steady state shows no nontrivial correlation function in the bulk, but displays spontaneous symmetry breaking order on the boundary, which is enforced by anomalies. We further establish the general relations between mixed-state anomalies and such unconventional boundary correlation. Finally, we explore the generalization of the ``anomaly inflow" mechanism in open quantum systems. We construct $(1+1)$-D and $(2+1)$-D Lindbladians whose steady states have mixed-state symmetry-protected-topological order in the bulk, with corresponding edge theories characterized by nontrivial anomalies.

\end{abstract}

\maketitle
\section{Introduction}   
Global symmetry plays a pivotal role in the realm of quantum many-body physics. One notable aspect is its associated 't Hooft anomalies \cite{tHooft:1979rat}, which arise from an obstruction in promoting the global symmetry to a gauge symmetry. At a microscopic level, anomalies also manifest as obstructions to implementing symmetry transformation on a subregion \cite{PhysRevB.90.235137}, or fractionalization on symmetry defects \cite{PhysRevB.87.104406,Barkeshli:2014cna,chen2015anomalous,tarantino2016symmetry,cheng2016translational,barkeshli2020reflection,barkeshli2020relative,bulmash2020absolute,PhysRevX.11.031043,kawagoe2021anomalies,Cheng:2022sgb,Delmastro:2022pfo}. From the ``anomaly matching" condition \cite{tHooft:1979rat}, the low-energy physics of systems with anomalous symmetries and the corresponding phase diagrams are strongly constrained, where a unique gapped ground state is forbidden. Instead, such systems must exhibit either gapless excitations or multiple degenerate ground states. 

Traditionally, anomalies are defined for pure states. However, real physical systems are inevitably subjected to noises and dissipation from the environment. These effects are especially important for modern quantum simulation platforms and devices, which render the system in a mixed state \cite{Saffman2010Rydberg,kjaergaard2020superconducting,Bruzewicz2019trapped}. Recently, there has been increasing interest and exciting developments about novel quantum phases and phase transitions in these open quantum systems \cite{dennis2002topological,diehl2008quantum,sieberer2016keldysh,diehl2010dynamical,altman2015two,coser2019classification,Lieu2020SB,wang2023topologically,liu2024dissipative,dai2023steady,rakovszky2023defining,lu2023mixed,mcginley2020fragility,deng2021stability,wang2021symmetry,de2022symmetry,ma2023average,ma2023topological,lee2022symmetry,zhang2022strange,su2023higher,fan2023diagnostics,bao2023mixed,lee2023quantum,wang2023intrinsic,chen2023separability,chen2023symmetry,chen2024unconventional,sang2023mixed,su2024tapestry,lyons2024understanding,li2024replica}. The goal of this work is to generalize the notion of anomaly to open quantum systems/mixed states, and to show that it plays an equally important role as in closed systems, by uncovering its various implications on mixed-state quantum phases.

One unique aspect of open quantum systems is the enriched meaning of symmetries, i.e., one needs to distinguish the following two types of symmetries: 1. Weak symmetries, which means that the system and environment as a whole have the symmetry, but they are allowed to exchange symmetry charges. 2. Strong symmetries, meaning that the system itself has the symmetry, that is, the symmetry transformation acts trivially on the environment \cite{Buca:2012symmetry}. The different notions of symmetries render the characterization of anomalies in open quantum systems both challenging and compelling. In a very recent paper that appears during the process of this work \cite{lessa2024mixed}, it is shown that anomalies of strong symmetries imply multipartite non-separability. However, a systematic definition or characterization of anomalies encompassing both strong and weak symmetries remains elusive. Also, more direct diagnostics of anomalies, especially those related to physical observables, are desired. 

Moreover, in closed systems, it is known that anomaly can arise on the boundary of symmetry-protected topological (SPT) phases, leading to intriguing edge physics, which is known as ``anomaly inflow” mechanism \cite{callan1985anomalies,PhysRevB.83.035107,PhysRevB.84.235141,chen2012symmetry,PhysRevB.86.115109,PhysRevB.87.155114,freed2014anomalies,PhysRevLett.112.231602,Kapustin:2014tfa,kapustin2014anomalies,Senthil_2015,witten2016fermion,yonekura2016dai}. In recent studies, interesting mixed-state SPT phases involving both strong and weak symmetries are identified \cite{deGroot2022spt,ma2023aspt,lee2022symmetry,zhang2022strange,ma2023topological,ma2024symmetry,guo2024locally,xue2024tensor}. However, the edge physics of these novel mixed-state topological phases remains unclear. One reason is, as stated above, the lack of understanding of anomaly with the presence of both strong and weak symmetries. Besides, it is tricky to ubiquitously define the edge theory of mixed-state SPT due to the absence of a Hamiltonian. 

In this paper, we address all the above issues. Since there is no widely accepted way of gauging in open quantum systems, we take the obstruction to localizing symmetry transformations as the definition of anomaly, where the symmetry transformation is generally represented by some superoperators. This approach enables us to directly compute the anomaly index of lattice systems and establish a classification scheme. Moreover, we find that a nontrivial anomaly index indeed indicates nontrivial mixed-state quantum phases, and it also plays a significant role in the bulk-boundary correspondence of mixed-state SPT phases. Most surprisingly, by constructing lattice models with anomalous symmetries, we find a new exotic mixed-state quantum phase, with all correlation functions appearing trivial in the bulk, but exhibiting spontaneous symmetry breaking (SSB) order on the boundary. 

\subsection{Summary of Results}
Here we give an overview of the rest of the paper, highlighting the main results.
\begin{enumerate}
\item Section \ref{sec:preliminaries} is mainly a review of basic definitions of symmetries in open quantum systems and the definition of mixed-state quantum phases. We also introduce the symmetry superoperator representation for later use.
\item In Section \ref{sec:anomaly}, we systematically define and characterize mixed-state anomalies for bosonic systems when the full symmetry takes the form $\Gamma=K\times G$, where $K$ is the strong symmetry and $G$ is the weak symmetry. By generalizing the Else-Nayak approach to mixed states, we illustrate how to extract the anomaly index for open quantum systems. We show that the anomalies in $d$ spatial dimensions are classified by $H^{d+2}(\Gamma,U(1))/H^{d+2}(G,U(1))$. Particularly, there is no anomaly if only the weak symmetry is present. These findings are in exact correspondence to previous results on mixed-state SPT in one higher dimension.
\item  Next, we show in Section \ref{sec:anomaly constrain} that anomaly has strong constraining power on mixed-state quantum phases. Specifically, we prove the following theorem: A state with anomalous symmetry $K\times G$ cannot be prepared via a $G$-symmetric finite depth local quantum channel starting from a $G$-symmetric product state. Here ``$G$-symmetric" refers to the weak symmetry condition.  
\item After the general abstract discussion above, we construct concrete lattice models with anomalies in Section \ref{sec:lattice model}. To diagnose more physical consequences of anomalies, we focus on open systems governed by Lindbladians. As a counterpart of ``anomaly matching" in open quantum systems, we find that the long-time dynamics must be nontrivial when the Lindbladian preserves anomalous symmetries. Specifically, the steady state(s) must belong to a nontrivial phase, which implies that either the steady states are degenerate or the relaxation time is divergent, or both. For example, the steady states may spontaneously break the symmetry. Yet we also find a more exotic scenario that has no pure-state counterpart, where there is no nontrivial correlation function in the bulk, but under open boundary conditions (OBC), the steady states exhibit nontrivial boundary correlation, which is enforced by anomalies. In Section \ref{sec:boundary cor} we show generalities of such anomaly-enforced boundary correlations in open quantum systems. 
\item Finally, in Section \ref{sec:ssaspt}, we discuss the ``anomaly inflow" mechanism in open quantum systems. That is, mixed-state anomalies can be realized on the edge of mixed-state SPT phases, including those jointly protected by strong and weak symmetries. Using the decorated domain wall (DDW) method \cite{chen2014symmetry}, we construct symmetric Lindbladians whose steady states realize such novel SPT phases, which allow us to gain a clearer perspective of the anomalous symmetry action in their edge theories. Moreover, the anomaly on the boundary/interface enables us to prove the separation of mixed-state phases for $(2+1)$-D mixed-state SPT.     

\end{enumerate}
\section{Preliminaries\label{sec:preliminaries}}
\subsection{Symmetries in open quantum systems\label{sec:sym}}
The purpose of this section is to review the strong and weak symmetry conditions for mixed states as well as quantum channels and Lindbladians, and meanwhile we also introduce the notion of symmetry superoperators for later convenience. 

For simplicity, throughout this paper we consider bosonic systems with internal symmetry groups $\Gamma$ of the product form $\Gamma=K\times G$, where $K(G)$ is the strong (weak) symmetry group. See below for the precise definition.

\textbf{Definition 1} (Symmetries of mixed states).  A density matrix $\rho$ has the $\Gamma$ symmetry iff:
\begin{equation}
\begin{aligned}
\mathscr{U}(g)[\rho]\equiv U(g)\rho U^\dagger(g)=\rho,\forall g\in G& \\
\text{ (weak symmetry condition)}&,\\
\mathscr{U}(k)[\rho]\equiv U(k)\rho= \lambda(k)\rho,\forall k \in K & \\
\text{ (strong symmetry condition)}&, 
\end{aligned}
\end{equation}
where $\lambda(k)$ is a $U(1)$ phase factor that forms a representation of $K$ \cite{ma2023aspt}, and $\mathscr{U}$ denotes symmetry superoperators which are linear maps in the space of operators, i.e., $\mathscr{U}$ maps operators to operators. 

Following the above definition, the action of a generic group element $\gamma=k\cdot g\in \Gamma$ is represented by the following superoperator: 
\begin{equation}
\mathscr{U}(\gamma)[\rho]=U(k\cdot g)\rho U^\dagger(g). 
\end{equation}
Here $U(\gamma)$ are unitary operators that form linear representations of $\Gamma$. Then it is straightforward to show that $\mathscr{U(\gamma)}$ also furnishes a linear representation of the symmetry group $\Gamma$ on the vector space of operators. That is, $\mathscr{U}(\gamma_1)\circ\mathscr{U}(\gamma_2) =\mathscr{U}(\gamma_1\gamma_2)$ ($\gamma_1\gamma_2$ is shorthand for $\gamma_1\cdot\gamma_2$, and we will omit the ``$\cdot$" hereafter). These linear maps define symmetry actions on mixed states.

Finite-depth local channels (FDLC) are a natural generalization of finite-depth local unitary circuits (FDLUC), which describe a generic locality-preserving evolution of mixed states. As proposed in reference \cite{hastings2011topological}, a general FDLC transformation can be constructed in the following steps: $a.$ Introduce additional qubits on each site (the environment), which defines an enlarged Hilbert space $\mathcal{H}^S_i\otimes \mathcal{H}^E_i$ on each site. The environment is initialized in some product state $|e\rangle_E$. $b.$ Apply an FDLUC to the total system $S\cup E$. $c.$ Trace out the environment. Thus a FDLC $\mathcal{N}$ can be written as
\begin{equation}
\mathcal{N}[\rho]=\Tr_E\{D\rho\otimes|e\rangle_E\langle e|_ED^\dagger\},
\end{equation}
where $D$ is a FDLUC. We note that the same channel can be realized using different purification schemes with different choices of $|e\rangle_E$ and $D$. Following \cite{deGroot2022spt,ma2023aspt,ma2023topological},  we define symmetries of local quantum channels as below.

\textbf{Definition 2} (Symmetries of local quantum channels). A local channel $\mathcal{N}$ has $\Gamma$ symmetry iff there exists a purification scheme, such that:
\begin{equation}
\begin{aligned}
[U(g)\otimes U_E(g),D]&=0,U_E(g)|e\rangle_E=\lambda_E(g)|e\rangle_E,\forall g\in G\\
&\text{(weak symmetry condition)},\\
[U(k)\otimes I_E,D]&=0,\forall k\in K & \\
&\text{(strong symmetry condition)},
\end{aligned}
\end{equation}
where $U_E(g)$ is a unitary representation of $G$, and $\lambda_E(g)$ is a $U(1)$ phase factor that forms a $1d$ representation of $G$. 

Under the above condition, it is easy to check that $\mathcal{N}\circ \mathscr{U}(\gamma)=\mathscr{U}(\gamma)\circ\mathcal{N}$, so $\mathcal{N}$ preserves the symmetry of states.

 A large class of open quantum systems can be effectively described under Markov approximation, which basically assumes that the environment has no memory. Then the evolution of density matrices is governed by the celebrated Lindblad equation (also known as the quantum master equation):

\begin{equation}
\begin{aligned}
\frac{d}{dt}\rho&=\mathcal{L}[\rho]\\
&\equiv-i[H,\rho]+\sum_\mu l_\mu\rho l_\mu -\frac{1}{2}\{l_\mu^\dagger l_\mu,\rho\}.
\end{aligned}
\end{equation}
Here the superoperator $\mathcal{L}$, known as the Lindbladian, is the generator of the nonequilibrium dynamics. $H$ is the Hamiltonian, and $l_\mu$, known as jump operators, describe coupling to the environment. Following \cite{Buca:2012symmetry,Lieu2020SB}, the symmetries of Lindbladians are defined as below. 

\textbf{Definition 3} (Symmetries of Lindbladians). A Lindbladian $\mathcal{L}$ has the $\Gamma$ symmetry iff $\mathcal{L}\circ\mathscr{U}(\gamma)=\mathscr{U}(\gamma)\circ\mathcal{L},\forall\gamma\in\Gamma$. 

It is often very useful to use the following sufficient conditions to diagnose the symmetry: The Lindbladian has the $\Gamma$ symmetry if
\begin{equation}
\begin{aligned}
&\mathcal{L} \text{ is invariant under } H\rightarrow U(g)HU^\dagger(g),\\
&l_\mu\rightarrow Ul_\mu U^\dagger(g),\forall g\in G \text{ (weak symmetry condition)}.\\
&[U(k),H]=0,[U(k),l_\mu]=0,\forall k\in K, \\
& \qquad\qquad\qquad\qquad\qquad\text{ (strong symmetry condition)}.
\end{aligned}
\end{equation}
Notably, the weak symmetry condition is both sufficient and necessary. 

As a final remark, we would like to emphasize one crucial difference between weak and strong symmetries: strong symmetries lead to conserved quantities, which are just the strong symmetry charges (generators); weak symmetries, on the other hand, are not tied with any conservation laws.
\subsection{Mixed-state quantum phases}
In this section we review the definition of mixed-state quantum phases proposed in recent papers \cite{ma2023aspt,ma2023topological,sang2023mixed}. The ideas stem from Hastings' work in 2011 \cite{hastings2011topological}. First, we consider the classification of phases without symmetries.

\textbf{Definition 4} (mixed-state quantum phases). Two mixed states $\rho_1$, $\rho_2$ belongs to the same phase if they are two-way connected by FDLCs, that is, $\exists$ a pair FDLCs $\mathcal{N}_{12},\mathcal{N}_{21}$, s.t. $\rho_2=\mathcal{N}_{12}[\rho_1],\rho_1=\mathcal{N}_{21}[\rho_2]$. Particularly, a state belongs to a (non)trivial phase if it is (not) two-way connected to a product state.

The above definition is clearly motivated by the classification of ground-state quantum phases of local gapped Hamiltonian under equivalence of FDLUC \cite{PhysRevB.82.155138,PhysRevB.83.035107}, and is thus natural for classification of mixed states with finite correlation length.  One subtlety is that an FDLC generally has no inverse, so we must require the two-way connectivity. Furthermore, just as symmetries can enrich the variety of quantum phases in pure states, we can also impose symmetry conditions on the above definition of mixed-state quantum phases.

\textbf{Definition 5} (Symmetric mixed-state quantum phases). Two mixed states with the $\Gamma$ symmetry belong to the same symmetric phase iff:
\begin{enumerate}
\item They are two-way connected by FDLCs with the $\Gamma$ symmetry. 
\item Each layer of the FDLCs satisfies the $\Gamma$ symmetry.
\end{enumerate}
Particularly, a state belongs to a (non)trivial $\Gamma$-symmetric phase if it can (not) be two-way connected to a $\Gamma$-symmetric product state. 

Interesting mixed-state topological phases including SPT and symmetry enriched topological order (SET) have been investigated under this framework \cite{ma2023aspt,ma2023topological}. Generically, we only consider onsite internal symmetries for mixed-state SPT and SETs, just like studies in pure states. One main reason is that for non-onsite internal symmetries, a symmetric product state may not even exist, so there is no natural choice of trivial state to start with. However, if we only impose weak symmetry conditions, $\Gamma=G$, then in principle we can allow it to have non-onsite action, and use the maximally mixed state to represent the trivial phase, which is indeed a symmetric product state.   

\section{Microscopic definition, calculation, and classification of anomalies for mixed states\label{sec:anomaly}}
In this section we perform the Else-Nayak type analysis of symmetry actions on bosonic mixed-states. In this approach, anomalies manifest as some obstructions to implementing the symmetry transformation on a subregion. Compared to other approaches of diagnosing anomalies, this approach has several advantages:
\begin{enumerate}
\item  Although the original definition of anomalies is an obstruction to gauging, the notion of gauging is still unclear for general open quantum systems. Hence we would like to avoid introducing gauge fields for general characterization. 
\item In this approach, anomaly is manifestly a purely kinematic property of the system, as it should be. Namely, it can be determined by given locality, a tensor-product Hilbert space, and the form of symmetry generators. No further information of the dynamics, including the form of the Hamiltonian/Lindbladian/partition function is needed. This allows us to apply the power of anomaly to various types of open quantum systems. For example, the mixed states may arise as a steady state of dissipative dynamics, which can be Markovian or non-Markovian, or it may arise from many-body states subjected to (short-time) decoherence or weak measurement.   
\end{enumerate}

Due to the above consideration, we believe that the Else-Nayak approach is the most natural and straightforward generalization from closed systems to open systems, which requires minimum assumption. Thus we take the obstruction to implementing symmetry transformations on subregions as the definition of anomalies for mixed states.

\subsection{The Else-Nayak approach to anomaly in pure states\label{sec:Nayak}}
To begin with, we briefly review the Else-Nayak approach in $(1+1)$-D bosonic closed systems with internal symmetry $\Gamma$ \cite{PhysRevB.90.235137}. This approach only works for symmetries with local unitary representation, i.e. the symmetry operator $U(\gamma)$ is an FDLUC. Let us first define restrictions of the symmetry transformations to a subregion $M$: $U_M\approx U$ inside $M$, i.e., $U_M$ and $U$ are the same in the interior of $M$, but are ambiguous up to some local unitaries near the boundary $\partial M$. Given the FDLUC nature of $U$, we have 
\begin{equation}
U_M(\gamma_1)U_M(\gamma_2)=W(\gamma_1,\gamma_2)U_M(\gamma_1\gamma_2),
\label{eq:W}
\end{equation}
for some local unitary $W(\gamma_1,\gamma_2)$ acting on $\partial M$. The associativity of $U_M(\gamma_1)U_M(\gamma_2)U_M(\gamma_3)$ dictates that $W$ should satisfy
\[\label{eq:associativity pure state}
W(\gamma_1,\gamma_2)W(\gamma_1\gamma_2,\gamma_3)=^{U_M(\gamma_1)}W(\gamma_2,\gamma_3)W(\gamma_1,\gamma_2\gamma_3),
\]
where we use the shorthand notation $^{U_M}W\equiv U_M W U_M^{-1}$. 
Then we further restrict $W$ to the left and right end of $M$:
\begin{equation}
W(\gamma_1,\gamma_2)=W_l(\gamma_1,\gamma_2)W_r(\gamma_1,\gamma_2).
\label{eq:Wl}
\end{equation}
$W_l$ and $W_r$ satisfies the consistent condition \eqref{eq:associativity pure state} up to a phase factor $\omega$:
\[
\begin{split}
&W_{l}(\gamma_1,\gamma_2)W_l(\gamma_1\gamma_2,\gamma_3)\\
=&\omega(\gamma_1,\gamma_2,\gamma_3)^{U_M(\gamma_1)}W_l(\gamma_2,\gamma_3)W_l(\gamma_1,\gamma_2\gamma_3),\\
&W_{r}(\gamma_1,\gamma_2)W_r(\gamma_1\gamma_2,\gamma_3)\\
=&\omega^{-1}(\gamma_1,\gamma_2,\gamma_3)^{U_M(\gamma_1)}W_r(\gamma_2,\gamma_3)W_r(\gamma_1,\gamma_2\gamma_3).
\end{split}
\label{eq:omega}
\]
From the associativity relation of $W_l(\gamma_1,\gamma_3)W_l(\gamma_1\gamma_2,\gamma_3)W_l(\gamma_1\gamma_2\gamma_3,\gamma_4)$, it can be shown that $\omega$ must satisfy the following 3-cocycle condition:
\begin{equation}
\begin{split}
&\omega(\gamma_1,\gamma_2,\gamma_3)\omega^{-1}(\gamma_1\gamma_2,\gamma_3,\gamma_4)\omega(\gamma_1,\gamma_2\gamma_3,\gamma_4)\\
&\omega^{-1}(\gamma_1,\gamma_1,\gamma_3\gamma_4)\omega(\gamma_2,\gamma_3,\gamma_4)=1.
\label{eq:cocycle pure state}
\end{split}
\end{equation}
Furthermore, as $W$ is invariant under 
\[
\begin{split}
W_l(\gamma_1,\gamma_2)&\rightarrow \beta(\gamma_1,\gamma_2)W_l(\gamma_1,\gamma_2),\\
W_r(\gamma_1,\gamma_2)&\rightarrow \beta(\gamma_1,\gamma_2)^{-1}W_r(\gamma_1,\gamma_2),
\end{split}
\label{eq:ambiguity}
\] 
$\omega$ is only uniquely defined modulo a 2-coboundary, i.e.,
\begin{equation}
\begin{split}
\omega(\gamma_1,\gamma_2,\gamma_3)\sim &\omega(\gamma_1,\gamma_2,\gamma_3)\beta(\gamma_1,\gamma_2)\beta(\gamma_1\gamma_2,\gamma_3)\\&\beta^{-1}(\gamma_2,\gamma_3)\beta^{-1}(\gamma_1,\gamma_2\gamma_3).
\label{eq:coboundary pure state}
\end{split}
\end{equation}
Consequently, the anomaly of internal symmetry $\Gamma
$ in (1+1)-D can be classified by $[\omega^\Gamma]$ ($[\cdot]$ denotes an equivalence class defined in \eqref{eq:coboundary pure state}), representing an element in the third cohomology group $H^3(\Gamma, U(1))$.

In higher dimensions, this approach only works for symmetries of the form 
\[
U(\gamma)=\sum_\alpha e^{i\mathcal{F}(\gamma)[\alpha]}|\gamma\alpha\rangle\langle \alpha|, 
\label{eq:sym_higherdim}
\]
where $|\alpha\rangle$ are product states that form a complete basis of the Hilbert space in $d$ spatial dimensions, and $\alpha\rightarrow\gamma\alpha$ defines an onsite symmetry action of $\Gamma$. Under such assumptions, the similar reduction procedure can be repeated, eventually leading to a $(d+2)$-cocycle $\omega(\gamma_1,\gamma_2,\cdots,\gamma_{d+2})$ with equivalence modulo a $(d+1)$-coboundary. Therefore anomalies of the $\Gamma$ symmetry in $d+1$ dimensions can be classified by $H^{d+2}(\Gamma,U(1))$.

\subsection{Generalization to mixed states\label{sec:Nayakmixed}}
\begin{figure}[htb]
 \centering
\includegraphics[width=1\linewidth]{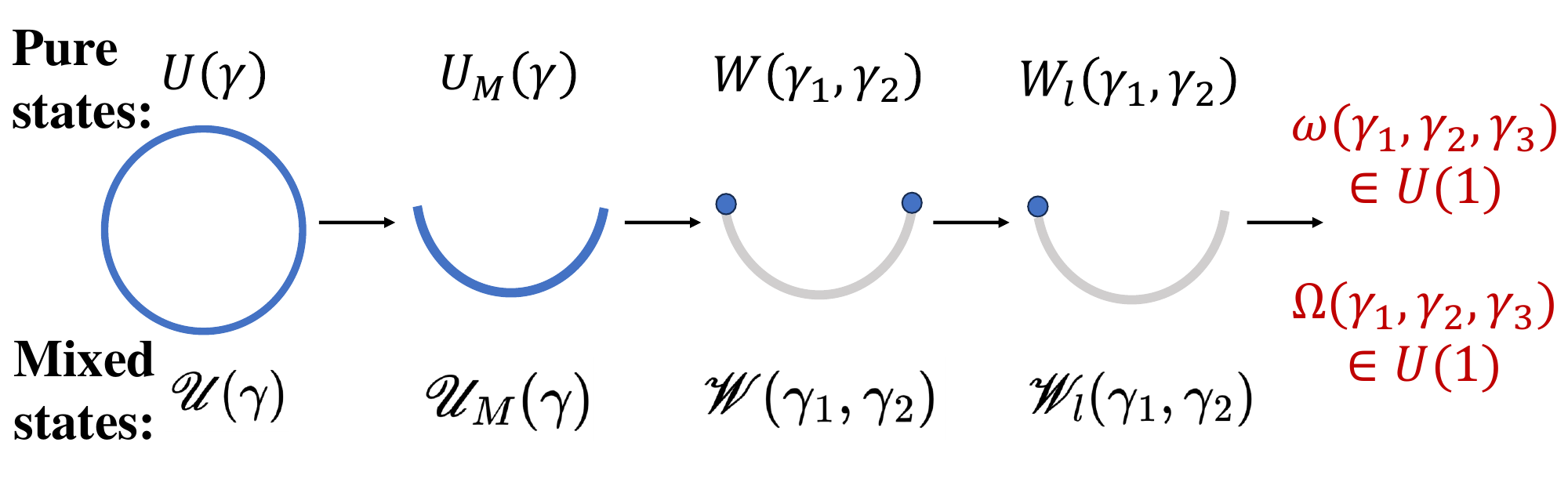}

\caption{The dimensional reduction procedure to calculate the anomaly in $(1+1)$-D.}

\label{fig:reduction}
\end{figure}

In this section, we further apply the Else-Nayak approach to mixed states. Here we consider bosonic systems with internal symmetries of the form $\Gamma=K\times G$, where $K,G$ denotes the strong and weak symmetry groups, respectively. 
In this case, the superoperator representation of symmetries we introduced in \ref{sec:sym} turns out to be very useful, which indicates a neat way of generalizing the Else-Nayak approach to mixed states. To start with, we consider $(1+1)$ dimensions and define restrictions of the symmetry transformations to a subregion $M$. $\mathscr{U}_M\approx \mathscr{U}$ inside $M$. More specifically, $\mathscr{U}_M(k\cdot g)[\rho]=U_M(k\cdot g)\rho U_M^\dagger(g)$, where $U_M$ is defined in Section \ref{sec:Nayak}. Then \eqref{eq:W}, \eqref{eq:associativity pure state}, \eqref{eq:Wl}, \eqref{eq:omega} can be naturally generalized into the superoperator version, as depicted in Fig.\ref{fig:reduction}. 
\begin{equation}
\mathscr{U}_M(\gamma_1)\circ\mathscr{U}_M(\gamma_2)=\mathscr{W}(\gamma_1,\gamma_2)\circ\mathscr{U}_M(\gamma_1 \gamma_2), 
\end{equation}
\begin{equation}\label{eq:associtivity mixed}
\mathscr{W}(\gamma_1,\gamma_2)\circ\mathscr{W}(\gamma_1\gamma_2,\gamma_3)=^{\mathscr{U}_M(\gamma_1)}\mathscr{W}(\gamma_2,\gamma_3)\circ\mathscr{W}(\gamma_1,\gamma_2\gamma_3),
\end{equation}
\begin{equation}
\mathscr{W}(\gamma_1,\gamma_2)=\mathscr{W}_l(\gamma_1,\gamma_2)\circ\mathscr{W}_r(\gamma_1\gamma_2),
\end{equation}
\[
\begin{split}
&\mathscr{W}_l(\gamma_1,\gamma_2)\circ\mathscr{W}_l(\gamma_1\gamma_2,\gamma_3)\\
&=\Omega(\gamma_1,\gamma_2,\gamma_3)^{\mathscr{U}_M(\gamma_1)}\mathscr{W}_l(\gamma_2,\gamma_3)\circ\mathscr{W}_l(\gamma_1,\gamma_2\gamma_3).
\end{split}
\label{eq:Omega_def}
\]
We use the shorthand notation $^{\mathscr{U}_M}\mathscr{W}\equiv \mathscr{U}_M\circ\mathscr{W}\circ\mathscr{U}^{-1}_M$ in Eq.~\eqref{eq:associtivity mixed}. Naturally, the phase factor $\Omega(\gamma_1,\gamma_2,\gamma_3)$ in \eqref{eq:Omega_def} can be viewed as the anomaly indicator of the mixed states, as a generalization of the $3$-cocycle $\omega(\gamma_1,\gamma_2,\gamma_3)$. Actually, $\Omega$ and $\omega$ are closely related, as we show below.

Since $\mathscr{U}_M$ is inherited from $U_M$, $\mathscr{W}$ is also determined by $W$:
\begin{equation}
\begin{split}
&\mathscr{W}(\gamma_1,\gamma_2)[O]\\
&=W(\gamma_1,\gamma_2)O W^\dagger(g_1,g_2),\forall \text{ operator }O\\
&=W_l(\gamma_1,\gamma_2)W_r(\gamma_1,\gamma_2)O W'^\dagger_r(g_1,g_2)W'^\dagger_l(g_1,g_2),
\end{split}
\label{eq:superW}
\end{equation}
where $W=W_lW_r=W'_lW'_r$. $W_{l(r)}$ and $W'_{l(r)}$ are identical as a function of $G$ up to a phase factor. Thus
\begin{equation}
\begin{split}
&\mathscr{W}_{l(r)}(\gamma_1,\gamma_2)[O]\\
&=W_{l(r)}(\gamma_1,\gamma_2)O W_{l(r)}'^\dagger(g_1,g_2),\forall \text{ operator }O,
\end{split}
\label{eq:superW}
\end{equation}
where we denote $\gamma_i=k_i g_i$. From \eqref{eq:omega} we can extract the cocycle $\omega,\omega'$ from $W_l,W'_l$ , respectively, and \eqref{eq:Omega_def} leads to
\begin{equation}
\Omega(\gamma_1,\gamma_2,\gamma_3)=\omega(\gamma_1,\gamma_2,\gamma_3)\omega'^{-1}(g_1,g_2,g_3).
\label{eq:Omega}
\end{equation}
Eq.~\eqref{eq:coboundary pure state} leads to the following equivalence relation of $\Omega$:
\begin{equation}
\begin{split}
&\Omega(\gamma_1,\gamma_2,\gamma_3)\sim \\
&\Omega(\gamma_1,\gamma_2,\gamma_3)\beta'^{-1}(g_1,g_2)\beta'^{-1}(g_1g_2,g_3)\beta'(g_2,g_3)\beta'(g_1,g_2g_3)\\
&\beta(\gamma_1,\gamma_2)\beta(\gamma_1\gamma_2,\gamma_3)\beta^{-1}(\gamma_2,\gamma_3)\beta^{-1}(\gamma_1,\gamma_2\gamma_3).\\
\end{split}
\end{equation}
In other words, $[\Omega^\Gamma]=[\omega^\Gamma][\omega^G]^{-1}$ (using the fact that $[\omega^G]=[\omega'^G]$). 

Recall that $[\omega^\Gamma],[\omega^G]$ are elements in $H^3(\Gamma,U(1)),H^3(G,U(1))$, respectively. From the K{\"u}nneth formula, the group cohomology $H^3(\Gamma=K\times G,U(1))$ can be decomposed as:
\[
H^3(\Gamma,U(1))=\sum^3_{k=0}(H^k(G,H^{3-k}(K,U(1)))),
\]
which contains $H^3(G,U(1))$ as a normal subgroup in the direct sum decomposition. Thus 

\[
\begin{split}
[\Omega^\Gamma] &\in H^3(\Gamma,U(1))/H^3(G,U(1))\\
&=\sum^2_{k=0}H^k(G,H^{3-k}(K,U(1))).
\end{split}
\]
Furthermore, for symmetries of the form in \eqref{eq:sym_higherdim}, we can perform the same analysis to extract the anomaly cocycle in higher dimensions
\[
\begin{split}
&\Omega(\gamma_1,\gamma_2,\cdots,\gamma_{d+2})\\
=&\omega(\gamma_1,\gamma_2,\cdots,\gamma_{d+2})\omega'^{-1}(g_1,g_2,\cdots,g_{d+2}).
\end{split}
\]
Then 
\[
\begin{split}
[\Omega^\Gamma] &\in H^{d+2}(\Gamma,U(1))/H^{d+2}(G,U(1))\\
&=\sum^{d+1}_{k=0}(H^k(G,H^{d+2-k}(K,U(1)))).
\end{split}
\]
In this way we arrive at the following definition.

\textbf{Definition 6.} (Mixed-state anomaly) The symmetry action of $\Gamma=K\times G$ on a mixed state $\rho$ is anomalous iff $[\Omega^\Gamma]$ is a nontrivial element in $H^{d+2}(\Gamma,U(1))/H^{d+2}(G,U(1))$, or more explicitly, it is $U(1)$ phase factor that cannot be removed by redefinition of $W_l,W_r$ as in \eqref{eq:ambiguity}.

The subgroup with $k=0$ corresponds to the pure anomalies of strong symmetry $K$ while the other subgroups with $1\leq k\leq d+1$ correspond to the mixed anomalies between $K$ and $G$. Physically, this mixed anomaly can be manifested by strong symmetry fractionalization on weak symmetry
defects, which we will discuss in Section~\ref{sec:lattice model}. Particularly, the absence of the $k=d+2$ term indicates that there is no anomaly with only weak symmetries. Indeed, $[\Omega]$ is always trivial when $\Gamma=G$. We will further justify this conclusion in the next section.

\section{Anomaly as a tool to constrain mixed-state quantum phases}\label{sec:anomaly constrain}

In this section, we will show that anomaly is powerful in constraining mixed-state quantum phases, which indicates that our definition of mixed-state anomalies is not only a natural one but also a useful one. 

 \textbf{Theorem 1.} A state $\rho$ with anomalous symmetry $K\times G$ cannot be prepared via a $G$-symmetric finite depth local quantum channel starting from a $G$-symmetric product state (by ``$G$-symmetric" we always mean weakly symmetric). Then, according to Definition 4, $\rho$ belongs to a nontrivial $G$-symmetric phase. 
 
{\it Proof.} We prove the contrapositive statement of the above theorem. Namely, we show that any state prepared via a $G$-symmetric FDLC from a $G$-symmetric pure product state $\rho_0$ must be anomaly free. Denote $\rho_0=\bigotimes_i\rho_i$, where $i$ is the site index. Then $\rho_0$ can be purified into the following product state: $|\psi_{SA}\rangle=\bigotimes_i|\psi_i\rangle=\bigotimes_i\sqrt{\rho_i}\otimes I^i_A(\sum_m|m^i\rangle_S\otimes|m^i\rangle_A),\rho_0=\Tr_A(|\psi_{SA}\rangle\langle\psi_{SA}|)$, where we introduce an ancilla Hilbert space $\mathcal{H}^i_A$ at each site $i$, which is isomorphic to the system Hilbert space $\mathcal{H}^i_S$ spanned by the complete orthonormal basis $\{|m^i\rangle\}$ \cite{lessa2024mixed}. Since $\rho_0$ is weakly symmetric, $U(g)\rho_0 U^\dagger(g)=\rho_0,\forall g\in G$, $|\psi_{SA}\rangle$ also has the $G$ symmetry:
\begin{equation}
U(g)\otimes U^*(g)|\psi_{SA}\rangle=|\psi_{SA}\rangle.
\end{equation}
Now suppose $\rho$ is prepared via the following channel:
\begin{equation}
\rho=\mathcal{N}[\rho_0]=\Tr_E(D_{SE}\rho_0\otimes|0\rangle_E\langle0|_ED_{SE}^\dagger),
\end{equation}
where $|0_E\rangle$ is a $G$-symmetric product state, $U_E(g)|0\rangle_E=e^{i\theta(g)}|0\rangle_E$, and $D_{SE}$ is some FDLUC with $G$ symmetry: $[D_{SE},U(g)\otimes U_E(g)]=0,\forall g\in G$. 

Then $\rho$ can be purified with the help of both the ancilla (A) and the environment (E):
\begin{equation}
\rho=\Tr_{E,A}(|\Psi\rangle\langle\Psi|),|\Psi\rangle=(D_{SE}\otimes I_A)|\psi\rangle_{SA}\otimes |0\rangle_E.
\end{equation}
It is easy to check that $|\Psi\rangle$ also satisfy the $G$ symmetry $U_{\text{tot}}(g)|\Psi\rangle=e^{i\theta(g)}|\Psi\rangle$, with the representation of $G$ on the total Hilbert space $\mathcal{H}^S\otimes \mathcal{H}^A\otimes\mathcal{H}^E$:
\begin{equation}
U_{\text{tot}}(g)=U(g)\otimes U^*(g)\otimes U_E(g).
\end{equation}

Next, if we further require $\rho$ has strong symmetry $K$, i.e., $U(k)\rho=\lambda(k)\rho$, then $|\Psi\rangle$ also has the same $K$ symmetry charge: $U(k)\otimes I_{AE}|\Psi\rangle=\lambda(k)|\Psi\rangle$, where $AE$ is shorthand for $A\cup E$. This can be easily verified by performing the Schmidt decomposition of $|\Psi\rangle$ under the bipartition $S\cup AE$. Therefore, $|\Psi\rangle$ has the full $\Gamma=K\times G$ symmetry, where the symmetry transformation is represented by 
$U_{\text{tot}}(k\cdot g)=U(k\cdot g)\otimes U^*(g)\otimes U_E(g)$. We assume $U_{\text{tot}}(\gamma)$ is an FDLUC supported on a $d$-dimensional spatial region. 
We can extract the $(d+2)$-cocycle $\omega_E(g_1,g_2,\cdots g_{d+2}), \omega_{\text{tot}}(g_1,g_2,\cdots g_{d+2})$ for $U_{E}$ and $U_{\text{tot}}$, respectively, in exactly the same way as we did for $U$ in Section \ref{sec:Nayak}. Due to the tensor product form of $U_{\text{tot}}$,the cocycles must satisfy the following relation:
\begin{equation}
\begin{aligned}
&\omega_{\text{tot}}(k_1g_1,k_2g_2,\cdots,k_{d+2}g_{d+2})\\
=&\omega(k_1g_1,k_2g_2,\cdots k_{d+2}g_{d+2})\\
&\omega^{-1}(g_1,g_2,\cdots,g_{d+2})\omega_E^{-1}(g_1,g_2,\cdots g_{d+2}).
\end{aligned}
\label{eq:omegatot}
\end{equation}

Then, since $|\psi_{SA}\rangle\otimes|0_E\rangle$ is a product state and $D$ is a FDLUC, $|\Psi\rangle$ is by definition short-range entangled (SRE), and thus must be anomaly free, so we can take $\omega_{\text{tot}}=1$. Taking $k_1=k_2=k_3=\cdots=k_{d+2}$ as the identity element of $K$ in \eqref{eq:omegatot}, we obtain $\omega_E=1$. Then comparing \eqref{eq:omegatot} and \eqref{eq:Omega}, we get $[\Omega]=[\omega_{\text{tot}}]=[1]$, so $\rho$ is anomaly free.$\qed$

We conclude this section with some remarks on the above theorem.
\begin{enumerate}
\item If the strong symmetry itself is anomalous, i.e., $\Omega(k_1,k_2,\cdots k_{d+2})(k_i\in K)$ represents a nontrivial element in $H^{d+2}(K,U(1))$, then we can take $G=1$ in the above theorem, which leads to the conclusion that the mixed state cannot be prepared via any FDLC from any product state. It is a straightforward generalization of the statement that a pure state preserving anomalous symmetry must be long-range entangled (LRE). We note that in this particular case a stronger result is proved in \cite{lessa2024mixed}.
\item As mentioned at the end of Section \ref{sec:Nayakmixed}, if $\rho$ only have weak symmetry $(K=1)$, it has no anomaly. Indeed, weak symmetry has little power in constraining mixed-state quantum phases --the maximally mixed state $\rho=I$, though a trivial product state, satisfies all types of weak symmetries, $U^\dagger I U=I$ \footnote{Nevertheless, in some recent papers anomalies of weak symmetries are discussed from other perspectives \cite{hsin2023anomalies,zang2023detecting}. There the anomaly has very different meanings than our notion.}.   
\item The most intriguing case is when the strong and weak symmetries have mixed anomalies, which is a phenomenon unique to open quantum systems. In this case, the naive generalization of "anomaly$\Rightarrow$LRE" fails——some mixed states with mixed anomaly can still be prepared (from product states) via an FDLC, and we give such an example in the next section. Nevertheless, as proved above, such preparation is prohibited once the weak symmetry condition is imposed. We also note that even for non-onsite weak symmetry $G$, a $G$-symmetric product state always exists in the mixed state, of which the maximally mixed state is a perfect example. Thus the generality of the above theorem is guaranteed.
\item In the above proof we only assume $D$ as a whole (not each layer of it) has the $G$ symmetry. Thus the conclusion is a bit stronger than a nontrivial $G$-symmetric phase. 
\item Finally but perhaps most importantly, in the above discussion a nontrivial phase is defined by impossibility of preparation via finite-depth local (symmetric) channels, based on which we establish the theorem ``anomaly $\Rightarrow$ nontrivial phase ". Then several questions arise naturally. What nontrivial phases are out there? How to diagnose them in a more direct way? For example, are there any physical quantities like correlation functions that can be used to distinguish them from trivial phases? We provide answers to these questions in the next question.
\end{enumerate}

\section{Lattice models with anomalous symmetry in open quantum systems}\label{sec:lattice model}
In the last section, we prove that mixed states preserving anomalous symmetries must belong to a nontrivial phase. As pointed out at the end of last section, it is desirable to find concrete examples of such anomaly-enforced phases and provide a more complete and physical description of them. For this purpose, we need to go beyond the kinematic-level discussion and further input the information of dynamics. For example, in studies of quantum phases in closed systems, of particular interest is the ground state of local Hamiltonians. Then it is known that if the Hamiltonian has anomalous symmetries, the ground states are guaranteed to be nontrivial, including the following possibilities: 
\begin{enumerate}[label=\alph*.]
\item Spontaneous symmetry breaking.
\item  Some local operators have a power-law correlation. In this case, the Hamiltonian has a gapless energy spectrum.
\item   Topological order. This is only possible in $(2+1)$-D and higher dimensions.

\end{enumerate}

In this context, we aim to establish a similar paradigm for open quantum systems. We focus on a prototypical type of dynamics, known as Markov dynamics, which means the environment is memoryless. As introduced in Section \ref{sec:sym} , such dynamics can be described by a Lindbladian $\mathcal{L}$. As a natural generalization of the paradigm in closed systems, we study several lattice models described by Lindbladians with anomalous symmetries, and show that as a consequence of Theorem 1, the system must have nontrivial long-time dynamics. 

The exact meaning of the ``nontrivial long-time dynamics" is twofold. Firstly, the steady states (defined as the zero modes of $\mathcal{L},\mathcal{L}[\rho^{ss}]=0$) 
must belong to some nontrivial mixed-state phases. Surprisingly, we identify a new type of nontrivial mixed-state phase enforced by anomalies, where correlation functions (including correlations of local operators and string order) in the bulk are all short ranged, but under OBC, the boundary states spontaneously break the symmetry. This is distinct from both scenarios in 1D pure states mentioned above. Also, it is distinct from the recently discovered mixed-state ASPT due to the lack of string order. Secondly, in each strong symmetry sector, either the steady states are degenerate or the relaxation time is divergent, or both. This actually directly follows from the nontrivialness of the steady state. Suppose the steady state is unique, then it must have the anomalous symmetry. Given that the Lindbladian can be viewed a continuous-time version of a $G$-symmetric quantum channel, with the evolution time replacing the channel depth, the relaxation time to reach the steady state must be divergent in the thermodynamic limit (since one can take the initial state to be a $G$-symmetric product state, in which case the conclusion is just a restatement of Theorem 1).

Below, we provide several examples of anomalous symmetries in open quantum systems. In all the examples below, we consider spin chains with a spin-1/2 degree of freedom on each site. In each example, we first define the symmetry action and demonstrate the anomaly according to our microscopic definition, and we also discuss some manifestations of anomalies. Then we construct Lindbladians with anomalous symmetries and discuss their properties.

{\it Example 1.} $K=U(1),G=\Z^X_2$.

Firstly, we consider the mixed anomaly between strong $U(1)$ symmetry and weak $\Z_2^X$ symmetry, generated by
\begin{equation}
\begin{aligned}
Q&:=\frac{1}{4}\sum_i(1-\sigma^z_i\sigma^z_{i+1}),\\
X&:=\prod_i\sigma^x_i,
\end{aligned}
\end{equation}
respectively. Notably, the generator $Q$ counts the number of domain walls of the $\Z_2^X$ symmetry, and the prefactor $\frac{1}{4}$ is to ensure that $Q\in \mathbb{Z}$ under periodic boundary conditions (PBC).

To calculate the anomaly cocycle, we focus on the $\Z_2$ subgroup of U(1) symmetry which is generated by: 
\[
U(\text{DW})=\exp(i\pi Q)=\exp[\frac{\pi i}{4}\sum_i(1-\sigma^z_i\sigma^z_{i+1})],
\]
where DW stands for domain walls.
We restrict it to a subregion $M$
\[
U_{M}(\text{DW})=\exp[\frac{\pi i}{4}\sum^{k-1}_{i=j}(1-\sigma^z_i\sigma^z_{i+1})],
\]
where $M$ includes sites ${j\leq i\leq k}$. Then we find
\[
(U_M(\text{DW}))^2=\sigma^z_j\sigma^z_k,
\]
which implies $W(\text{DW},\text{DW})=\sigma^z_j\sigma^z_k$ and $W_l(\text{DW},\text{DW})=\sigma^z_j$. Similarly we define $X_M=\prod^k_{i=j}\sigma^x_i$ and $U_M(X\text{DW})=X_MU_M(\text{DW})$. Then it follows $W_l(X\text{DW},X)=W_l(X,\text{DW})=W_l(I,X)=W_l(I,DW)=I,W_l(X\text{DW},X\text{DW})=W_l(X\text{DW},\text{DW})=\sigma^z_j$.
As the weak symmetry is onsite, its cocycle $\omega^G$ must be in the trivial class. To determine whether $[\Omega]$ belongs to a nontrivial element in $H^3(K\times G,U(1))/H^3(G,U(1))$, we compute the following gauge invariant combination (meaning that it is invariant under \eqref{eq:coboundary pure state}) of $\omega^\Gamma$:
\begin{equation}\label{eq:gauge-invar-DW}
\begin{aligned}
&\omega(X,\text{DW},\text{DW})\omega(X\text{DW},X,\text{DW})\omega(X\text{DW},X\text{DW},X)\cdot\\
&\omega(X,I,X)=-1\times1\times1\times 1=-1.
\end{aligned}
\end{equation}
Since the result is not unity and strong symmetry itself is anomaly free, it corresponds to the nontrivial cocycle in $H^2(G,H^1(K,U(1))$, signaling the mixed anomaly between $K$ and $G$.


One important and intuitive manifestation of anomaly in pure states is symmetry fractionalization on symmetry defects \cite{10.21468/SciPostPhys.15.2.051}. Here we show that this intriguing phenomenon also shows up for mixed states, even in the case of mixed anomaly between strong and weak symmetries (the generalization to mixed states is more obvious for anomalies of purely strong symmetries), with some minor modification. Specifically, the strong symmetry fractionalizes on weak symmetry defects, which can be viewed as a physical interpretation of a nontrivial cocycle $\Omega$. 

 To see this, we can start with a mixed state that spontaneously breaks the weak $\Z_2^X $ symmetry:
\begin{equation}
\rho=|\uparrow\uparrow\uparrow\cdots\uparrow\rangle\langle \uparrow\uparrow\uparrow\cdots\uparrow|+|\downarrow\downarrow\downarrow\cdots\downarrow\rangle\langle \downarrow\downarrow\downarrow\cdots\downarrow|.
\end{equation}
$\rho$ is invariant under the weak symmetry, $X\rho X=\rho$. However, it exhibits spontaneous weak symmetry breaking, reflected in the long range correlation $\langle \sigma_i^z\sigma_j^z\rangle=1$. $\rho$ is free of $\Z_2^X$ domain walls and $Q=0$. Now can we introduce weak symmetry defect (domain walls) by a string operator $\mathcal{S}=\prod_{j\leq i\leq k}\sigma^x_i$, which is a restriction of the symmetry transformation $X$ on the segment ${j\leq i\leq k}$.
\begin{widetext}
\begin{equation}
\begin{aligned}
&\rho\rightarrow \rho'=\mathcal{S}\rho\mathcal{S}\\
=&|\uparrow\cdots\uparrow_j\downarrow_{j+1}\cdots\downarrow_{k}\uparrow_{k+1}\cdots\uparrow\rangle\langle \uparrow\cdots\uparrow_j\downarrow_{j+1}\cdots\downarrow_{k}\uparrow_{k+1}\cdots\uparrow|\\
+&|\downarrow\cdots\downarrow_j\uparrow_{j+1}\cdots\uparrow_{k}\downarrow_{k+1}\cdots\downarrow\rangle\langle \downarrow\cdots\downarrow_j\uparrow_{j+1}\cdots\uparrow_{k}\downarrow_{k+1}\cdots\uparrow|.
\end{aligned}
\end{equation}
\end{widetext}
The string operator creates one $\Z_2^X$ defect at each end. It is easy to check that compared to $\rho$, the total change of $U(1)$ charge $Q$ is one. Thus each weak symmetry defect carries a half $U(1)$ charge. We note that although both the weak symmetry defect and strong symmetry charge are well-defined, there is no well-defined charge associated with weak symmetry. Thus it is questionable whether one can reversely consider weak symmetry fractionalization on strong symmetry defect. This is one major difference to anomalies in pure states. 

Below, we construct Lindbladians with the anomalous $U(1)\times \Z_2^X$ symmetry.
\begin{equation}
\begin{aligned}
H&=\sum_i{J(\sigma^x_i-\sigma^z_{i-1}\sigma^x_i\sigma^z_{i+1}})-\lambda \sigma^z_i\sigma^z_{i+1},\\
l_i&=\sqrt{r}\sigma^z_i.
\end{aligned}
\label{eq:example1}
\end{equation}
The Hamiltonian is a prototypical realization of the edge theory of the $(2+1)$-D Levin-Gu model \cite{PhysRevB.86.115109,10.21468/SciPostPhys.15.2.051}. The first term describes hopping of $\Z_2^X$ domain walls and the second term is the energy penalty of domain walls. It is invariant under both the $U(1)$ symmetry and $\Z_2^X$ symmetry. Here we consider the simplest form of jump operator $l_i$, whose effect is simply dephasing on each site. From the algebra $[Q,\sigma^z_i]=0,\{X,\sigma^z_i\}=0$, it is clear that the jump operator keeps the strong $U(1)$ symmetry and (partially) breaks the $\Z_2^X$ symmetry from strong to weak. Due to the strong $U(1)$ symmetry, the number of domain walls is conserved (see the end of section III). As a consequence, there must be at least one steady state in each charge sector $ Q=q$ (of the $U(1)$ symmetry). Below we investigate the properties of steady state(s) in each charge sector to reveal the consequence of anomalies.

First, we discuss the case $q=0$. This scenario is extremely simple and yet illuminating. Note that there are only two states in this sector, i.e., $|\text{all up}\rangle$ and $|\text{all down}\rangle$. With dephasing, the steady states are $2$-fold: $\rho^{ss}_1=|\text{all up}\rangle\langle\text{all up}|,\rho^{ss}_2=|\text{all down}\rangle\langle\text{all down}|$. The key point is that the two steady states spontaneously break the weak $\Z_2^X$ symmetry, and are ferromagnetically ordered. Their symmetric combination, $\rho_{+}=\frac{|\text{all up}\rangle\langle\text{all up}|+|\text{all down}\rangle\langle \text{all down}|}{2}$, is also a valid steady state, and satisfies the weak symmetry $X\rho_{+}X=\rho_{+}$. Still, the weak symmetry breaking is reflected in the long-range correlation $\lim_{|i-j|\rightarrow\infty}\langle \sigma^z_i\sigma^z_j\rangle-\langle\sigma^z_i\rangle\langle\sigma^z_j\rangle=1$, given that $\sigma^z_i$ is odd under the $\Z_2^X$ transformation. From this perspective, $\rho_{+}$ can be viewed as the mixed-state counterpart of the GHZ state. The case $q=q_{\text{max}}$ (for a periodic chain with size $L$, $q_{\text{max}}=L/2$) is almost identical to $q=0$, except the steady states have Neel order in this case.

Now we turn to the more nontrivial case with intermediate densities of domain wall $q=\alpha L(0<\alpha<1/2)$. Under PBC, the steady state is unique, which is the maximally mixed state in each sector, $\rho^{ss}\propto P_{Q=q}$, where $P_{Q=q}$ is the projector to the eigenspace of $Q$ with eigenvalue $q$. It is easy to check that $\mathcal{L}[P_{Q=q}]=0$ indeed, by noting that $[P_{Q=q},l_i]=0,[P_{Q=q},H]=0$. From Theorem 1, $\rho^{ss}$ must belong to a nontrivial $\Z_2^X$-symmetric phase, and the relaxation time to reach the steady state must be divergent in the thermodynamic limit (see the general discussion at the beginning of this section). On the other hand, in this steady state, all local operators are short-range correlated, fitting into none of the cases listed at the beginning of this section. Then how is this a nontrivial phase?  

To answer this question, we investigate this model under OBC. We take
\begin{equation}
\begin{aligned}
Q_{\text{OBC}}&=\frac{1}{4}\sum_{i=1}^{L-1}(1-\sigma^z_i\sigma^z_{i+1}),\\
H_{\text{OBC}}&=J\sum_{i=2}^{L-1}(\sigma^x_i-\sigma^z_{i-1}\sigma^x_i\sigma^z_{i+1})-\lambda\sum_{i=1}^{L-1}\sigma^z_i\sigma^z_{i+1}.\\
\end{aligned}
\end{equation}
In this case the steady states are 2-fold degenerate for each $q$-sector.
\begin{equation}
\begin{aligned}
 &\rho^{ss}_{\text{OBC},1}\propto P_{Q_{\text{OBC}}=q}(I+
 \sigma^z_1)(I+(-1)^{2q}\sigma^z_L),\\
 &\rho^{ss}_{\text{OBC},2}=X\rho^{ss}_{\text{OBC},1}X\propto P_{Q_{\text{OBC}}=q}(I-\sigma^z_1)(I-(-1)^{2q}\sigma^z_L).
 \end{aligned}
 \end{equation}
Both $\rho^{ss}_{1},\rho^{ss}_2$ spontaneously breaks the weak $\Z^X_2$ symmetry on the boundary with nonzero magnetization, but there is no magnetization away from the boundary (in the thermodynamic limit $L\rightarrow \infty$). We can also consider their symmetric combination
 \begin{equation}
 \begin{aligned}
 \rho^{ss}_{\text{OBC},+}&=\frac{\rho_1+\rho_2}{2}\propto P_{Q=q}(I+(-1)^{2q}\sigma^z_1\sigma^z_L),\\
 X\rho^{ss}_{\text{OBC},+}X&=\rho^{ss}_{\text{OBC},+}.
 \end{aligned}
 \end{equation}
Then the boundary SSB is shown in the long-range correlation between the boundary spins $\langle \sigma^z_1\sigma^z_L\rangle=(-1)^{2q}$. 

Although in the above we consider a particular choice of $H_{\text{OBC}}$, i.e., simply dropping terms near the boundary, we show below that the boundary SSB is actually an unavoidable consequence of the anomalous symmetry. Consider any state $\rho$ that has the full $\Gamma$ symmetry: $Q_{\text{OBC}}\rho=q\rho,X\rho X=\rho$. Then the boundary correlation is guaranteed:
\begin{equation}
\begin{aligned}
\langle\sigma^z_1\sigma^z_L\rangle =\langle \exp(2\pi i Q_{\text{OBC}})\rangle=(-1)^{2q}.
\end{aligned}
\end{equation}
Due to the boundary SSB, for Lindbladians with the $\Gamma$ symmetry, the steady states in each $q$-sector must be at least 2-fold degenerate under OBC. Later, we will see more examples of this boundary SSB phenomenon.

{\it Example 2. $K=\Z^{\text{CZ}}_2,G=\Z^X_2$.}

The second example is the mixed anomaly between strong $\Z^{\text{CZ}}_2$ symmetry and weak $\Z_2^X$ symmetry. The strong symmetry generator is
\[
U(\text{CZ})=\prod_{i}\text{CZ}_{i,i+1}=\prod_{i}\exp[\frac{\pi i}{4}(1-\sigma^z_{i+1})(1-\sigma^z_{i+1})],
\]
which is the product of controlled-$Z$ gates on each pair of neighboring sites. We have $[U(\text{CZ})]^2=1$ and $[U(\text{CZ}),X]=0$ under PBC.

To calculate the anomaly cocycle, we restrict both strong and weak symmetries to a subregion $M$
\[
\begin{split}
U_M(\text{CZ})=\exp[\sum^{k-1}_{i=j}\frac{\pi i}{4}(1-\sigma^z_{i})(1-\sigma^z_{i+1})], X_M=\prod^k_{i=j}\sigma^x_i.
\end{split}
\]
It is easy to check $U_M(\text{CZ})X_M=(-1)^{k-j}\sigma^z_j \sigma^z_k  X_MU_M(\text{CZ})$. To identify $W$, we can restrict the combination of $U(\text{CZ})$ and $X$ to $M$ as
\[
U_M(\text{CZ}X)=U_M(X\text{CZ})= U_M(\text{CZ})X_M.
\]
Thus $W(\text{CZ},X)=I$ and $W(X,\text{CZ})=W(X,\text{CZ}X)=(-1)^{k-j}\sigma^z_j \sigma^z_k$. After further restriction to the left end, we can take $W_l(\text{CZ}X,X)=I$ and $W_l(X,\text{CZ})=W_l(X,\text{CZ}X)=(-1)^j\sigma^z_j$. Similarly, we also find $W_l(\text{CZ},\text{CZ})=W_l(\text{CZ},\text{CZ}X)=W_l(X,X)=W_l(I,X)=I$ and $W_l(\text{CZ}X,\text{CZ})=W_l(\text{CZ}X,\text{CZ}X)=(-1)^j\sigma^z_j $. Then we compute the following gauge invariant combination of $\omega^\Gamma$ where we replace the DW in eq~\eqref{eq:gauge-invar-DW} with CZ:
\begin{equation}\label{eq:gauge-invar-CZ}
\begin{aligned}
&\omega(X,\text{CZ},\text{CZ})\omega(\text{CZ}X,X,\text{CZ})\omega(\text{CZ}X,\text{CZ}X,X)\cdot\\
&\omega(X,I,X)=1\times(-1)\times1\times 1=-1.
\end{aligned}
\end{equation} 
This nontrivial  phase also signals the mixed anomaly cocycle which belongs to $H^2(G,H^1(K,U(1))$. As in the previous example, the mixed anomaly between strong $\Z^{\text{CZ}}_2$ and weak $\Z_2^X$ symmetry also leads to boundary SSB. Consider the symmetric state $
\rho$ under OBC:
\begin{equation}
U_{\text{OBC}}(\text{CZ})\rho=c\rho,X\rho X=\rho,(c=\pm 1),
\end{equation}
where $U_{\text{OBC}}(\text{CZ})=\exp[\sum^{L-1}_{i=1}\frac{\pi i}{4}(1-\sigma^z_{i})(1-\sigma^z_{i+1})]$. First, note that $U_{\text{OBC}}(\text{CZ})$ no longer commutes with $X$. Instead,
\begin{equation}
U_{\text{OBC}}(\text{CZ})XU_{\text{OBC}}(\text{CZ})=(-1)^{L-1}\sigma^z_1\sigma^z_LX.
\end{equation}
Thus
\begin{equation}
\begin{aligned}
&\langle \sigma^z_1 \sigma^z_L\rangle\equiv\Tr(\sigma^z_1\sigma^z_L\rho)=\Tr(\sigma^z_1\sigma^z_LX\rho X)\\
=&(-1)^{L-1}\Tr[U_{\text{OBC}}(\text{CZ})XU_{\text{OBC}}(\text{CZ})\rho X]\\
=&(-1)^{L-1}.
\end{aligned}
\label{eq:BSSB2}
\end{equation}
Therefore, in the presence of $\Z_2^{\text{CZ}}$ symmetry, the boundary spin must break the weak $\Z_2^X$ symmetry.  
For example, we consider the following Lindbladian with the anomalous  $\Z^{\text{CZ}}_2\times \Z_2^X$ symmetry \footnote{The $J_1$ term has a global $U(1)$ symmetry generated by $\sum_i(-1)^i\sigma^z_i\sigma^z_{i+1}$which contains the $Z_2^{\text{CZ}}$ subgroup. We break the $U(1)$ symmetry down to $Z_2$ by adding the $J_2$ term.}:
\begin{equation}
\begin{aligned}
H&=\sum_i{J_1(\sigma^x_i+\sigma^z_{i-1}\sigma^x_i\sigma^z_{i+1}})\\
&+J_2(\sigma^x_{i-1}\sigma^x_{i+1}+\sigma^z_{i-2}\sigma^x_{i-1}\sigma^x_{i+1}\sigma^z_{i+2}),\\
l_i&=\sqrt{r}\sigma^z_i.
\end{aligned}
\end{equation}
 Without loss of generality, we investigate the steady state in the even $\Z_2^{\text{CZ}}$ sector, $U(\text{CZ})\rho=\rho$ and we take $L\in 2\mathbb{Z}$ . Under PBC, there is only one steady state (in this sector), which is the maximally mixed state in this sector:
 \begin{equation}
 \rho^{ss}_{\text{PBC}}\propto I+U(\text{CZ}).
 \end{equation}
Under OBC (as in the previous example, we define $H_{\text{OBC}}$ by dropping terms cut by the boundary), on the other hand, the steady states are $4$-fold:
\begin{equation}
\begin{aligned}
\rho^{ss}_{\text{OBC},1}&\propto (I+\sigma^z_1)(I+\sigma^z_L)[I+U_{\text{OBC}}(\text{CZ})],\\
\rho^{ss}_{\text{OBC},2}&\propto (I-\sigma^z_1)(I-\sigma^z_L)[I+U_{\text{OBC}}(\text{CZ})],\\
\rho^{ss}_{\text{OBC},3}&\propto (I+\sigma^z_1)(I-\sigma^z_L)[I+U_{\text{OBC}}(\text{CZ})],\\
\rho^{ss}_{\text{OBC},4}&\propto (I-\sigma^z_1)(I+\sigma^z_L)[I+U_{\text{OBC}}(\text{CZ})].\\
\end{aligned}
\end{equation}
They all break the $\Z_2^X$ symmetry on the boundary.  Only the symmetric combination of $\rho^{ss}_{\text{OBC},3}$ and $\rho^{ss}_{\text{OBC},4}$ leads to a $\Z_2^X$ symmetric state $\rho^{ss}_{\text{OBC},+}\propto(I-\sigma^z_1\sigma^z_L)[I+U_{\text{OBC}}(\text{CZ})]$, and the boundary SSB is manifested by the boundary correlation $\langle \sigma^z_1\sigma^z_L\rangle_{+}=-1$. 

Despite the boundary SSB, any physical observables have trivial correlations in the bulk. To show this, we calculate the reduced density matrix for a segment  $M=[\alpha L,(1-\alpha)L]$ in the bulk ($0<\alpha<\frac{1}{2}$), and find that $\rho_M\propto I$ in the limit $L\rightarrow \infty$. Thus the steady states of this model realize a new many-body phase, which, as far as we know, has not been discussed in previous literature. First, it resembles none of the three types of (pure-state) phases with anomalous symmetry listed at the beginning of this section. Furthermore, it is also very different from the recently proposed mixed-state SPT due to the following reasons: 1. There is no string order in the bulk; 2. According to the classification in \cite{ma2023aspt,ma2023topological}, there is no nontrivial SPT solely protected by weak symmetries, but based on the theorem in last section, $\rho^{ss}_{\text{PBC}}$ must belong to a nontrivial mixed-state quantum phase when the weak $\Z_2^X$ symmetry is imposed.

{\it Example 3. $K=\Z_2^X,G=\Z_2^{\text{CZ}}$}

In this case, as $\Z^X_2$ is strong symmetry while $\Z_2^{\text{CZ}}$ is weak symmetry, we simply need to exchange their positions in gauge invariant combination \eqref{eq:gauge-invar-CZ}.
That is
\begin{equation}
\begin{aligned}
&\omega(\text{CZ},X,X)\omega(\text{CZ}X,\text{CZ},X)\omega(\text{CZ}X,\text{CZ}X,\text{CZ})\cdot\\
&\omega(\text{CZ},I,\text{CZ})=1\times 1\times (-1)\times 1=-1.
\end{aligned}
\end{equation} 
Since the cocycle of $\Z_2^{\text{CZ}}$ is also trivial, this nontrivial phase signals the mixed anomaly between $K=\Z^X_2$ and $G=\Z_2^{\text{CZ}}$.
In this case, the boundary must break the strong $\Z_2^X$ symmetry under OBC. Consider the symmetric state $
\rho$,
\begin{equation}
U_{\text{OBC}}(\text{CZ})\rho U_{\text{OBC}}=\rho,X\rho =c\rho,(c=\pm 1).
\end{equation}
Then 
\begin{equation}
\begin{aligned}
\Tr(\sigma^z_1\sigma^z_L\rho)&=(-1)^{L-1}\Tr(XU_{\text{OBC}}(\text{CZ})XU_{\text{OBC}}(\text{CZ})\rho)\\
&=(-1)^{L-1}\Tr(XU_{\text{OBC}}(\text{CZ})\rho XU_{\text{OBC}}(\text{CZ}))\\
&=(-1)^{L-1}.
\end{aligned}
\label{eq:BSSB3}
\end{equation}
Below we construct a Lindbladian with the anomalous $\Z_2^X\times \Z_2^{\text{CZ}}$ symmetry:

\begin{equation}
\begin{aligned}
H&=\sum_i{J(\sigma^x_i+\sigma^z_{i-1}\sigma^x_i\sigma^z_{i+1}}),\\
l_i&=\sqrt{r}(\sigma^x_i-\sigma^z_{i-1}\sigma^x_i\sigma^z_{i+1}).
\end{aligned}
\end{equation}
The jump operator breaks the strong $\Z_2^{\text{CZ}}$ symmetry but preserves a residual weak $\Z_2^{\text{CZ}} $ symmetry, because $\{l_i,U(\text{CZ})\}=0$. Below we investigate the steady state in the even $\Z_2^X$ sector $X=1$, and take $L\in 2\Z$. 

Under PBC, the steady state is the maximally mixed state in this sector $\rho^{ss}_{\text{PBC}}= I+X$. This is an example where the naive generalization of ``anomaly$\Rightarrow$LRE" fails, as promised in last section. Despite the anomalous $\Z_2^X\times \Z_2^{\text{CZ}}$ symmetry, $\rho^{ss}_{\text{PBC}}$ can be prepared via an FDLC, for example, through the following procedure: 
\begin{enumerate}
\item Introduce an additional qubit $\tau_i$ on each site $i$, and initialize all spins to the $x$ direction, $\tau^x_i=\sigma^x_i=1$.
\item   Apply the controlled-Z gates on onsite pairs $(\sigma_i,\tau_i)$ and neighboring pairs $(\tau_i,\sigma_{i+1})$, which is a depth-$2$ LUC. 
\item  Trace out the $\tau$ qubits. 
\end{enumerate}
However, in the above construction, the initial states break the weak $\Z_2^{\text{CZ}}$ symmetry. After imposing the weak symmetry condition (to both the initial states and channels), $\rho^{ss}_{\text{PBC}}$ can no longer be prepared by FDLC. This justifies the necessity of the weak symmetry condition in Theorem 1.

Under OBC, the steady states are $2$-fold degenerate in the $X=1$ sector (and likewise in the sector $X=-1$).
\begin{equation}
\begin{aligned}
\rho^{ss}_{\text{OBC},+}&= (I+\sigma^z_1\sigma^z_L)(I+X),\\
\rho^{ss}_{\text{OBC},-}&= (I-\sigma^z_1\sigma^z_L)(I+X).
\end{aligned}
\end{equation}
They both show boundary SSB of the strong $\Z_2^X$ symmetry, $\langle\sigma^z_1\sigma^z_L\rangle_{\pm}=\pm 1$. Moreover, $
\rho^{ss}_{\text{OBC},+}$ also spontaneously break the weak $\Z_2^{\text{CZ}}$ symmetry, $U_{\text{OBC}}(\text{CZ})\rho^{ss}_{\text{OBC},+}U_{\text{OBC}}(\text{CZ})=(I+\sigma^z_1\sigma^z_L)(I-X)$, which is a steady state in the $X=-1$ sector. Thus the boundary weak $\Z_2^{\text{CZ}}$ SSB is responsible for the $2$-fold degeneracy in both sectors $X=\pm 1$. 

{\it Discussions.}
As a nontrivial consequence of anomaly, in all three examples under PBC, the $\Gamma$-symmetric steady states cannot be prepared using a $G$-symmetric FDLC from a $G$-symmetric product state. For {\it Example 3} we explicitly show that such preparation exists if the weak symmetry condition is relaxed. However, for {\it Example 1, 2} we do not find any way to prepare $\rho^{ss}_{\text{PBC}}$ using an FDLC from any product state. Here we conjecture that for onsite symmetry $G$, Theorem 1 could be strengthened to the following form.

\textbf{Conjecture.}   A state $\rho$ with anomalous symmetry $K\times G$ with onsite symmetry action of $G$ cannot be prepared via any FDLC from any product state.

This conjecture is supported by the results from \cite{ma2023aspt,ma2023topological} that there is no SPT protected only by (onsite) weak symmetries. If any counter example of the conjecture exists, then that state can be prepared by an FDLC but is a nontrivial phase solely protected by onsite weak symmetry $G$, which would be extremely intriguing and beyond the decorated domain wall picture.  

In the three examples, we leave the scenario of critical states with power law correlation. This is actually also possible for mixed states with anomalies. For example, one can start with the ground state of the Hamiltonian in \eqref{eq:example1}. In the case $-1<\lambda< 1$, it lies in a critical phase. Then we can apply a dephasing channel (with Kraus operator $K_i\propto \sigma^z_i$) to that state, breaking the strong $\Z_2^X$ symmetry to a weak symmetry, which still has a mixed anomaly with the strong $U(1)$ symmetry. In this case, the mixed states exhibit power law
correlation, $\langle \sigma^z_i\sigma^z_j\rangle\sim \frac{1}{|i-j|^\alpha}(\alpha>0)$, since the dephasing channel does not change correlations between $\sigma^z$.

Finally, in all three examples, we encounter the boundary SSB phenomena. By making a natural choice of the symmetry operators under OBC, we show that the boundary SSB is inevitable as long as the symmetry is preserved. Then it is desirable to find out whether bondary SSB is a generic consequence of anomaly, and does not depend on how the symmetry operator is defined under OBC. For pure states, this kind of result has been proved in the context of boundary conformal field theories \cite{Wang:2013yta,PhysRevB.96.125105,Jensen:2017eof,Numasawa:2017crf,Li:2022drc,Choi:2023xjw}, where anomaly is an obstruction for the existence of symmetric conformal boundary conditions. 
Moreover, this kind of result has been also studied in higher dimensional quantum field theories by considering fusions of symmetry defects \cite{Thorngren:2020yht}, where anomalous symmetries must be explicitly or spontaneously broken for boundary theories.  In the next section, we aim to focus on the lattice framework on $(1+1)$-D and generalize the discussion to generic mixed states.   

\section{Generalities of anomaly-enforced boundary correlation}\label{sec:boundary cor}
In this section, we (partially) resolve the issue raised at the end of last section. Here we focus on $(1+1)$-D systems, and prove under certain conditions that anomalies would enforce nontrivial boundary correlation.
First we illustrate the general ideas. Recall that anomalies are characterized by an obstruction to implementing the (truncated) symmetry transformation $\mathscr{U}_M$ on a subregion $M$. Under OBC, however, the symmetry (super)operator is inevitably truncated by the boundary, and thus $\mathscr{U}_{\text{OBC}}$ is just a special case of $\mathscr{U}_M$, and only furnish a representation of the symmetry up to some boundary obstruction. Under many circumstances, such boundary obstruction leads to nontrivial correlations between the left and right edges. 

Most strikingly, we find that the above statement holds generally for anomalous strong symmetries.

\textbf{Theorem 2.} For an arbitrary $(1+1)$-D state $\rho$ with anomalous strong symmetry $K$, there must exist long-range boundary correlation (under OBC), i.e., $\exists$ $O_l,O_r$ supported near the left and right edges, respectively, s.t. $\langle O_lO_r\rangle-\langle O_l\rangle\langle O_r\rangle\neq 0$, even in the thermodynamic limit.

{\it Proof.} As pointed out above, we can repeat the analysis in Section \ref{sec:Nayak}, with $U_M$ replaced by $U_{\text{OBC}}$: 
\[
U_{\text{OBC}}(k_1)U_{\text{OBC}}(k_2)=W(k_1,k_2)U_{\text{OBC}}(k_1\cdot k_2),
\label{eq:OBC1}
\]

\[
{U}_{\text{OBC}}(k_1){U}_{\text{OBC}}(k_2)={W}(k_1,k_2){U}_{\text{OBC}}(k_1\cdot k_2),
\label{eq:OBC2}
\]

\[
 {W}(k_1,k_2)={W}_l(k_1,k_2){W}_r(k_1k_2),
 \label{eq:OBC3}
\]

\[
\begin{split}
&W_l(k_1,k_2) {W}_l(k_1k_2,k_3)\\
&=\omega(k_1,k_2,k_3)^{{U}_{\text{OBC}}(k_1)}{W}_l(k_2,k_3){W}_l(k_1,k_2k_3).
\end{split}
\label{eq:W_l}
\]
The strong symmetry of $\rho$ states that ${U}_{\text{OBC}}(k)\rho=\lambda(k)\rho\quad (\forall k \in  K)$ for some $U(1)$ phase factor $\lambda(k)$, so $\rho$ must be strongly symmetric under ${W}$: ${W}(k_1,k_2)\rho=\lambda(k_1)\lambda(k_2)\lambda^{-1}(k_1\cdot k_2)\rho$. 

Next, we prove by contradiction  that $\exists k, k'$, s.t. $\rho$ cannot be strongly symmetric under ${W}_l$. Assume that ${W}_l(k,k')\rho=\beta(k,k')\rho,\forall k,k'\in K$, for some U(1) phase factors $\beta$. Then by applying the left hand side of \eqref{eq:W_l} on $\rho$, we obtain 
$\omega(k_1,k_2,k_3)=\beta(k_1,k_2)\beta(k_1k_2,k_3)\beta^{-1}(k_1,k_2k_3)\beta^{-1}(k_2,k_3)$, which is a 2-coboundary. It then contradicts the condition that $K$ is anomalous. 

Furthermore, due to the unitarity of $W_l$, all of its eigenvalues are distributed on a unit circle on the complex plane, so we can conclude from the above that $\exists k,k'\in K,| \langle W_l(k,k')\rangle|\equiv |\Tr(W_l(k,k')\rho)|<1.$ The same goes for $W_r$. Therefore,

\[
|\langle W_l(k,k')W_r(k,k')\rangle-\langle W_l(k,k')\rangle\langle W_r(k,k')\rangle|>0,
\]
which proves the long-range correlation between left and right edges by construction. $\qed$

Some comments regarding the above theorem are in order. Firstly, for pure states, all symmetries are automatically strong symmetries, so the above theorem can be generally applied to pure states with anomalous symmetries \footnote{Although the main goal of this paper is to study novel aspects of anomalies in mixed states, we believe the implication of Theorem 2 on pure states has also not been obtained before, and is interesting in its own right.}.
Secondly, although in Section \ref{sec:lattice model} we focus on models with strong-weak mixed anomalies, the analysis regarding boundary SSB, e.g., \eqref{eq:BSSB2}, \eqref{eq:BSSB3} still go through when we promote the $G$ symmetry also to a strong symmetry. For example, we can replace the dissipator in {\it Example 1, Example 2} with $l_i=\sigma^z_i\sigma^z_{i+1}$, thus preserving the strong $\Z_2^X$ symmetry. Then under OBC the strong $\Z_2^X$ symmetry must break spontaneously on the boundary, with qubits on the two edges forming an EPR pair.

For the case with strong-weak mixed anomalies, we no longer have a proof of complete generality. Instead, we can prove the following weaker versions of the theorem.

\textbf{Theorem 3.} For a $(1+1)$-D state $\rho$ with anomalous symmetry $\Gamma=K\times G$, if the weak symmetry $G$ has onsite action, there must exist long-range boundary correlation, $\langle O_lO_r\rangle-\langle O_l\rangle\langle O_r\rangle\neq 0$, even in the thermodynamic limit. 

The proof of Theorem 3 is almost identical to Theorem 2, and is left to Appendix \ref{app:boundary}. The boundary weak $\Z^X_2$ SSB in {\it Example 1, 2} are both consequences of theorem 3. For generic weak symmetry $G$, it remains unclear whether such strong conclusions still hold generally, and we are only able to show the existence of nontrivial boundary correlation in the form of Renyi-2 correlators, which is also encountered in the "strong-to-weak SSB" phenomenon \cite{lee2023quantum,ma2023topological}. See Appendix \ref{app:boundary} for discussions of this case.


\section{Steady-state average SPT and anomalies of the edge theory\label{sec:ssaspt}}

In closed systems, anomalies can manifest in the boundary theories of SPT phases, known as "anomaly inflow mechanism". Recently, SPT phases have been generalized to the open quantum systems with decoherence, under the notation of decohered average SPT (ASPT) phases \footnote{In \cite{ma2023aspt,ma2023topological} generalization of SPT to both disordered and decohered systems are investigated. For the latter case that is more relevant to this paper, exact(average) symmetry has the same meaning as strong(weak) symmetry.}. 
Using the decorated domain wall (DDW) approach, various types of decohered ASPT (or simply ASPT hereafter) phases are constructed and diagnosed, with a focus on their bulk properties \cite{ma2023aspt,lee2022symmetry,zhang2022strange,ma2023topological}. Notably, for symmetries of the form $\Gamma=K\times G$, the ASPT phases in $(d+2)$ spacetime dimensions are classified by $H^{d+2}(\Gamma,U(1))/H^{d+2}(G,U(1))$, which exactly coincides with our classification of mixed-state anomalies in $(d+1)$ dimensions. Therefore, it is more than natural to expect a correspondence between ASPTs and anomalies of its edge theory, generalizing the bulk-edge correspondence in closed systems. 

In this section, we apply the DDW construction to Lindbladians, thus realizing ASPT in the steady states of the quantum Markov dynamics. In this way, we can facilitate a clearer and more straightforward discussion of the boundary theory and its symmetries. As an illustration, we present constructions of 2+1-dimensional $\Z_2\times \Z_2$ ASPT states in Lindbladians whose edge theories possess the same anomalous symmetries as the {\it Example 2} and {\it Example 3} in Section \ref{sec:lattice model}. In appendix~\ref{app:1+1d ASPT}, we also study an example of $(1+1)$-D steady-state ASPT.

\subsection{$(1+1)$-D $\Z_2\times \Z_2$ average SPT }\label{sec:1+1dASPT}
As a warm up, we briefly review the construction of a $(1+1)$-D ASPT proposed in \cite{ma2023aspt,ma2023topological,zhang2022strange,lee2022symmetry}. Consider a spin-$\frac{1}{2}$ chain with strong (weak) $\Z_2$ symmetries generated by spin flips on even (odd) sites:
\[
U_{K}=\prod_{i}\sigma^x_{2i}, \quad U_{G}=\prod_{i}\tau^x_{2i-1},
\]
where we denote the spins on even (odd) sites as $\sigma(\tau)$. Then $\rho_{\text{trivial}}=\rho_\sigma^\rightarrow\otimes I_\tau$ is a trivial symmetric product state, where $\rho_\sigma^\rightarrow=\bigotimes_{i}|\sigma^x_{2i}=1\rangle\langle \sigma^x_{2i}=1|$, and $I_\tau$ is the maximally mixed state for $\tau$ spins. Similar to the construction of cluster state in closed systems, one can obtain a $\Z^{\text{strong}}_2\times \Z^{\text{weak}}_2$ ASPT by applying the controlled-$Z$ gates:
\[\label{eq:rhocluster}
\begin{split}
\rho_{\text{cluster}}=&U(\text{CZ})\rho_{\text{trivial}}U(\text{CZ})\\
=&\sum_{\{z_{2i-1}\}}\bigotimes_{i}|\tau^z_{2i-1}=z_{2i-1}\rangle\langle \tau^z_{2i-1}=z_{2i-1}|\otimes\\
&|\sigma^x_{2i}=z_{2i-1}z_{2i+1}\rangle\langle \sigma^x_{2i}=z_{2i-1}z_{2i+1}|.
\end{split}
\]
The above expansion indicates the DDW picture of this ASPT: a $\Z^{\text{strong}}_2$ charge is decorated on each $\Z_2^{\text{weak}}$ domain wall, which proliferates classically to restore the $\Z_2^{\text{weak}}$ symmetry. Using the DDW method, we construct a Lindbladian to realize $\rho_{\text{cluster}}$ as a steady state in Appendix \ref{app:1+1d ASPT}. There we show that under OBC, the $\Z^{\text{strong}}_2\times \Z^{\text{weak}}_2$ symmetry is realized projectively on the edge, which leads to a larger steady-state degeneracy compared to PBC. 

\subsection{$(2+1)$-D $\Z_2\times \Z_2\times \Z_2$ and $\Z_2\times \Z_2$ average SPT }\label{sec:2+1dASPT}
As pointed out by \cite{ma2023aspt,ma2023topological}, one route to construct $(2+1)$-D ASPT is to decorate the $(1+1)$-D ASPT on domain walls of some other symmetry, and proliferate the domain walls. In this section we construct $(2+1)$-D steady-state ASPT by applying this idea to Lindbladians. 

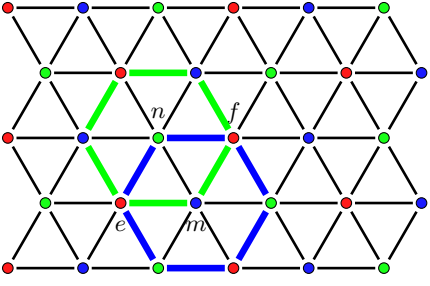
\begin{figure}
	\begin{tikzpicture}
	\node(b70) at (0, 2){} ;
	\draw[ fill=red!90] (1,2) circle (2pt);	
	\node(b00) at (1, 2){} ;
	\draw[fill=blue!90] (2,2) circle (2pt);
	\node(b01) at (2, 2) {};
	\draw[fill=green!90] (1.5,1.134) circle (2pt);
	\node(b10) at (1.5,1.134) {};	
	\draw[ fill=red!90] (2.5,1.134) circle (2pt);	
	\node(b11) at (2.5,1.134){} ;
	\draw[fill=green!90] (3,2) circle (2pt);
	\node(b02) at (3,2) {};	
	\draw[fill=blue!90] (3.5,1.134) circle (2pt);
	\node(b12) at (3.5,1.134) {};
	\draw[ fill=red!90] (4,2) circle (2pt);	
	\node(b03) at (4, 2){} ;
	\draw[fill=blue!90] (5,2) circle (2pt);
	\node(b04) at (5, 2) {};
	\draw[fill=green!90] (4.5,1.134) circle (2pt);
	\node(b13) at (4.5,1.134) {};	
	\draw[ fill=red!90] (5.5,1.134) circle (2pt);	
	\node(b14) at (5.5,1.134){} ;
	\draw[fill=green!90] (6,2) circle (2pt);
	\node(b05) at (6,2) {};	
	\draw[fill=blue!90] (6.5,1.134) circle (2pt);
	\node(b15) at (6.5,1.134) {};
	\draw[ fill=red!90] (1,0.268) circle (2pt);	
	\node(b20) at (1, 0.268){} ;
	\draw[fill=blue!90] (2,0.268) circle (2pt);
	\node(b21) at (2, 0.268) {};
	\draw[fill=green!90] (1.5,-0.598) circle (2pt);
	\node(b30) at (1.5,-0.598) {};	
	\draw[ fill=red!90] (2.5,-0.598) circle (2pt);	
	\node(b31) at (2.5,-0.598){} ;
 \node() at (2.5,-0.9){$e$};
	\draw[fill=green!90] (3,0.268) circle (2pt);
	\node(b22) at (3,0.268) {};	
	\draw[fill=blue!90] (3.5,-0.598) circle (2pt);
	\node(b32) at (3.5,-0.598) {};
	\draw[ fill=red!90] (4,0.268) circle (2pt);	
	\node(b23) at (4, 0.268){} ;
 \node() at (4,0.6){$f$};
	\draw[fill=blue!90] (5,0.268) circle (2pt);
	\node(b24) at (5, 0.268) {};
	\draw[fill=green!90] (4.5,-0.598) circle (2pt);
	\node(b33) at (4.5,-0.598) {};
 
	\draw[ fill=red!90] (5.5,-0.598) circle (2pt);	
	\node(b34) at (5.5,-0.598){} ;
	\draw[fill=green!90] (6,0.268) circle (2pt);
	\node(b25) at (6,0.268) {};	
	\draw[fill=blue!90] (6.5,-0.598) circle (2pt);
	\node(b35) at (6.5,-0.598) {};
	
	\draw[ fill=red!90] (1,-1.464) circle (2pt);	
	\node(b40) at (1,-1.464){} ;
 
	\draw[fill=blue!90] (2,-1.464) circle (2pt);
	\node(b41) at (2,-1.464) {};	
	\draw[fill=green!90] (3,-1.464) circle (2pt);
	\node(b42) at (3,-1.464) {};
	\draw[ fill=red!90] (4,-1.464) circle (2pt);	
	\node(b43) at (4,-1.464){} ;
 
	\draw[fill=blue!90] (5,-1.464) circle (2pt);
	\node(b44) at (5,-1.464) {};	
	\draw[fill=green!90] (6,-1.464) circle (2pt);
	\node(b45) at (6,-1.464) {};

	\path[line width=1pt] (b00) edge (b01);
	\path[line width=1pt] (b10) edge (b01);
	\path[line width=1pt] (b00) edge (b10);
	\path[line width=1pt] (b10) edge (b11);
	\path[line width=1pt] (b01) edge (b11);
	\path[line width=1pt] (b01) edge (b02);
	\path[line width=1pt] (b11) edge (b02);
	\path[color=green,line width=2.5pt] (b11) edge (b12);
	\path[line width=1pt] (b12) edge (b02);
	\path[line width=1pt] (b13) edge (b12);
	\path[line width=1pt] (b03) edge (b02);
	\path[line width=1pt] (b03) edge (b04);
	\path[line width=1pt] (b03) edge (b12);
	\path[line width=1pt] (b03) edge (b13);
	\path[line width=1pt] (b13) edge (b04);
	\path[line width=1pt] (b13) edge (b14);
	\path[line width=1pt] (b04) edge (b14);
	\path[line width=1pt] (b05) edge (b14);
	\path[line width=1pt] (b04) edge (b05);
	\path[line width=1pt] (b15) edge (b14);
	\path[line width=1pt] (b05) edge (b15);
	\path[line width=1pt] (b10) edge (b20);
	\path[line width=1pt] (b10) edge (b21);
	\path[color=green,line width=2.5pt] (b11) edge (b21);
	\path[line width=1pt] (b11) edge (b22);
	\path[line width=1pt] (b12) edge (b22);
	\path[color=green,line width=2.5pt] (b12) edge (b23);
	\path[line width=1pt] (b13) edge (b23);
	\path[line width=1pt] (b13) edge (b24);
	\path[line width=1pt] (b14) edge (b24);
	\path[line width=1pt] (b14) edge (b25);
	\path[line width=1pt] (b15) edge (b25);

	\path[line width=1pt] (b20) edge (b21);
	\path[line width=1pt] (b30) edge (b21);
	\path[line width=1pt] (b20) edge (b30);
	\path[line width=1pt] (b30) edge (b31);
	\path[color=green,line width=2.5pt] (b21) edge (b31);
	\path[line width=1pt] (b21) edge (b22);
	\path[color=blue!120,line width=2.5pt] (b31) edge (b22);
	\path[color=green,line width=2.5pt] (b31) edge (b32);
	\path[line width=1pt] (b32) edge (b22);
	\path[ line width=1pt] (b33) edge (b32);
	\path[color=blue!120,line width=2.5pt] (b23) edge (b22);
	\path[line width=1pt] (b23) edge (b24);
	\path[color=green,line width=2.5pt] (b23) edge (b32);
	\path[color=blue!120,line width=2.5pt] (b23) edge (b33);
	\path[line width=1pt] (b33) edge (b24);
	\path[line width=1pt] (b33) edge (b34);
	\path[line width=1pt] (b24) edge (b34);
	\path[line width=1pt] (b25) edge (b34);
	\path[line width=1pt] (b24) edge (b25);
	\path[line width=1pt] (b35) edge (b34);
	\path[line width=1pt] (b25) edge (b35);
	
	\path[line width=1pt] (b30) edge (b40);
	\path[line width=1pt] (b30) edge (b41);
	\path[line width=1pt] (b31) edge (b41);
	\path[color=blue!120,line width=2.5pt] (b31) edge (b42);
	\path[line width=1pt] (b32) edge (b42);
	\path[line width=1pt] (b32) edge (b43);
	\path[color=blue!120,line width=2.5pt] (b33) edge (b43);
	\path[line width=1pt] (b33) edge (b44);
	\path[line width=1pt] (b34) edge (b44);
	\path[line width=1pt] (b34) edge (b45);
	\path[line width=1pt] (b35) edge (b45);
	\path[line width=1pt] (b40) edge (b41);
	\path[ line width=1pt] (b42) edge (b41);
	\path[color=blue!120,line width=2.5pt] (b43) edge (b42);
	\path[line width=1pt] (b43) edge (b44);
	\path[line width=1pt] (b45) edge (b44);

	\node(c06) at (1.25,-0.53){};
	\node(c07) at (2,-1.8){};
	\node(c08) at (1.9,-1.65){};
	\node(c09) at (3.2,-1.65){};
	\node(c10) at (3,-1.8){};
	\node(c11) at (3.75,-0.5){};
	\node(c12) at (3.6,-0.68){};
	\node(c13) at (4.4,-0.68){};
	\node(c14) at (4.25,-0.53){};
	\node(c15) at (5,-1.8){};
	\node(c17) at (6.2,-1.65){};
	\node(c16) at (4.9,-1.65){};
	\node(c18) at (6,-1.8){};
	\node(c19) at (6.75,-0.5){};
	
	\node(c04) at (4.5,-2.5){};
	\node(c05) at (6.5,-2.5){};

	\node(d06) at (4,2.3){};
	\node(d07) at (3.2,1){};
	\node(d08) at (3.2,1.2){};
	\node(d09) at (4,-0.1){};
	\node(d10) at (3.9,0.1){};
	\node(d11) at (5.1,0.1){};
	\node(d12) at (4.9,0.1){};
	\node(d13) at (5.4,-0.8){};
	\node(d14) at (5.35,-0.65){};
	\node(d15) at (4.7,-1.664){};
	\node() at (3,0.6){$n$};
 \node() at (3.5,-0.9){$m$};
	\end{tikzpicture}
	\caption{$\sigma$, $\tau$ and $\mu$ spin are assigned with vertices of $A, B, C$ sublattices, colored by red, green and blue. The green loop $N_B$ is on the dual lattice of $B$ sublattice and $M_B$ involve single $B$ vertex in its bulk. The blue loop $N_C$ is on the dual lattice of $C$ sublattice and $M_C$ involves single $C$ vertex. The two loops have two intersection points $e$ and $f$.}\label{fig:tau-mudomain}
\end{figure}

To begin with, we assign a spin $\frac{1}{2}$ on each vertex of a triangular lattice. We define three global $\Z_2$ (strong or weak) symmetries generated by spin-flips on the $A, B, C$ sublattices, which are colored by red, green and blue in Fig~\ref{fig:tau-mudomain}:
\begin{eqnarray}
U_A= \prod_{v\in A} \sigma_v^x, ~~~~~ U_B= \prod_{v\in B} \tau_{v}^x, ~~~~~ U_C= \prod_{v\in C} \mu_{v}^x.
\end{eqnarray}

In the first example, we consider strong $\Z^A_2$ symmetry and weak $G=\Z^B_2\times \Z^C_2$ symmetry. The corresponding ASPT can be constructed by decorating the $\Z^B_2$ domain wall with $(1+1)$-D $\Z^A_2\times \Z^C_2$ ASPT, which is equivalent to decorating codimension-2 $\Z^B_2\times \Z^C_2$ defects with $\Z^A_2$ charge. Its cocycle is represented by the nontrivial element of $H^1(\Z^B_2, H^2(\Z^A_2 \times \Z^C_2,U(1))=H^2(\Z^B_2\times \Z^C_2, H^1(\Z^A_2,U(1))$.  

Let us start with a Lindbladian under PBC, where three types of spins are decoupled, with a trivial product state as its steady state: 
\[\label{eq:trivial Lindbladian}
\begin{split}
&H=0, \quad l^1_{i\in A}=\sum_{<i,j>\in A}\sigma^z_i \sigma^z_j\frac{1-\sigma^x_i}{2},\\
&l^2_{i\in B/C}=\tau^z_i/\mu^z_i,\quad l^3_{i\in B/C}=\tau^x_i/\mu^x_i,
\end{split}
\]
where $\langle i,j\rangle\in A$ denotes nearest neighbored sites $i,j$ belonging to the sublattice $A$. .
The effect of dissipators $l_1$ is to move or annihilate the excitation with $\sigma^x=-1$. Thus the steady state in the even $K$-sector has all $\sigma^x_i=1$,  while in the odd $K$-sector it is the maximally mixed
state of states with $\sigma^x=-1$ only on a single site. In other words, we have
\[\label{eq:sigma ss}
\begin{split}
&\rho^{ss}_{\sigma, +}=|\rightarrow \rightarrow\cdots \rightarrow\rangle\langle \rightarrow\rightarrow\cdots\rightarrow|\equiv\rho_{\sigma}^\rightarrow,\\
&\rho^{ss}_{\sigma, -}=\frac{1}{2^{N_{\sigma}}}\sum_{i\in A} |\rightarrow \rightarrow \cdots \leftarrow_i\cdots\rightarrow\rangle\langle\rightarrow \rightarrow \cdots \leftarrow_i\cdots\rightarrow|.
\end{split}
\]
The effect of $l^2$ is to dephase on $\tau$ and $\mu$ spins and to select diagonal configurations. Then the effect of $l^3$ is to ensure the probability of each $\tau$ and $\mu$ configuration in steady state to be the same. Hence the steady state of $\tau$ and $\mu$ is identity $\rho^{ss}_{\tau,\mu}=I_{\tau,\mu}$. In the following discussion, we will mainly focus on the steady state in the even $K$-sector where the steady state is a trivial symmetric product state $\rho_{\text{trivial}}=\rho_\sigma^\rightarrow\otimes I_{\tau}\otimes I_{\mu}$.

In closed systems, the decoration corresponding to the nontrivial class of $H^2(\Z^B_2\times\Z^C_2, H^1(\Z^A_2,U(1))$ can be realized by a unitary transformation $U(\text{CCZ})$ \cite{PhysRevB.93.155131,PhysRevB.106.224420}. More precisely, $U(\text{CCZ})$ relates trivially gapped phase and $\Z^A_2\times \Z^B_2\times \Z^C_2 $ SPT phase. Motivated by this result, we also conjugate the Lindbladian \eqref{eq:trivial Lindbladian} by $U(\text{CCZ})$. This unitary transformation is
\[
\begin{split}
&U(\text{CCZ})=\prod_{(i,j,k)\in\bigtriangleup}(\text{CCZ})_{i,j,k},\nonumber\\ 
&(\text{CCZ})_{i,j,k}|\sigma^z_i=\alpha,\tau^z_j=\beta,\mu^z_k=\gamma\rangle\\
&=(-1)^{\alpha\beta\gamma}|\sigma^z_i=\alpha,\tau^z_j=\beta,\mu^z_k=\gamma\rangle,
\end{split}
\]
where $\bigtriangleup$ represents the sets of all  triangles and CCZ is a unitary operator acting on each triple of spins belonging to one triangle. 

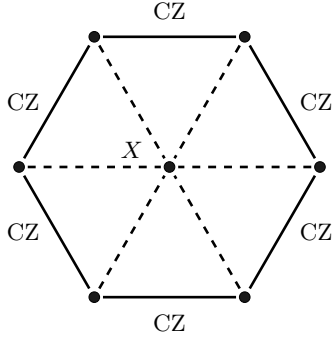
\begin{figure}
   
	\begin{tikzpicture}
	
	\draw[ fill=black!90] (4,0.268) circle (2pt);	
	\node(b11) at (4,0.268){} ;
	\draw[fill=black!90] (5,2) circle (2pt);
	\node(b02) at (5,2) {};	
	\draw[fill=black!90] (6,0.268) circle (2pt);
	\node(b12) at (6,0.268) {};
	\draw[ fill=black!90] (7,2) circle (2pt);	
	\node(b03) at (7, 2){} ;
	\draw[fill=black!90] (8,0.268) circle (2pt);
	\node(b13) at (8,0.268) {};	
	\draw[fill=black!90] (5,-1.464) circle (2pt);
	\node(b22) at (5,-1.464) {};	
	\draw[fill=black!90] (7, -1.464) circle (2pt);
	\node(b23) at (7, -1.464){} ;
	\node(c3) at (2, -1.464){};
	\node(b27)[label={[black]above:{CZ}}] at (6, 2) {};
	\node(b28)[label={[black]left:{CZ}}] at (4.5, 1.134) {};
	\node(b29)[label={[black]right:{CZ}}] at (7.5, 1.134) {};
	\node(b30)[label={[black]left:{CZ}}] at (4.5, -0.6) {};
	\node(b31)[label={[black]right:{CZ}}] at (7.5, -0.6) {};
	\node(b32)[label={[black]below:{CZ}}] at (6, -1.464) {};
	\node(b90) at (5.5,0.5) {$X$};

	\path[color=black,line width=1pt] (b11) edge (b02);
	\path[color=black,line width=1pt] (b03) edge (b02);
	\path[color=black,line width=1pt] (b13) edge (b03);
	\path[color=black,line width=1pt] (b13) edge (b23);
	\path[color=black,line width=1pt] (b22) edge (b23);
	\path[color=black,line width=1pt] (b22) edge (b11);
	\path[dashed,line width=1pt] (b11) edge (b12);
	\path[dashed,line width=1pt] (b02) edge (b12);
	\path[dashed,line width=1pt] (b03) edge (b12);
	\path[dashed,line width=1pt] (b13) edge (b12);
	\path[dashed,line width=1pt] (b22) edge (b12);  
	\path[dashed,line width=1pt] (b23) edge (b12);
	
	\end{tikzpicture}
	\caption{The local
		term $O_i$ where a Pauli $X$ operator at the vertex $i$ is decorated by CZ operators on all links surround this vertex. $X$ operator corresponds to the Pauli operator $\sigma^x$ or $\tau^x$ or $\mu^x$.  }\label{fig: CCZ Lindbladian}

\end{figure}

Then the Lindbladian after conjugated by $U_{\text{CCZ}}$ is
\[\label{eq:2+1d ASPT-1}
\begin{split}
&l^1_{i\in A}=\sum_{<i,j>\in A}\sigma^z_i \sigma^z_j\frac{1-O_i}{2},\\
&l^2_{i\in B/C}=\tau^z_i/\mu^z_i,\quad l^3_{i\in B/C}=O_i,
\end{split}
\]
with steady state $\rho^{ss}_{\text{ASPT}}=U(\text{CCZ})\rho_{\text{trivial}}U^{\dagger}(\text{CCZ})$.
Here $O_i$ operator is represented in Fig.~\ref{fig: CCZ Lindbladian} including CZ operators on all links $e$ surround the vertex $i$ (also known as the 1-links of vertex $i$):
\[
\begin{split}
& O_{i\in A}=\sigma^x_i \prod_{e\in 1-\text{link}(i)}\text{CZ}_e,\nonumber\\& O_{i\in B}=\tau^x_i \prod_{e\in 1-\text{link}(i)}\text{CZ}_e,\nonumber\\& O_{i\in C}=\mu^x_i \prod_{e\in 1-\text{link}(i)}\text{CZ}_e.
\end{split}
\]
Hence,  the dissipator $l^1$ enforces steady states satisfying $O_{i\in A}=1$, which we will study its SPT feature later. Next, the effect of $l^2$ is dephasing on $\tau,\sigma$ spins, which selects diagonal domain wall configuration. Finally, $l^3$ will incoherently proliferate the domain wall configuration in a way compatible with the decoration pattern, since $[l^3,O_{i\in A}]=0$.  


To detect the SPT feature of steady states, we consider a $\Z^B_2$ domain wall supported in a closed loop $N_B$ in the dual lattice of $B$, namely $A$ and $C$ sublattice.  Such the domain wall can be constructed by conjugating the steady state using $\prod_{i\in M_B}\tau^x_i$, where $M_B$ involves $B$ sites enclosed by $N_B$, as shown in Fig.~\ref{fig:tau-mudomain}. It is straightforward to check
\[\label{eq:ccz-decorate}
\begin{split}
&\prod_{M_B}\tau^x_i \rho^{ss}_{\text{ASPT}}\prod_{ M_B}\tau^x_i\\=&U(\text{CCZ})
 U_{N_B}(\text{CZ})\prod_{i\in M_B}\tau^x_i\rho_{\text{trivial}}\prod_{i\in M_B}\tau^x_i U^{\dagger}_{N_B}(\text{CZ})U^{\dagger}(\text{CCZ})\\
=&U(\text{CCZ})
 U_{N_B}(\text{CZ})\rho_{\text{trivial}} U^{\dagger}_{N_B}(\text{CZ})U^{\dagger}(\text{CCZ})
 \\=&U_{N_B}(\text{CZ})\rho^{ss}_{\text{ASPT}}U^{\dagger}_{N_B}(\text{CZ}).
\end{split}
\]
As shown in Section~\ref{sec:1+1dASPT}, $U(\text{CZ})$ is a duality between $(1+1)$-D trival mixed state and $\Z^{\text{strong}}_2\times \Z^{\text{weak}}_2$ ASPT. Thus Eq.~\eqref{eq:ccz-decorate} implies $\Z^B_2$ domain wall is decorated with a $(1+1)$-D $\Z^A_2\times \Z^C_2$ ASPT. 

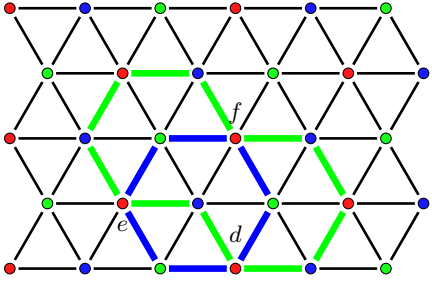
\begin{figure}
	\begin{tikzpicture}
	\node(b70) at (0, 2){} ;
	\draw[ fill=red!90] (1,2) circle (2pt);	
	\node(b00) at (1, 2){} ;
	\draw[fill=blue!90] (2,2) circle (2pt);
	\node(b01) at (2, 2) {};
	\draw[fill=green!90] (1.5,1.134) circle (2pt);
	\node(b10) at (1.5,1.134) {};	
	\draw[ fill=red!90] (2.5,1.134) circle (2pt);	
	\node(b11) at (2.5,1.134){} ;
	\draw[fill=green!90] (3,2) circle (2pt);
	\node(b02) at (3,2) {};	
	\draw[fill=blue!90] (3.5,1.134) circle (2pt);
	\node(b12) at (3.5,1.134) {};
	\draw[ fill=red!90] (4,2) circle (2pt);	
	\node(b03) at (4, 2){} ;
	\draw[fill=blue!90] (5,2) circle (2pt);
	\node(b04) at (5, 2) {};
	\draw[fill=green!90] (4.5,1.134) circle (2pt);
	\node(b13) at (4.5,1.134) {};	
	\draw[ fill=red!90] (5.5,1.134) circle (2pt);	
	\node(b14) at (5.5,1.134){} ;
	\draw[fill=green!90] (6,2) circle (2pt);
	\node(b05) at (6,2) {};	
	\draw[fill=blue!90] (6.5,1.134) circle (2pt);
	\node(b15) at (6.5,1.134) {};
	\draw[ fill=red!90] (1,0.268) circle (2pt);	
	\node(b20) at (1, 0.268){} ;
	\draw[fill=blue!90] (2,0.268) circle (2pt);
	\node(b21) at (2, 0.268) {};
	\draw[fill=green!90] (1.5,-0.598) circle (2pt);
	\node(b30) at (1.5,-0.598) {};	
	\draw[ fill=red!90] (2.5,-0.598) circle (2pt);	
	\node(b31) at (2.5,-0.598){} ;
	\draw[fill=green!90] (3,0.268) circle (2pt);
	\node(b22) at (3,0.268) {};	
	\draw[fill=blue!90] (3.5,-0.598) circle (2pt);
	\node(b32) at (3.5,-0.598) {};
	\draw[ fill=red!90] (4,0.268) circle (2pt);	
	\node(b23) at (4, 0.268){} ;
	\draw[fill=blue!90] (5,0.268) circle (2pt);
	\node(b24) at (5, 0.268) {};
	\draw[fill=green!90] (4.5,-0.598) circle (2pt);
	\node(b33) at (4.5,-0.598) {};	
	\draw[ fill=red!90] (5.5,-0.598) circle (2pt);	
	\node(b34) at (5.5,-0.598){} ;
	\draw[fill=green!90] (6,0.268) circle (2pt);
	\node(b25) at (6,0.268) {};	
	\draw[fill=blue!90] (6.5,-0.598) circle (2pt);
	\node(b35) at (6.5,-0.598) {};
	
	\draw[ fill=red!90] (1,-1.464) circle (2pt);	
	\node(b40) at (1,-1.464){} ;
	\draw[fill=blue!90] (2,-1.464) circle (2pt);
	\node(b41) at (2,-1.464) {};	
	\draw[fill=green!90] (3,-1.464) circle (2pt);
	\node(b42) at (3,-1.464) {};
	\draw[ fill=red!90] (4,-1.464) circle (2pt);	
	\node(b43) at (4,-1.464){} ;
	\draw[fill=blue!90] (5,-1.464) circle (2pt);
	\node(b44) at (5,-1.464) {};	
	\draw[fill=green!90] (6,-1.464) circle (2pt);
	\node(b45) at (6,-1.464) {};

	\path[line width=1pt] (b00) edge (b01);
	\path[line width=1pt] (b10) edge (b01);
	\path[line width=1pt] (b00) edge (b10);
	\path[line width=1pt] (b10) edge (b11);
	\path[line width=1pt] (b01) edge (b11);
	\path[line width=1pt] (b01) edge (b02);
	\path[line width=1pt] (b11) edge (b02);
	\path[color=green,line width=2.5pt] (b11) edge (b12);
	\path[line width=1pt] (b12) edge (b02);
	\path[line width=1pt] (b13) edge (b12);
	\path[line width=1pt] (b03) edge (b02);
	\path[line width=1pt] (b03) edge (b04);
	\path[line width=1pt] (b03) edge (b12);
	\path[line width=1pt] (b03) edge (b13);
	\path[line width=1pt] (b13) edge (b04);
	\path[line width=1pt] (b13) edge (b14);
	\path[line width=1pt] (b04) edge (b14);
	\path[line width=1pt] (b05) edge (b14);
	\path[line width=1pt] (b04) edge (b05);
	\path[line width=1pt] (b15) edge (b14);
	\path[line width=1pt] (b05) edge (b15);
	\path[line width=1pt] (b10) edge (b20);
	\path[line width=1pt] (b10) edge (b21);
	\path[color=green,line width=2.5pt] (b11) edge (b21);
	\path[line width=1pt] (b11) edge (b22);
	\path[line width=1pt] (b12) edge (b22);
	\path[color=green,line width=2.5pt] (b12) edge (b23);
	\path[line width=1pt] (b13) edge (b23);
	\path[line width=1pt] (b13) edge (b24);
	\path[line width=1pt] (b14) edge (b24);
	\path[line width=1pt] (b14) edge (b25);
	\path[line width=1pt] (b15) edge (b25);

	\path[line width=1pt] (b20) edge (b21);
	\path[line width=1pt] (b30) edge (b21);
	\path[line width=1pt] (b20) edge (b30);
	\path[line width=1pt] (b30) edge (b31);
	\path[color=green,line width=2.5pt] (b21) edge (b31);
	\path[line width=1pt] (b21) edge (b22);
	\path[color=blue,line width=2.5pt] (b31) edge (b22);
	\path[color=green,line width=2.5pt] (b31) edge (b32);
	\path[line width=1pt] (b32) edge (b22);
	\path[ line width=1pt] (b33) edge (b32);
	\path[color=blue,line width=2.5pt] (b23) edge (b22);
	\path[color=green,line width=2.5pt] (b23) edge (b24);
	\path[line width=1pt] (b23) edge (b32);
	\path[color=blue,line width=2.5pt] (b23) edge (b33);
	\path[line width=1pt] (b33) edge (b24);
	\path[line width=1pt] (b33) edge (b34);
	\path[color=green,line width=2.5pt] (b24) edge (b34);
	\path[line width=1pt] (b25) edge (b34);
	\path[line width=1pt] (b24) edge (b25);
	\path[line width=1pt] (b35) edge (b34);
	\path[line width=1pt] (b25) edge (b35);
	
	\path[line width=1pt] (b30) edge (b40);
	\path[line width=1pt] (b30) edge (b41);
	\path[line width=1pt] (b31) edge (b41);
	\path[color=blue,line width=2.5pt] (b31) edge (b42);
	\path[line width=1pt] (b32) edge (b42);
	\path[color=green,line width=2.5pt] (b32) edge (b43);
	\path[color=blue,line width=2.5pt] (b33) edge (b43);
	\path[line width=1pt] (b33) edge (b44);
	\path[color=green,line width=2.5pt] (b34) edge (b44);
	\path[line width=1pt] (b34) edge (b45);
	\path[line width=1pt] (b35) edge (b45);
	\path[line width=1pt] (b40) edge (b41);
	\path[ line width=1pt] (b42) edge (b41);
	\path[color=blue,line width=2.5pt] (b43) edge (b42);
	\path[color=green,line width=2.5pt] (b43) edge (b44);
	\path[line width=1pt] (b45) edge (b44);

	\node(c06) at (1.25,-0.53){};
	\node(c07) at (2,-1.8){};
	\node(c08) at (1.9,-1.65){};
	\node(c09) at (3.2,-1.65){};
	\node(c10) at (3,-1.8){};
	\node(c11) at (3.75,-0.5){};
	\node(c12) at (3.6,-0.68){};
	\node(c13) at (4.4,-0.68){};
	\node(c14) at (4.25,-0.53){};
	\node(c15) at (5,-1.8){};
	\node(c17) at (6.2,-1.65){};
	\node(c16) at (4.9,-1.65){};
	\node(c18) at (6,-1.8){};
	\node(c19) at (6.75,-0.5){};
	
	\node(c04) at (4.5,-2.5){};
	\node(c05) at (6.5,-2.5){};

	\node(d06) at (4,2.3){};
	\node(d07) at (3.2,1){};
	\node(d08) at (3.2,1.2){};
	\node(d09) at (4,-0.1){};
	\node(d10) at (3.9,0.1){};
	\node(d11) at (5.1,0.1){};
	\node(d12) at (4.9,0.1){};
	\node(d13) at (5.4,-0.8){};
	\node(d14) at (5.35,-0.65){};
	\node(d15) at (4.7,-1.664){};
\node() at (2.5,-0.9){$e$};
\node() at (4,0.6){$f$};
\node() at (4,-1){$d$};
	\end{tikzpicture}
	\caption{ The subregion $M_{B}$ involves two $B$ vertices and $M_C$ involves single $C$ vertex. $N_{B}$ and $N_C$ are colored by green and blue respectively and have three intersection points  $d$, $e$, $f$ where $s_d=s_e=1, s_f=2$. }\label{fig:taudomain}
\end{figure} 

Furthermore, one can continue to construct the $\mu$ domain walls on the closed loop $N_C$ in the dual lattice of $C$ sublattice, i.e., $A$ and $B$ sublattice. This corresponds to conjugating the steady state by $\mu$ spin flip in the subregion $M_C$ which involves the $C$ vertices enclosed by $N_C$. Let us first consider $M_B$ and $M_C$ includes two nearest neighboured sites $m, n$ as shown in Fig.~\ref{fig:tau-mudomain}, then we have
\[\label{eq:ccz-decorate-1}
\begin{split}
&\mu^x_m \tau^x_n \rho^{ss}_{\text{ASPT}}\tau^x_n\mu^x_m\\=&\sigma^z_e \sigma^z_f
 U_{N_B\cup N_C}(\text{CZ})\rho^{ss}_{\text{ASPT}} U^{\dagger}_{N_B\cup N_C}(\text{CZ})\sigma^z_e \sigma^z_f.
\end{split}
\]
This can be easily generalized to general $M_B$ and $M_C$ as follows.  For each $A$ vertex $i\in N_B\cap N_C$, if one of its 1-links, i.e., a $BC$ link surround $i$, has two endpoints belonging to $M_B$ and $M_C$ respectively, the steady state will obtain a $\sigma^z_i$ after conjugating the spin flip on the two endpoints. In other words, we have   
\[\label{eq:ccz-decorate-15}
\begin{split}
&\prod_{ M_C}\mu^x_j\prod_{ M_B}\tau^x_k \rho^{ss}_{\text{ASPT}}\prod_{ M_B}\tau^x_k\prod_{ M_C}\mu^x_j\\=&\prod_{ N_B\cap N_C}(\sigma^z_i)^{s_i}
 U_{N_B\cup N_C}(\text{CZ})\rho^{ss}_{\text{ASPT}} U^{\dagger}_{N_B\cup N_C}(\text{CZ})\prod_{N_B\cap N_C}(\sigma^z_i)^{s_i}.
\end{split}
\]
Here $s_i$ is the number of 1-links of $A$ vertex $i$ whose endpoints are in $M_B$ and $M_C$. An example is shown in Fig.~\ref{fig:taudomain}.

The result \eqref{eq:ccz-decorate-15} implies the codimension-2 defects of the weak symmetry, i.e.,  intersection points of $\Z^B_2$ and $\Z^C_2$ domain walls with odd $s_i$, are decorated with $\Z^A_2$ charge. 

Moreover, we show that the same model also realizes a nontrivial ASPT even when we only consider the sub-symmetry group $K=\Z^A_2,G=\Z^{BC}_2$, with $\Z_2^{BC}$ generated by $U_BU_C$. In this case $\rho^{ss}_{\text{ASPT}}$ corresponds to the nontrivial class in $H^2(\Z^{BC}_2, H^1(\Z^A_2,U(1))$ in the DDW classification, as we demonstrate below.  

\begin{figure}
	
	\begin{tikzpicture}
	\node(b70) at (0, 2){} ;
	\draw[ fill=red!90] (1,2) circle (2pt);	
	\node(b00) at (1, 2){} ;
	\draw[fill=blue!90] (2,2) circle (2pt);
	\node(b01) at (2, 2) {};
	\draw[fill=green!90] (1.5,1.134) circle (2pt);
	\node(b10) at (1.5,1.134) {};	
	\draw[ fill=red!90] (2.5,1.134) circle (2pt);	
	\node(b11) at (2.5,1.134){} ;
	\draw[fill=green!90] (3,2) circle (2pt);
	\node(b02) at (3,2) {};	
	\draw[fill=blue!90] (3.5,1.134) circle (2pt);
	\node(b12) at (3.5,1.134) {};
	\draw[ fill=red!90] (4,2) circle (2pt);	
	\node(b03) at (4, 2){} ;
	\draw[fill=blue!90] (5,2) circle (2pt);
	\node(b04) at (5, 2) {};
	\draw[fill=green!90] (4.5,1.134) circle (2pt);
	\node(b13) at (4.5,1.134) {};	
	\draw[ fill=red!90] (5.5,1.134) circle (2pt);	
	\node(b14) at (5.5,1.134){} ;
	\draw[fill=green!90] (6,2) circle (2pt);
	\node(b05) at (6,2) {};	
	\draw[fill=blue!90] (6.5,1.134) circle (2pt);
	\node(b15) at (6.5,1.134) {};
	\draw[ fill=red!90] (1,0.268) circle (2pt);	
	\node(b20) at (1, 0.268){} ;
	\draw[fill=blue!90] (2,0.268) circle (2pt);
	\node(b21) at (2, 0.268) {};
	\draw[fill=green!90] (1.5,-0.598) circle (2pt);
	\node(b30)[label={[black]}] at (1.5,-0.598) {};	
	\draw[ fill=red!90] (2.5,-0.598) circle (2pt);	
	\node(b31) at (2.5,-0.598){} ;
	\draw[fill=green!90] (3,0.268) circle (2pt);
	\node(b22) at (3,0.268) {};	
	\draw[fill=blue!90] (3.5,-0.598) circle (2pt);
	\node(b32) at (3.5,-0.598) {};
	\draw[ fill=red!90] (4,0.268) circle (2pt);	
	\node(b23) at (4, 0.268){} ;
	\draw[fill=blue!90] (5,0.268) circle (2pt);
	\node(b24) at (5, 0.268) {};
	\draw[fill=green!90] (4.5,-0.598) circle (2pt);
	\node(b33) at (4.5,-0.598) {};	
	\draw[ fill=red!90] (5.5,-0.598) circle (2pt);	
	\node(b34) at (5.5,-0.598){} ;
	\draw[fill=green!90] (6,0.268) circle (2pt);
	\node(b25) at (6,0.268) {};	
	\draw[fill=blue!90] (6.5,-0.598) circle (2pt);
	\node(b35) at (6.5,-0.598) {};
	
	\draw[ fill=red!90] (1,-1.464) circle (2pt);	
	\node(b40)[label={[black]}] at (1,-1.464){} ;
	\draw[fill=blue!90] (2,-1.464) circle (2pt);
	\node(b41)[label={[black]}] at (2,-1.464) {};	
	\draw[fill=green!90] (3,-1.464) circle (2pt);
	\node(b42) at (3,-1.464) {};
	\draw[ fill=red!90] (4,-1.464) circle (2pt);	
	\node(b43) at (4,-1.464){} ;
	\draw[fill=blue!90] (5,-1.464) circle (2pt);
	\node(b44) at (5,-1.464) {};	
	\draw[fill=green!90] (6,-1.464) circle (2pt);
	\node(b45) at (6,-1.464) {};

	\path[line width=1pt] (b00) edge (b01);
	\path[line width=1pt] (b10) edge (b01);
	\path[line width=1pt] (b00) edge (b10);
	\path[line width=1pt] (b10) edge (b11);
	\path[line width=1pt] (b01) edge (b11);
	\path[line width=1pt] (b01) edge (b02);
	\path[line width=1pt] (b11) edge (b02);
	\path[line width=1pt] (b11) edge (b12);
	\path[line width=1pt] (b12) edge (b02);
	\path[line width=1pt] (b13) edge (b12);
	\path[line width=1pt] (b03) edge (b02);
	\path[line width=1pt] (b03) edge (b04);
	\path[line width=1pt] (b03) edge (b12);
	\path[line width=1pt] (b03) edge (b13);
	\path[line width=1pt] (b13) edge (b04);
	\path[line width=1pt] (b13) edge (b14);
	\path[line width=1pt] (b04) edge (b14);
	\path[line width=1pt] (b05) edge (b14);
	\path[line width=1pt] (b04) edge (b05);
	\path[line width=1pt] (b15) edge (b14);
	\path[line width=1pt] (b05) edge (b15);
	\path[line width=1pt] (b10) edge (b20);
	\path[line width=1pt] (b10) edge (b21);
	\path[line width=1pt] (b11) edge (b21);
	\path[line width=1pt] (b11) edge (b22);
	\path[line width=1pt] (b12) edge (b22);
	\path[line width=1pt] (b12) edge (b23);
	\path[line width=1pt] (b13) edge (b23);
	\path[line width=1pt] (b13) edge (b24);
	\path[line width=1pt] (b14) edge (b24);
	\path[line width=1pt] (b14) edge (b25);
	\path[line width=1pt] (b15) edge (b25);

	\path[line width=1pt] (b20) edge (b21);
	\path[line width=1pt] (b30) edge (b21);
	\path[line width=1pt] (b20) edge (b30);
	\path[line width=1pt] (b30) edge (b31);
	\path[line width=1pt] (b21) edge (b31);
	\path[line width=1pt] (b21) edge (b22);
	\path[line width=1pt] (b31) edge (b22);
	\path[line width=1pt] (b31) edge (b32);
	\path[line width=1pt] (b32) edge (b22);
	\path[line width=1pt] (b33) edge (b32);
	\path[line width=1pt] (b23) edge (b22);
	\path[line width=1pt] (b23) edge (b24);
	\path[line width=1pt] (b23) edge (b32);
	\path[line width=1pt] (b23) edge (b33);
	\path[line width=1pt] (b33) edge (b24);
	\path[line width=1pt] (b33) edge (b34);
	\path[line width=1pt] (b24) edge (b34);
	\path[line width=1pt] (b25) edge (b34);
	\path[line width=1pt] (b24) edge (b25);
	\path[line width=1pt] (b35) edge (b34);
	\path[line width=1pt] (b25) edge (b35);
	
	\path[line width=1pt] (b30) edge (b40);
	\path[line width=1pt] (b30) edge (b41);
	\path[line width=1pt] (b31) edge (b41);
	\path[line width=1pt] (b31) edge (b42);
	\path[line width=1pt] (b32) edge (b42);
	\path[line width=1pt] (b32) edge (b43);
	\path[line width=1pt] (b33) edge (b43);
	\path[line width=1pt] (b33) edge (b44);
	\path[line width=1pt] (b34) edge (b44);
	\path[line width=1pt] (b34) edge (b45);
	\path[line width=1pt] (b35) edge (b45);
	\path[line width=1pt] (b40) edge (b41);
	\path[line width=1pt] (b42) edge (b41);
	\path[line width=1pt] (b43) edge (b42);
	\path[line width=1pt] (b43) edge (b44);
	\path[line width=1pt] (b45) edge (b44);

 \path[dashed,color=red,line width=2.5pt] (b11) edge (b31);
 \path[dashed,color=red,line width=2.5pt] (b43) edge (b31);
	
 \path[dashed,color=red,line width=2.5pt] (b23) edge (b43);
  \path[dashed,color=red,line width=2.5pt] (b23) edge (b11);
	 \node() at (2.5,-0.9){$e$};
   \node() at (4,0.6){$f$};
	\end{tikzpicture}
	\caption{ The $\Z_2^{BC}$ domain walls form loops on the $A$ sublattice. The red dashed loop in the figure is an example of $\Z_2^{BC}$ domain wall, with $\tau^z=\mu^z=-1(+1)$ for the three vertices inside the loop and $\tau^z=\mu^z=-1(+1)$ outside. The condition $O_{i\in A}=1$ means the $\sigma^x=-1$ on vertices $e,f$ and $\sigma^x=1$ elsewhere.} \label{fig: ZBC domain wall}
\end{figure}
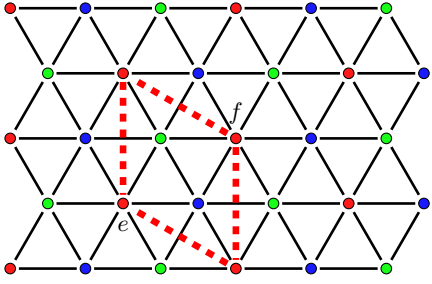

First, due to the dissipator $l^1$, steady states satisfy $O_{i\in A}=1$. This indicates that the $K$ charge distribution $\{\sigma^x_i\}$ is determined by the domain wall configuration of $\Z_2^{BC}$, which is a loop $N_{BC}$ on the $A$ sublattice. More specifically, a $K$-charge ($\sigma^x_i=-1$) is decorated on each $120^\circ$ (or equivalently, $240^\circ$) corner of $\Z_2^{BC}$ domain walls, which we denote as a codimension-2 defect of the $\Z_2^{BC}$ symmetry. See Fig. \ref{fig: ZBC domain wall} for an illustration. Next, $l^2$ and $l^3$ will still dephase on $\tau,\sigma$ spins and incoherently proliferate the domain wall configuration.



To check the above DDW picture, we apply the following truncated $\Z_2^{BC}$ transformation on a subregion $M_{BC}$ and construct a $\Z_2^{BC}$ domain wall on $N_{BC}$, where $M_{BC}$ involves vertices enclosed by $N_{BC}$. Following the same calculation, we have
\[\label{eq:ccz-decorate-2}
\begin{split}
&\prod_{M_{BC}}\mu^x_{j}\tau^x_{k} \rho^{ss}_{\text{ASPT}}\prod_{M_{BC}}\mu^x_{j}\tau^x_{k}\\
=&\prod_{ N_{BC}}(\sigma^z_i)^{s_i}
 U_{N'}(\text{CZ})\rho^{ss}_{\text{ASPT}} U^{\dagger}_{N'}(\text{CZ})\prod_{N_{BC}}(\sigma^z_i)^{s_i}\\
=&\prod_{\text{odd } s_i, N_{BC}}\sigma^z_i
 U_{N'}(\text{CZ})\rho^{ss}_{\text{ASPT}} U^{\dagger}_{N'}(\text{CZ})\prod_{\text{odd } s_i,N_{BC}}\sigma^z_i,
\end{split}
\]
where $N'$ is a closed loop that is not important and $s_i$ is the number of 1-links of $A$ vertex $i$ whose end points are in $M_{BC}$. We also observe that each $120^\circ$ (or equivalently, $240^\circ$) corner of $\Z_2^{BC}$ domain walls has odd $s$ and the other cases have even $s$. Thus Eq.~\eqref{eq:ccz-decorate-2} is consistent with the DDW picture above.


\begin{figure}
	
	\begin{tikzpicture}
	\node(b70) at (0, 2){} ;
	
	\draw[fill=blue!90] (2,2) circle (2pt);
	\node(b01) at (2, 2) {};
	\draw[fill=green!90] (1.5,1.134) circle (2pt);
	\node(b10) at (1.5,1.134) {};	
	\draw[ fill=red!90] (2.5,1.134) circle (2pt);	
	\node(b11) at (2.5,1.134){} ;
	\draw[fill=green!90] (3,2) circle (2pt);
	\node(b02) at (3,2) {};	
	\draw[fill=blue!90] (3.5,1.134) circle (2pt);
	\node(b12) at (3.5,1.134) {};
	
	\draw[fill=blue!90] (5,2) circle (2pt);
	\node(b04) at (5, 2) {};
	\draw[fill=green!90] (4.5,1.134) circle (2pt);
	\node(b13) at (4.5,1.134) {};	
	\draw[ fill=red!90] (5.5,1.134) circle (2pt);	
	\node(b14) at (5.5,1.134){} ;
	\draw[fill=green!90] (6,2) circle (2pt);
	\node(b05) at (6,2) {};	
	\draw[fill=blue!90] (6.5,1.134) circle (2pt);
	\node(b15) at (6.5,1.134) {};
	
	\draw[fill=blue!90] (2,0.268) circle (2pt);
	\node(b21) at (2, 0.268) {};
	\draw[fill=green!90] (1.5,-0.598) circle (2pt);
	
	\draw[ fill=red!90] (2.5,-0.598) circle (2pt);	
	\node(b31) at (2.5,-0.598){} ;
	\draw[fill=green!90] (3,0.268) circle (2pt);
	\node(b22) at (3,0.268) {};	
	\draw[fill=blue!90] (3.5,-0.598) circle (2pt);
	\node(b32) at (3.5,-0.598) {};
	\draw[ fill=red!90] (4,0.268) circle (2pt);	
	\node(b23) at (4, 0.268){} ;
	\draw[fill=blue!90] (5,0.268) circle (2pt);
	\node(b24) at (5, 0.268) {};
	\draw[fill=green!90] (4.5,-0.598) circle (2pt);
	\node(b33) at (4.5,-0.598) {};	
	\draw[ fill=red!90] (5.5,-0.598) circle (2pt);	
	\node(b34) at (5.5,-0.598){} ;
	\draw[fill=green!90] (6,0.268) circle (2pt);
	\node(b25) at (6,0.268) {};	
	\draw[fill=blue!90] (6.5,-0.598) circle (2pt);
	\node(b35) at (6.5,-0.598) {};

\draw[fill=blue!90] (2,-1.464) circle (2pt);
	\draw[fill=green!90] (3,-1.464) circle (2pt);
	\node(b42) at (3,-1.464) {};
	
	\draw[fill=blue!90] (5,-1.464) circle (2pt);
	\node(b44) at (5,-1.464) {};	
	\draw[fill=green!90] (6,-1.464) circle (2pt);
	\node(b45) at (6,-1.464) {};

	\path[line width=1pt] (b10) edge (b01);
 
	\path[line width=1pt] (b10) edge (b11);
	\path[line width=1pt] (b01) edge (b11);
	\path[line width=1pt] (b01) edge (b02);
	\path[line width=1pt] (b11) edge (b02);
	\path[line width=1pt] (b11) edge (b12);
	\path[line width=1pt] (b12) edge (b02);
	\path[line width=1pt] (b13) edge (b12);

	\path[line width=1pt] (b13) edge (b04);
	\path[line width=1pt] (b13) edge (b14);
	\path[line width=1pt] (b04) edge (b14);
	\path[line width=1pt] (b05) edge (b14);
	\path[line width=1pt] (b04) edge (b05);
	\path[line width=1pt] (b15) edge (b14);
	\path[line width=1pt] (b05) edge (b15);
	
	\path[line width=1pt] (b10) edge (b21);
	\path[line width=1pt] (b11) edge (b21);
	\path[line width=1pt] (b11) edge (b22);
	\path[line width=1pt] (b12) edge (b22);
	\path[line width=1pt] (b12) edge (b23);
	\path[line width=1pt] (b13) edge (b23);
	\path[line width=1pt] (b13) edge (b24);
	\path[line width=1pt] (b14) edge (b24);
	\path[line width=1pt] (b14) edge (b25);
	\path[line width=1pt] (b15) edge (b25);

	\path[line width=1pt] (b30) edge (b21);
	
	\path[line width=1pt] (b30) edge (b31);
	\path[line width=1pt] (b21) edge (b31);
	\path[line width=1pt] (b21) edge (b22);
	\path[line width=1pt] (b31) edge (b22);
	\path[line width=1pt] (b31) edge (b32);
	\path[line width=1pt] (b32) edge (b22);
	\path[line width=1pt] (b33) edge (b32);
	\path[line width=1pt] (b23) edge (b22);
	\path[line width=1pt] (b23) edge (b24);
	\path[line width=1pt] (b23) edge (b32);
	\path[line width=1pt] (b23) edge (b33);
	\path[line width=1pt] (b33) edge (b24);
	\path[line width=1pt] (b33) edge (b34);
	\path[line width=1pt] (b24) edge (b34);
	\path[line width=1pt] (b25) edge (b34);
	\path[line width=1pt] (b24) edge (b25);
	\path[line width=1pt] (b35) edge (b34);
	\path[line width=1pt] (b25) edge (b35);

	\path[line width=1pt] (b30) edge (b41);
	\path[line width=1pt] (b31) edge (b41);
	\path[line width=1pt] (b31) edge (b42);
	\path[line width=1pt] (b32) edge (b42);
	
	\path[line width=1pt] (b33) edge (b44);
	\path[line width=1pt] (b34) edge (b44);
	\path[line width=1pt] (b34) edge (b45);
	\path[line width=1pt] (b35) edge (b45);
	
	\path[line width=1pt] (b42) edge (b41);
	
	\path[line width=1pt] (b45) edge (b44);

	\end{tikzpicture}
	\caption{ The open triangular lattice where $\sigma$ spins are all in the bulk and boundary only consists of $\tau$ and $\mu$ spins. }\label{fig: Triangle lattice OBC}
\end{figure}
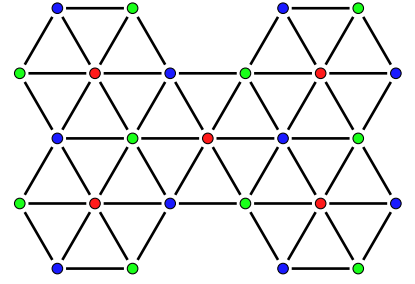

Now let us put this model on an open lattice as shown in Fig~\ref{fig: Triangle lattice OBC}, where the boundary only consists of $\tau$ and $\mu$ spins, and we consider Lindblad $\mathcal{L}=\mathcal{L}_{\text{bulk}}+\mathcal{L}_{\text{edge}}$. Here the $\mathcal{L}_{\text{bulk}}$ involves dissipators in \eqref{eq:2+1d ASPT-1} fully supported on this open lattice and the $\mathcal{L}_{\text{edge}}$ involves dissipators localized near the edges.  Let us focus on the subspace $\mathcal{C}$ which involves steady states of $\mathcal{L}_{\text{bulk}}$:
\[
\mathcal{C}=\{\rho_j: \mathcal{L}_{\text{bulk}}[\rho_j]=0\}.
\] For simplicity, we can assume boundary Lindblad preserves this subspace. Thus the subspace $\mathcal{C}$ includes the edge degree of freedom (DOF). 

To identify the subspace $\mathcal{C}$ and find associated operators which characterize the edge DOF, we can apply $U(\text{CCZ})$, which is truncated at the edge, to make bulk and edge decoupled, thus recovering the Lindbladian \eqref{eq:trivial Lindbladian} without boundary terms under OBC. For $\tau$ and $\mu$ spins, the bulk steady states are also the maximally mixed states and the edge density matrix is free to be chosen.  For $\sigma$ spins, there are also two $\sigma$ steady states, which are same as \eqref{eq:sigma ss} with $(\prod_i \sigma^x_i) \rho^{ss}_{\sigma,\pm}=\pm 1$. 
Thus we can define $\rho_{j,\pm}=U(\text{CCZ})\rho^{ss}_{\sigma,\pm}\otimes\rho^{\text{edge}}_{\tau,\mu,j}U^{\dagger}(\text{CCZ})$. Then, it is easy to check the following dressed edge operators preserve $\mathcal{C}$ and composes a complete set of Pauli operators for edge DOFs in $\mathcal{C}$:
\[\label{eq:dressed operators}
\begin{split}
\tilde{\tau}^{\alpha}_{j\in \text{edge}}=U(\text{CCZ}) \tau^{\alpha}_{j\in \text{edge}}U^{\dagger}(\text{CCZ}),\\
\tilde{\mu}^{\alpha}_{j\in \text{edge}}=U(\text{CCZ})\ \mu^{\alpha}_{j\in \text{edge}}U^{\dagger}(\text{CCZ}),
\end{split}
\]
where $\alpha=x,y,z$. In particular, $\tilde{\tau}^{z}_{j\in \text{edge}}=\tau^{z}_{j\in \text{edge}},\quad \tilde{\mu}^{z}_{j\in \text{edge}}=\mu^{z}_{j\in \text{edge}}$.

We can identify that these dressed edge operators transform under the symmetry as follows:
\[\label{eq:edge symmetry}
\begin{split}
&U_A\tilde{\tau}^{\alpha}_{j\in \text{edge}}U^{\dagger}_A
\\=&U(\text{CCZ})U_{\text{edge}}(\text{CZ})  \tau^{\alpha}_{j\in \text{edge}}U^{\dagger}_{\text{edge}}(\text{CZ})U^{\dagger}(\text{CCZ})\\
=&\tilde{U}_{\text{edge}}(\text{CZ})\tilde{\tau}^{\alpha}_{j\in \text{edge}}\tilde{U}^{\dagger}_{\text{edge}}(\text{CZ}),\\
&U_B U_C \tilde{\tau}^{\alpha}_{j\in \text{edge}} U^{\dagger}_B U^{\dagger}_C\\
=&U(\text{CCZ})U_B U_C \tau^{\alpha}_{j\in \text{edge}} U^{\dagger}_B U^{\dagger}_C U^{\dagger}(\text{CCZ})\\
=&U(\text{CCZ})\prod_{i,k \in \text{edge}}\tau^x_i\mu^x_k \tau^{\alpha}_{j\in \text{edge}} (\prod_{i,k \in \text{edge}}\tau^x_i\mu^x_k)^{\dagger} U^{\dagger}(\text{CCZ})\\
=&\prod_{i,k \in \text{edge}}\tilde{\tau}^x_i\tilde{\mu}^x_k \tau^{\alpha}_{j\in \text{edge}} (\prod_{i,k \in \text{edge}}\tilde{\tau}^x_i\tilde{\mu}^x_k)^{\dagger}
\end{split}
\]
and the result for $\mu$ operators is the same as $\tau$ operators. Moreover, one can also check the symmetry action on the $\rho_{j,\pm}$ as
\[
\begin{split}
&U_A \rho_{j,\pm}=\pm \tilde{U}_{\text{edge}}(\text{CZ})\rho_{j,\pm},\\
&U_B U_C \rho_{j,\pm}U_{B}U_C
=\prod_{i,j \in \text{edge}}\tilde{\tau}^x_i\tilde{\mu}^x_j\rho_{j,\pm}\prod_{i,j \in \text{edge}}\tilde{\tau}^x_i\tilde{\mu}^x_j.
\end{split}
\]

Thus, strong $U_A$ and weak $U_{B}U_C$  restrict to strong $ \tilde{U}(\text{CZ})$ and weak $\prod_{i,j \in \text{edge}} \tilde{\tau}^x_i\tilde{\mu}^x_j$ on the edge DOFs, which is same as {\it Example 2} in section \ref{sec:lattice model}.

The idea above can also inspire the other example, where we consider the strong $K=\Z^B_2\times \Z^C_2$ and weak $G=\Z^A_2$ symmetry. The corresponding ASPT can be constructed by decorating the $\Z^A_2$ domain wall with the $(1+1)$-D $\Z^B_2\times \Z^C_2$ SPT, or equivalently, decorating codimension-2 $\Z^A_2\times\Z^B_2$ defects with $\Z^C_2$ charges, and thus is characterized by
the nontrivial element of $H^1(\Z^A_2, H^2(\Z^B_2\times\Z^C_2, U(1))=H^2(\Z^A_2\times \Z^B_2, H^1(\Z^C_2, U(1))$. 


On the lattice, the DDW construction can also be realized by conjugating a trivial mixed-state using $U(\text{CCZ})$. Hence we consider this Lindbladian with $H=0$ under PBC:
\[\label{eq:trivial Lindbladian-2}
\begin{split}
&l^1_{i\in B}=\sum_{<i,j>\in B}\tau^z_i \tau^z_j\frac{1-\tau^x_i}{2},\\
&l^1_{i\in C}=\sum_{<i,j>\in C}\mu^z_i \mu^z_j\frac{1-\mu^x_i}{2},\\
&l^2_{i\in A}=\sigma^z_i,\quad l^3_{i\in A}=\sigma^x_i,
\end{split}
\]
where its steady state in even $K$-sector is a trivial symmetric product state 
\[
\begin{split}
&\rho_{\text{trivial}}=\rho_{\tau}^\rightarrow\otimes \rho_{\mu}^\rightarrow\otimes I_{\sigma},\\
&\rho^{ss}_{\tau/\mu, +}=|\rightarrow \rightarrow\cdots \rightarrow\rangle\langle \rightarrow\rightarrow\cdots\rightarrow|\equiv\rho_{\tau/\mu}^\rightarrow,\\
&\rho^{ss}_{\sigma}=I_{\sigma}.
\end{split}
\]
After conjugating the Lindbladian \eqref{eq:trivial Lindbladian-2} by $U(\text{CCZ})$, we obtain
\[\label{eq:2+1d ASPT-2}
\begin{split}
&l^1_{i\in B}=\sum_{<i,j>\in B}\tau^z_i \tau^z_j\frac{1-O_i}{2},\\
&l^1_{i\in C}=\sum_{<i,j>\in C}\mu^z_i \mu^z_j\frac{1-O_i}{2},\\
&l^2_{i\in A}=\sigma^z_i,\quad l^3_{i\in A}=O_i.
\end{split}
\]
with steady state  $\rho^{ss}_{\text{ASPT}}=U(\text{CCZ})\rho_{\text{trivial}}U^{\dagger}(\text{CCZ})$. In appendix \ref{app:2+1d ASPT}, we show how to detect its SPT feature and edge theories
possess the same anomalous symmetry as  {\it Example 3} in Section \ref{sec:lattice model}, where the calculation is similar to that of the first example with $K=\Z^A_2$ and $G=\Z^B_2\times \Z^C_2$.

~
\\~\\

\subsection{Separation of mixed-state phases between $\rho_{\text{trivial}}$ and $\rho^{ss}_{\text{ASPT}}$}\label{sec:separation}
Above, we construct the steady state $\rho^{ss}_{\text{ASPT}}$ using the DDW method. From the spirit of the DDW approach and from the anomalies of edge theories, it is expected that the steady state realizes a nontrivial ASPT. In this section, we will show that $\rho^{ss}_{\text{ASPT}}$ indeed belongs to a distinct symmetric mixed-state quantum phase from the trivial product state. We note that in \cite{ma2023aspt} such a statement is proved for the $(1+1)$-D ASPT $\rho_{\text{cluster}}$, which relies on the string order. However, for generic $(2+1)$-D ASPT there is no string order, which makes it difficult to generalize their approach to $(2+1)$ dimensions. Here, with our results on mixed-state anomalies in Section \ref{sec:anomaly constrain} and the bulk-boundary correspondence established in Section \ref{sec:2+1dASPT}, we are able to prove the separation of mixed-state phases in $(2+1)$-D by generalizing the approach in \cite{PhysRevB.90.235137}.
\begin{figure}[htb]
 \centering
\includegraphics[width=0.65\linewidth]{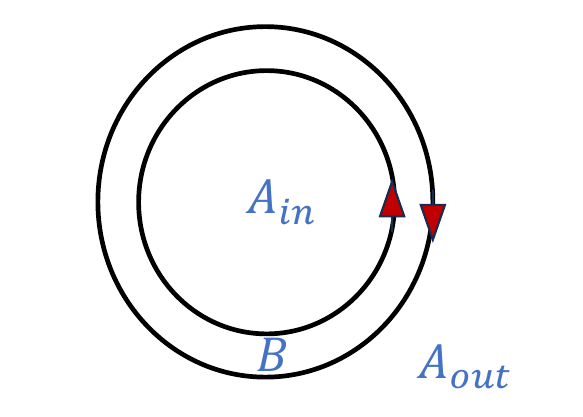}

\caption{Tripartition $A_{in} \cup B\cup A_{out}$ of a 2d lattice with no boundary. The red arrow depicts the orientation of $\partial A_{in}$ and $\partial A_{out}$. }
\label{fig:tripartition}
\end{figure}


We prove the above statement by contradiction. Assume that there is a symmetric FDLC $\mathcal{N}$, with each layer preserving the $\Gamma=K\times G$ symmetry, s.t. $\rho^{ss}_{\text{ASPT}}=\mathcal{N}[\rho_{\text{trivial}}]$. Notably, the following discussion applies for both the case $K=\Z_2^{BC},G=\Z_2^{A}$ and $K=\Z_2^A,G=\Z_2^{BC}$. Now take a 2D infinite lattice and take the tripartition $A_{in}\cup B\cup A_{out}$. $B$ is a finite-width strip separating the inner and outer region, and we require its width (in units of lattice spacing) to be large compared to the depth of $\mathcal{N}$. See Fig. \ref{fig:tripartition}. Since here both $K$ and $G$ are onsite symmetries, $\mathscr{U}(\gamma)=\mathscr{U}_{A_{in}}(\gamma)\circ\mathscr{U}_B(\gamma)\circ \mathscr{U}_{A_{out}}(\gamma)$. 

We define $\mathcal{N}_{in}$ to be a restriction of $\mathcal{N}$ to region $A'_{in}$ , with $A_{in}\subsetneq A'_{in}\subsetneq A_{in}\cup B$. Thus $\rho_{\text{inter}}=\mathcal{N}_{in}[\rho_{\text{trivial}}]$ looks like $\rho^{ss}_{\text{ASPT}}$ in $A_{in}$ but remains the trivial product state in $A_{out}$. From the analysis in Section \ref{sec:2+1dASPT}, we have 
\[
\mathscr{U}_{A_{in}}(\gamma)[\rho_{\text{inter}}]=\mathscr{V}_{\partial A_{in}}(\gamma)[\rho_{in}],
\]
where $\partial A_{in}$ is a strip near the boundary of $A_{in}$, and $\mathscr{V}_{\partial A_{in}}$ is the symmetry action on $\partial A_{in}$, which is anomalous. In addition, $\mathscr{U}_B(\gamma)[\rho_{\text{inter}}]=\rho_{\text{inter}}$. Thus
\[
\begin{split}
\mathscr{U}(\gamma)[\rho_{\text{inter}}]&=\mathscr{V}_{\partial A_{in}}(\gamma)\circ \mathscr U_B(\gamma)[\rho_{\text{inter}}]\\
&\equiv {\mathscr{I}}(\gamma)[\rho_{\text{inter}}]=\rho_{\text{inter}}.
\end{split}
\]
Therefore, $\mathscr{I}(\gamma)$ is an anomalous symmetry action, supported on the 1D interface between $A_{in}$ and $A_{out}$, which means that its action is characterized by a nontrivial element in $H^3(\Gamma,U(1))/H^3(G,U(1))$. However, this contradicts Theorem 1: Since $\rho_{\text{inter}}$ is obtained from the trivial product state from $\mathcal{N}_{in}$, a symmetric FDLC, it should be anomaly free. In conclusion, such FDLC connecting $\rho_{\text{trivial}}$ to $\rho^{ss}_{\text{ASPT}}$ does not exist. 

In the above, we hide some subtleties by abusing Theorem 1. Strictly speaking, since we are analyzing anomalies of the effective symmetry action on the interface, instead of the original global symmetry transformation, the validity of Theorem 1 needs to be verified in this context. We bridge this gap below. First, we can apply the same trick as in the proof of Theorem 1, i.e., purifying both the initial state $\rho_{\text{trivial}}$ and the quantum channel $\mathcal{N}_{in}$, thus arriving at a purification of $\rho_{\text{inter}}$, denoted as $|\psi_{\text{inter}}\rangle$, which should be SRE. For $|\psi_{\text{inter}}\rangle$, we can again localize the global symmetry transformation to the interface, where the effective symmetry action should be anomaly free (See Lemma 3 in \cite{PhysRevB.90.235137}). It then contradicts with the anomalous symmetry action on the interface for $\rho_{\text{inter}}$, with similar reasons to our original proof of Theorem 1.   

So far, we have not utilized any special properties of the two models we constructed, apart from the boundary anomaly, which we believe is a generic feature of ASPT. Thus the above argument can be applied to other ASPTs. It can also be generalized to prove the separation of mixed-state phases between two symmetric states (from the DDW construction) $\rho_1,\rho_2$ with different types of boundary anomalies, with the following adjustments: First, we still assume the existence of symmetric FDLC $\mathcal{N}$, $\rho_2=\mathcal{N}[\rho_1]$ and construct $\rho_{\text{inter}}=\mathcal{N}_{in}[\rho_1]$. Second, apply the same type of analysis, we will get $\mathscr{U}_{A_{out}}(\gamma)[\rho_{\text{inter}}]=\mathscr{V}_{\partial A_{in}}(\gamma)\circ \mathscr U_B(\gamma)\circ\mathscr{V}_{\partial A_{out}}(\gamma)[\rho_{\text{inter}}]\equiv {\mathscr{I}}(\gamma)[\rho_{\text{inter}}]$, where the anomalous action $\mathscr{V}_{\partial A_{in}},\mathscr{V}_{\partial A_{out}}$ are characterized by different nontrivial elements $[\Omega_1],[\Omega_2]$ in $H^3(\Gamma,U(1))/H^3(G,U(1))$. Thus $\mathscr{I}(\gamma)$ is characterized by $[\Omega_1][\Omega_2]^{-1}\neq [1]$. The inverse comes from the opposite orientations of $\partial A_{in}$ and $\partial A_{out}$. Finally, note that states constructed from the DDW method become trivial when we only have weak symmetries. Therefore, we can construct a FDLC $\varepsilon$ with the weak symmetry $G$ (each layer of $\varepsilon$ is $G$-symmetric) to prepare $\rho_1$ from the trivial symmetric product state, so $\rho_{\text{inter}}=\mathcal{N}_{in}\circ\varepsilon[\rho_{\text{trivial}}]$, which leads to contradiction with Theorem 1.

\section{Discussion}
In the end, we list some interesting future directions.

\begin{enumerate}
\item In this paper we only focus on symmetries of the direct product product type. In general the full symmetry group can be a group extension $1\rightarrow K\rightarrow \Gamma\rightarrow G\rightarrow 1$, where $K$ is a normal subgroup of $\Gamma$ and $G=\Gamma/K$. For example, for nontrivial group extension $\Gamma$, the so-called intrinsically mixed ASPT is constructed \cite{ma2023topological}, which is similar to intrinsically gapless SPT in closed systems \cite{PhysRevB.104.075132,PhysRevB.107.125158,li2022symmetry,10.21468/SciPostPhys.12.6.196,PhysRevB.107.245127,li2023intrinsically,Wen:2023otf}. It is intriguing to work out the bulk-edge correspondence in that case. 
\item   Here we have only investigated anomalies of $0$-form global symmetries. It is interesting to generalize the discussion to anomalies involving higher-form symmetries \cite{seiberg2014generalized}. Although we do not have a systematic treatment yet, we note that many aspects of the recently investigated mixed-state topological order in decohered toric code can be understood from 1-form symmetry anomaly. Recall that the toric code model has two 1-form symmetries, $\Z_2^{(1),e}$ and $\Z_2^{(1),m}$, whose defects are $e,m$ anyons, respectively. The nontrivial mutual statistics between $e,m$ indicates a mixed anomaly between $\Z_2^{(1),e}$ and $\Z_2^{(1),m}$ \cite{wen2019emergent}. By adding some single-qubit phase errors (or bit-flip errors), one of the 1-form symmetries becomes a weak symmetry. Still, the remaining symmetry has strong-weak mixed anomalies, which can be easily seen using the superoperator representation. The classical memory \cite{fan2023diagnostics,bao2023mixed,lee2023quantum} and anyon braiding statistics \cite{wang2023intrinsic,liu2023efficient} can both be viewed as consequences of the anomaly. Models with anomalous strong 1-form symmetries are also investigated in \cite{wang2023intrinsic,sohal2024noisy,cheng2024towards}, where anomalies lead to a landscape of intrinsic mixed-state topological order. 
\item It is also interesting to extend the discussion to mixed anomalies between internal symmetries and spatial symmetries, e.g., translation symmetries, generalizing the Lieb-Schultz-Mattis (LSM) theorem \cite{Lieb:1961aa,Oshikawa:2000aa, Hastings:2004ab,Fuji-SymmetryProtection-PRB2016,Watanabe:2015aa,PhysRevB.96.205106,Yao:2019aa,PhysRevX.8.011040,Yao:2021aa,Yao:2023bnj,10.21468/SciPostPhys.15.2.051} to open quantum systems. Various versions of the LSM theorem in open quantum systems have already been proposed recently \cite{PhysRevLett.132.070402,zhou2023reviving,hsin2023anomalies,zang2023detecting,lessa2024mixed}, but the general LSM constraint on mixed-state quantum phases remains to be uncovered.
\item Besides, the investigation of anomalous global symmetry in this paper is based on anomaly cocycle of the Else-Nayak method. Recent studies have proved that in (1+1)-D closed systems, the anomaly cocycle must be trivial to enable consistent lattice gauging \cite{Seifnashri:2023dpa}.  Thus it would be interesting to further explore the understanding of anomaly in open systems and its constraint on mixed-state phases from the view of the obstruction to gauging. 

\item   We only briefly discuss anomalies in higher dimensions without giving examples. It is desirable to see what types of nontrivial phases with anomalies can be realized in higher dimensions, and what is the higher dimensional generalization of the anomaly-enforced boundary correlation discussed in Section \ref{sec:lattice model} and \ref{sec:boundary cor}.
\end{enumerate}

\begin{acknowledgements}
{\it Acknowledgments}.---We thank Meng Cheng, Zhong Wang, Ruochen Ma, Jian-Hao Zhang and Liang Mao for helpful discussions. We are especially grateful to Zhengzhi Wu for numerous discussions.
This work is supported by NSFC under Grant No. 12125405.

{\it Note added:} In completing this manuscript, we became aware of several independent related works on mixed-state SPT phases, which appeared on arXiv in the same week. Ref.~\cite{ma2024symmetry,xue2024tensor} investigate mixed-state SPT phases from the perspective of doubled space. Ref.~\cite{guo2024locally} discusses mixed-state SPT phases using locally purifiable density operators. 
\end{acknowledgements}
\bibliography{bib}

\begin{thebibliography}{113}%
\makeatletter
\providecommand \@ifxundefined [1]{%
 \@ifx{#1\undefined}
}%
\providecommand \@ifnum [1]{%
 \ifnum #1\expandafter \@firstoftwo
 \else \expandafter \@secondoftwo
 \fi
}%
\providecommand \@ifx [1]{%
 \ifx #1\expandafter \@firstoftwo
 \else \expandafter \@secondoftwo
 \fi
}%
\providecommand \natexlab [1]{#1}%
\providecommand \enquote  [1]{``#1''}%
\providecommand \bibnamefont  [1]{#1}%
\providecommand \bibfnamefont [1]{#1}%
\providecommand \citenamefont [1]{#1}%
\providecommand \href@noop [0]{\@secondoftwo}%
\providecommand \href [0]{\begingroup \@sanitize@url \@href}%
\providecommand \@href[1]{\@@startlink{#1}\@@href}%
\providecommand \@@href[1]{\endgroup#1\@@endlink}%
\providecommand \@sanitize@url [0]{\catcode `\\12\catcode `\$12\catcode `\&12\catcode `\#12\catcode `\^12\catcode `\_12\catcode `\%12\relax}%
\providecommand \@@startlink[1]{}%
\providecommand \@@endlink[0]{}%
\providecommand \url  [0]{\begingroup\@sanitize@url \@url }%
\providecommand \@url [1]{\endgroup\@href {#1}{\urlprefix }}%
\providecommand \urlprefix  [0]{URL }%
\providecommand \Eprint [0]{\href }%
\providecommand \doibase [0]{http://dx.doi.org/}%
\providecommand \selectlanguage [0]{\@gobble}%
\providecommand \bibinfo  [0]{\@secondoftwo}%
\providecommand \bibfield  [0]{\@secondoftwo}%
\providecommand \translation [1]{[#1]}%
\providecommand \BibitemOpen [0]{}%
\providecommand \bibitemStop [0]{}%
\providecommand \bibitemNoStop [0]{.\EOS\space}%
\providecommand \EOS [0]{\spacefactor3000\relax}%
\providecommand \BibitemShut  [1]{\csname bibitem#1\endcsname}%
\let\auto@bib@innerbib\@empty
\bibitem [{\citenamefont {'t~Hooft}(1980)}]{tHooft:1979rat}%
  \BibitemOpen
  \bibfield  {author} {\bibinfo {author} {\bibfnamefont {Gerard}\ \bibnamefont {'t~Hooft}},\ }\bibfield  {title} {\enquote {\bibinfo {title} {{Naturalness, chiral symmetry, and spontaneous chiral symmetry breaking}},}\ }\href {\doibase 10.1007/978-1-4684-7571-5_9} {\bibfield  {journal} {\bibinfo  {journal} {NATO Sci. Ser. B}\ }\textbf {\bibinfo {volume} {59}},\ \bibinfo {pages} {135--157} (\bibinfo {year} {1980})}\BibitemShut {NoStop}%
\bibitem [{\citenamefont {Else}\ and\ \citenamefont {Nayak}(2014)}]{PhysRevB.90.235137}%
  \BibitemOpen
  \bibfield  {author} {\bibinfo {author} {\bibfnamefont {Dominic~V.}\ \bibnamefont {Else}}\ and\ \bibinfo {author} {\bibfnamefont {Chetan}\ \bibnamefont {Nayak}},\ }\bibfield  {title} {\enquote {\bibinfo {title} {Classifying symmetry-protected topological phases through the anomalous action of the symmetry on the edge},}\ }\href {\doibase 10.1103/PhysRevB.90.235137} {\bibfield  {journal} {\bibinfo  {journal} {Phys. Rev. B}\ }\textbf {\bibinfo {volume} {90}},\ \bibinfo {pages} {235137} (\bibinfo {year} {2014})}\BibitemShut {NoStop}%
\bibitem [{\citenamefont {Essin}\ and\ \citenamefont {Hermele}(2013)}]{PhysRevB.87.104406}%
  \BibitemOpen
  \bibfield  {author} {\bibinfo {author} {\bibfnamefont {Andrew~M.}\ \bibnamefont {Essin}}\ and\ \bibinfo {author} {\bibfnamefont {Michael}\ \bibnamefont {Hermele}},\ }\bibfield  {title} {\enquote {\bibinfo {title} {Classifying fractionalization: Symmetry classification of gapped ${\mathbb{z}}_{2}$ spin liquids in two dimensions},}\ }\href {\doibase 10.1103/PhysRevB.87.104406} {\bibfield  {journal} {\bibinfo  {journal} {Phys. Rev. B}\ }\textbf {\bibinfo {volume} {87}},\ \bibinfo {pages} {104406} (\bibinfo {year} {2013})}\BibitemShut {NoStop}%
\bibitem [{\citenamefont {Barkeshli}\ \emph {et~al.}(2019)\citenamefont {Barkeshli}, \citenamefont {Bonderson}, \citenamefont {Cheng},\ and\ \citenamefont {Wang}}]{Barkeshli:2014cna}%
  \BibitemOpen
  \bibfield  {author} {\bibinfo {author} {\bibfnamefont {Maissam}\ \bibnamefont {Barkeshli}}, \bibinfo {author} {\bibfnamefont {Parsa}\ \bibnamefont {Bonderson}}, \bibinfo {author} {\bibfnamefont {Meng}\ \bibnamefont {Cheng}}, \ and\ \bibinfo {author} {\bibfnamefont {Zhenghan}\ \bibnamefont {Wang}},\ }\bibfield  {title} {\enquote {\bibinfo {title} {{Symmetry Fractionalization, Defects, and Gauging of Topological Phases}},}\ }\href {\doibase 10.1103/PhysRevB.100.115147} {\bibfield  {journal} {\bibinfo  {journal} {Phys. Rev. B}\ }\textbf {\bibinfo {volume} {100}},\ \bibinfo {pages} {115147} (\bibinfo {year} {2019})},\ \Eprint {http://arxiv.org/abs/1410.4540} {arXiv:1410.4540 [cond-mat.str-el]} \BibitemShut {NoStop}%
\bibitem [{\citenamefont {Chen}\ \emph {et~al.}(2015)\citenamefont {Chen}, \citenamefont {Burnell}, \citenamefont {Vishwanath},\ and\ \citenamefont {Fidkowski}}]{chen2015anomalous}%
  \BibitemOpen
  \bibfield  {author} {\bibinfo {author} {\bibfnamefont {Xie}\ \bibnamefont {Chen}}, \bibinfo {author} {\bibfnamefont {Fiona~J}\ \bibnamefont {Burnell}}, \bibinfo {author} {\bibfnamefont {Ashvin}\ \bibnamefont {Vishwanath}}, \ and\ \bibinfo {author} {\bibfnamefont {Lukasz}\ \bibnamefont {Fidkowski}},\ }\bibfield  {title} {\enquote {\bibinfo {title} {Anomalous symmetry fractionalization and surface topological order},}\ }\href@noop {} {\bibfield  {journal} {\bibinfo  {journal} {Physical Review X}\ }\textbf {\bibinfo {volume} {5}},\ \bibinfo {pages} {041013} (\bibinfo {year} {2015})}\BibitemShut {NoStop}%
\bibitem [{\citenamefont {Tarantino}\ \emph {et~al.}(2016)\citenamefont {Tarantino}, \citenamefont {Lindner},\ and\ \citenamefont {Fidkowski}}]{tarantino2016symmetry}%
  \BibitemOpen
  \bibfield  {author} {\bibinfo {author} {\bibfnamefont {Nicolas}\ \bibnamefont {Tarantino}}, \bibinfo {author} {\bibfnamefont {Netanel~H}\ \bibnamefont {Lindner}}, \ and\ \bibinfo {author} {\bibfnamefont {Lukasz}\ \bibnamefont {Fidkowski}},\ }\bibfield  {title} {\enquote {\bibinfo {title} {Symmetry fractionalization and twist defects},}\ }\href@noop {} {\bibfield  {journal} {\bibinfo  {journal} {New Journal of Physics}\ }\textbf {\bibinfo {volume} {18}},\ \bibinfo {pages} {035006} (\bibinfo {year} {2016})}\BibitemShut {NoStop}%
\bibitem [{\citenamefont {Cheng}\ \emph {et~al.}(2016)\citenamefont {Cheng}, \citenamefont {Zaletel}, \citenamefont {Barkeshli}, \citenamefont {Vishwanath},\ and\ \citenamefont {Bonderson}}]{cheng2016translational}%
  \BibitemOpen
  \bibfield  {author} {\bibinfo {author} {\bibfnamefont {Meng}\ \bibnamefont {Cheng}}, \bibinfo {author} {\bibfnamefont {Michael}\ \bibnamefont {Zaletel}}, \bibinfo {author} {\bibfnamefont {Maissam}\ \bibnamefont {Barkeshli}}, \bibinfo {author} {\bibfnamefont {Ashvin}\ \bibnamefont {Vishwanath}}, \ and\ \bibinfo {author} {\bibfnamefont {Parsa}\ \bibnamefont {Bonderson}},\ }\bibfield  {title} {\enquote {\bibinfo {title} {Translational symmetry and microscopic constraints on symmetry-enriched topological phases: A view from the surface},}\ }\href@noop {} {\bibfield  {journal} {\bibinfo  {journal} {Physical Review X}\ }\textbf {\bibinfo {volume} {6}},\ \bibinfo {pages} {041068} (\bibinfo {year} {2016})}\BibitemShut {NoStop}%
\bibitem [{\citenamefont {Barkeshli}\ \emph {et~al.}(2020)\citenamefont {Barkeshli}, \citenamefont {Bonderson}, \citenamefont {Cheng}, \citenamefont {Jian},\ and\ \citenamefont {Walker}}]{barkeshli2020reflection}%
  \BibitemOpen
  \bibfield  {author} {\bibinfo {author} {\bibfnamefont {Maissam}\ \bibnamefont {Barkeshli}}, \bibinfo {author} {\bibfnamefont {Parsa}\ \bibnamefont {Bonderson}}, \bibinfo {author} {\bibfnamefont {Meng}\ \bibnamefont {Cheng}}, \bibinfo {author} {\bibfnamefont {Chao-Ming}\ \bibnamefont {Jian}}, \ and\ \bibinfo {author} {\bibfnamefont {Kevin}\ \bibnamefont {Walker}},\ }\bibfield  {title} {\enquote {\bibinfo {title} {Reflection and time reversal symmetry enriched topological phases of matter: path integrals, non-orientable manifolds, and anomalies},}\ }\href@noop {} {\bibfield  {journal} {\bibinfo  {journal} {Communications in Mathematical Physics}\ }\textbf {\bibinfo {volume} {374}},\ \bibinfo {pages} {1021--1124} (\bibinfo {year} {2020})}\BibitemShut {NoStop}%
\bibitem [{\citenamefont {Barkeshli}\ and\ \citenamefont {Cheng}(2020)}]{barkeshli2020relative}%
  \BibitemOpen
  \bibfield  {author} {\bibinfo {author} {\bibfnamefont {Maissam}\ \bibnamefont {Barkeshli}}\ and\ \bibinfo {author} {\bibfnamefont {Meng}\ \bibnamefont {Cheng}},\ }\bibfield  {title} {\enquote {\bibinfo {title} {Relative anomalies in (2+ 1) d symmetry enriched topological states},}\ }\href@noop {} {\bibfield  {journal} {\bibinfo  {journal} {SciPost Physics}\ }\textbf {\bibinfo {volume} {8}},\ \bibinfo {pages} {028} (\bibinfo {year} {2020})}\BibitemShut {NoStop}%
\bibitem [{\citenamefont {Bulmash}\ and\ \citenamefont {Barkeshli}(2020)}]{bulmash2020absolute}%
  \BibitemOpen
  \bibfield  {author} {\bibinfo {author} {\bibfnamefont {Daniel}\ \bibnamefont {Bulmash}}\ and\ \bibinfo {author} {\bibfnamefont {Maissam}\ \bibnamefont {Barkeshli}},\ }\bibfield  {title} {\enquote {\bibinfo {title} {Absolute anomalies in (2+ 1) d symmetry-enriched topological states and exact (3+ 1) d constructions},}\ }\href@noop {} {\bibfield  {journal} {\bibinfo  {journal} {Physical Review Research}\ }\textbf {\bibinfo {volume} {2}},\ \bibinfo {pages} {043033} (\bibinfo {year} {2020})}\BibitemShut {NoStop}%
\bibitem [{\citenamefont {Zou}\ \emph {et~al.}(2021)\citenamefont {Zou}, \citenamefont {He},\ and\ \citenamefont {Wang}}]{PhysRevX.11.031043}%
  \BibitemOpen
  \bibfield  {author} {\bibinfo {author} {\bibfnamefont {Liujun}\ \bibnamefont {Zou}}, \bibinfo {author} {\bibfnamefont {Yin-Chen}\ \bibnamefont {He}}, \ and\ \bibinfo {author} {\bibfnamefont {Chong}\ \bibnamefont {Wang}},\ }\bibfield  {title} {\enquote {\bibinfo {title} {Stiefel liquids: Possible non-lagrangian quantum criticality from intertwined orders},}\ }\href {\doibase 10.1103/PhysRevX.11.031043} {\bibfield  {journal} {\bibinfo  {journal} {Phys. Rev. X}\ }\textbf {\bibinfo {volume} {11}},\ \bibinfo {pages} {031043} (\bibinfo {year} {2021})}\BibitemShut {NoStop}%
\bibitem [{\citenamefont {Kawagoe}\ and\ \citenamefont {Levin}(2021)}]{kawagoe2021anomalies}%
  \BibitemOpen
  \bibfield  {author} {\bibinfo {author} {\bibfnamefont {Kyle}\ \bibnamefont {Kawagoe}}\ and\ \bibinfo {author} {\bibfnamefont {Michael}\ \bibnamefont {Levin}},\ }\bibfield  {title} {\enquote {\bibinfo {title} {Anomalies in bosonic symmetry-protected topological edge theories: Connection to f symbols and a method of calculation},}\ }\href@noop {} {\bibfield  {journal} {\bibinfo  {journal} {Physical Review B}\ }\textbf {\bibinfo {volume} {104}},\ \bibinfo {pages} {115156} (\bibinfo {year} {2021})}\BibitemShut {NoStop}%
\bibitem [{\citenamefont {Cheng}\ and\ \citenamefont {Seiberg}(2023{\natexlab{a}})}]{Cheng:2022sgb}%
  \BibitemOpen
  \bibfield  {author} {\bibinfo {author} {\bibfnamefont {Meng}\ \bibnamefont {Cheng}}\ and\ \bibinfo {author} {\bibfnamefont {Nathan}\ \bibnamefont {Seiberg}},\ }\bibfield  {title} {\enquote {\bibinfo {title} {{Lieb-Schultz-Mattis, Luttinger, and 't Hooft - anomaly matching in lattice systems}},}\ }\href {\doibase 10.21468/SciPostPhys.15.2.051} {\bibfield  {journal} {\bibinfo  {journal} {SciPost Phys.}\ }\textbf {\bibinfo {volume} {15}},\ \bibinfo {pages} {051} (\bibinfo {year} {2023}{\natexlab{a}})},\ \Eprint {http://arxiv.org/abs/2211.12543} {arXiv:2211.12543 [cond-mat.str-el]} \BibitemShut {NoStop}%
\bibitem [{\citenamefont {Delmastro}\ \emph {et~al.}(2023)\citenamefont {Delmastro}, \citenamefont {Gomis}, \citenamefont {Hsin},\ and\ \citenamefont {Komargodski}}]{Delmastro:2022pfo}%
  \BibitemOpen
  \bibfield  {author} {\bibinfo {author} {\bibfnamefont {Diego~Gabriel}\ \bibnamefont {Delmastro}}, \bibinfo {author} {\bibfnamefont {Jaume}\ \bibnamefont {Gomis}}, \bibinfo {author} {\bibfnamefont {Po-Shen}\ \bibnamefont {Hsin}}, \ and\ \bibinfo {author} {\bibfnamefont {Zohar}\ \bibnamefont {Komargodski}},\ }\bibfield  {title} {\enquote {\bibinfo {title} {{Anomalies and symmetry fractionalization}},}\ }\href {\doibase 10.21468/SciPostPhys.15.3.079} {\bibfield  {journal} {\bibinfo  {journal} {SciPost Phys.}\ }\textbf {\bibinfo {volume} {15}},\ \bibinfo {pages} {079} (\bibinfo {year} {2023})},\ \Eprint {http://arxiv.org/abs/2206.15118} {arXiv:2206.15118 [hep-th]} \BibitemShut {NoStop}%
\bibitem [{\citenamefont {Saffman}\ \emph {et~al.}(2010)\citenamefont {Saffman}, \citenamefont {Walker},\ and\ \citenamefont {M\o{}lmer}}]{Saffman2010Rydberg}%
  \BibitemOpen
  \bibfield  {author} {\bibinfo {author} {\bibfnamefont {M.}~\bibnamefont {Saffman}}, \bibinfo {author} {\bibfnamefont {T.~G.}\ \bibnamefont {Walker}}, \ and\ \bibinfo {author} {\bibfnamefont {K.}~\bibnamefont {M\o{}lmer}},\ }\bibfield  {title} {\enquote {\bibinfo {title} {Quantum information with rydberg atoms},}\ }\href {\doibase 10.1103/RevModPhys.82.2313} {\bibfield  {journal} {\bibinfo  {journal} {Rev. Mod. Phys.}\ }\textbf {\bibinfo {volume} {82}},\ \bibinfo {pages} {2313--2363} (\bibinfo {year} {2010})}\BibitemShut {NoStop}%
\bibitem [{\citenamefont {Kjaergaard}\ \emph {et~al.}(2020)\citenamefont {Kjaergaard}, \citenamefont {Schwartz}, \citenamefont {Braumüller}, \citenamefont {Krantz}, \citenamefont {Wang}, \citenamefont {Gustavsson},\ and\ \citenamefont {Oliver}}]{kjaergaard2020superconducting}%
  \BibitemOpen
  \bibfield  {author} {\bibinfo {author} {\bibfnamefont {Morten}\ \bibnamefont {Kjaergaard}}, \bibinfo {author} {\bibfnamefont {Mollie~E.}\ \bibnamefont {Schwartz}}, \bibinfo {author} {\bibfnamefont {Jochen}\ \bibnamefont {Braumüller}}, \bibinfo {author} {\bibfnamefont {Philip}\ \bibnamefont {Krantz}}, \bibinfo {author} {\bibfnamefont {Joel I.-J.}\ \bibnamefont {Wang}}, \bibinfo {author} {\bibfnamefont {Simon}\ \bibnamefont {Gustavsson}}, \ and\ \bibinfo {author} {\bibfnamefont {William~D.}\ \bibnamefont {Oliver}},\ }\bibfield  {title} {\enquote {\bibinfo {title} {Superconducting qubits: Current state of play},}\ }\href {\doibase https://doi.org/10.1146/annurev-conmatphys-031119-050605} {\bibfield  {journal} {\bibinfo  {journal} {Annual Review of Condensed Matter Physics}\ }\textbf {\bibinfo {volume} {11}},\ \bibinfo {pages} {369--395} (\bibinfo {year} {2020})}\BibitemShut {NoStop}%
\bibitem [{\citenamefont {Bruzewicz}\ \emph {et~al.}(2019)\citenamefont {Bruzewicz}, \citenamefont {Chiaverini}, \citenamefont {McConnell},\ and\ \citenamefont {Sage}}]{Bruzewicz2019trapped}%
  \BibitemOpen
  \bibfield  {author} {\bibinfo {author} {\bibfnamefont {Colin~D.}\ \bibnamefont {Bruzewicz}}, \bibinfo {author} {\bibfnamefont {John}\ \bibnamefont {Chiaverini}}, \bibinfo {author} {\bibfnamefont {Robert}\ \bibnamefont {McConnell}}, \ and\ \bibinfo {author} {\bibfnamefont {Jeremy~M.}\ \bibnamefont {Sage}},\ }\bibfield  {title} {\enquote {\bibinfo {title} {{Trapped-ion quantum computing: Progress and challenges}},}\ }\href {\doibase 10.1063/1.5088164} {\bibfield  {journal} {\bibinfo  {journal} {Applied Physics Reviews}\ }\textbf {\bibinfo {volume} {6}},\ \bibinfo {pages} {021314} (\bibinfo {year} {2019})},\ \Eprint {http://arxiv.org/abs/https://pubs.aip.org/aip/apr/article-pdf/doi/10.1063/1.5088164/19742554/021314\_1\_online.pdf} {https://pubs.aip.org/aip/apr/article-pdf/doi/10.1063/1.5088164/19742554/021314\_1\_online.pdf} \BibitemShut {NoStop}%
\bibitem [{\citenamefont {Dennis}\ \emph {et~al.}(2002)\citenamefont {Dennis}, \citenamefont {Kitaev}, \citenamefont {Landahl},\ and\ \citenamefont {Preskill}}]{dennis2002topological}%
  \BibitemOpen
  \bibfield  {author} {\bibinfo {author} {\bibfnamefont {Eric}\ \bibnamefont {Dennis}}, \bibinfo {author} {\bibfnamefont {Alexei}\ \bibnamefont {Kitaev}}, \bibinfo {author} {\bibfnamefont {Andrew}\ \bibnamefont {Landahl}}, \ and\ \bibinfo {author} {\bibfnamefont {John}\ \bibnamefont {Preskill}},\ }\bibfield  {title} {\enquote {\bibinfo {title} {{Topological quantum memory}},}\ }\href {\doibase 10.1063/1.1499754} {\bibfield  {journal} {\bibinfo  {journal} {Journal of Mathematical Physics}\ }\textbf {\bibinfo {volume} {43}},\ \bibinfo {pages} {4452--4505} (\bibinfo {year} {2002})}\BibitemShut {NoStop}%
\bibitem [{\citenamefont {Diehl}\ \emph {et~al.}(2008)\citenamefont {Diehl}, \citenamefont {Micheli}, \citenamefont {Kantian}, \citenamefont {Kraus}, \citenamefont {B{\"u}chler},\ and\ \citenamefont {Zoller}}]{diehl2008quantum}%
  \BibitemOpen
  \bibfield  {author} {\bibinfo {author} {\bibfnamefont {S.}~\bibnamefont {Diehl}}, \bibinfo {author} {\bibfnamefont {A.}~\bibnamefont {Micheli}}, \bibinfo {author} {\bibfnamefont {A.}~\bibnamefont {Kantian}}, \bibinfo {author} {\bibfnamefont {B.}~\bibnamefont {Kraus}}, \bibinfo {author} {\bibfnamefont {H.~P.}\ \bibnamefont {B{\"u}chler}}, \ and\ \bibinfo {author} {\bibfnamefont {P.}~\bibnamefont {Zoller}},\ }\bibfield  {title} {\enquote {\bibinfo {title} {Quantum states and phases in driven open quantum systems with cold atoms},}\ }\href {\doibase 10.1038/nphys1073} {\bibfield  {journal} {\bibinfo  {journal} {Nature Physics}\ }\textbf {\bibinfo {volume} {4}},\ \bibinfo {pages} {878--883} (\bibinfo {year} {2008})}\BibitemShut {NoStop}%
\bibitem [{\citenamefont {Sieberer}\ \emph {et~al.}(2016)\citenamefont {Sieberer}, \citenamefont {Buchhold},\ and\ \citenamefont {Diehl}}]{sieberer2016keldysh}%
  \BibitemOpen
  \bibfield  {author} {\bibinfo {author} {\bibfnamefont {L~M}\ \bibnamefont {Sieberer}}, \bibinfo {author} {\bibfnamefont {M}~\bibnamefont {Buchhold}}, \ and\ \bibinfo {author} {\bibfnamefont {S}~\bibnamefont {Diehl}},\ }\bibfield  {title} {\enquote {\bibinfo {title} {Keldysh field theory for driven open quantum systems},}\ }\href {\doibase 10.1088/0034-4885/79/9/096001} {\bibfield  {journal} {\bibinfo  {journal} {Reports on Progress in Physics}\ }\textbf {\bibinfo {volume} {79}},\ \bibinfo {pages} {096001} (\bibinfo {year} {2016})}\BibitemShut {NoStop}%
\bibitem [{\citenamefont {Diehl}\ \emph {et~al.}(2010)\citenamefont {Diehl}, \citenamefont {Tomadin}, \citenamefont {Micheli}, \citenamefont {Fazio},\ and\ \citenamefont {Zoller}}]{diehl2010dynamical}%
  \BibitemOpen
  \bibfield  {author} {\bibinfo {author} {\bibfnamefont {Sebastian}\ \bibnamefont {Diehl}}, \bibinfo {author} {\bibfnamefont {Andrea}\ \bibnamefont {Tomadin}}, \bibinfo {author} {\bibfnamefont {Andrea}\ \bibnamefont {Micheli}}, \bibinfo {author} {\bibfnamefont {Rosario}\ \bibnamefont {Fazio}}, \ and\ \bibinfo {author} {\bibfnamefont {Peter}\ \bibnamefont {Zoller}},\ }\bibfield  {title} {\enquote {\bibinfo {title} {Dynamical phase transitions and instabilities in open atomic many-body systems},}\ }\href {\doibase 10.1103/PhysRevLett.105.015702} {\bibfield  {journal} {\bibinfo  {journal} {Phys. Rev. Lett.}\ }\textbf {\bibinfo {volume} {105}},\ \bibinfo {pages} {015702} (\bibinfo {year} {2010})}\BibitemShut {NoStop}%
\bibitem [{\citenamefont {Altman}\ \emph {et~al.}(2015)\citenamefont {Altman}, \citenamefont {Sieberer}, \citenamefont {Chen}, \citenamefont {Diehl},\ and\ \citenamefont {Toner}}]{altman2015two}%
  \BibitemOpen
  \bibfield  {author} {\bibinfo {author} {\bibfnamefont {Ehud}\ \bibnamefont {Altman}}, \bibinfo {author} {\bibfnamefont {Lukas~M.}\ \bibnamefont {Sieberer}}, \bibinfo {author} {\bibfnamefont {Leiming}\ \bibnamefont {Chen}}, \bibinfo {author} {\bibfnamefont {Sebastian}\ \bibnamefont {Diehl}}, \ and\ \bibinfo {author} {\bibfnamefont {John}\ \bibnamefont {Toner}},\ }\bibfield  {title} {\enquote {\bibinfo {title} {Two-dimensional superfluidity of exciton polaritons requires strong anisotropy},}\ }\href {\doibase 10.1103/PhysRevX.5.011017} {\bibfield  {journal} {\bibinfo  {journal} {Phys. Rev. X}\ }\textbf {\bibinfo {volume} {5}},\ \bibinfo {pages} {011017} (\bibinfo {year} {2015})}\BibitemShut {NoStop}%
\bibitem [{\citenamefont {Coser}\ and\ \citenamefont {P{\'{e}}rez-Garc{\'{i}}a}(2019)}]{coser2019classification}%
  \BibitemOpen
  \bibfield  {author} {\bibinfo {author} {\bibfnamefont {Andrea}\ \bibnamefont {Coser}}\ and\ \bibinfo {author} {\bibfnamefont {David}\ \bibnamefont {P{\'{e}}rez-Garc{\'{i}}a}},\ }\bibfield  {title} {\enquote {\bibinfo {title} {Classification of phases for mixed states via fast dissipative evolution},}\ }\href {\doibase 10.22331/q-2019-08-12-174} {\bibfield  {journal} {\bibinfo  {journal} {{Quantum}}\ }\textbf {\bibinfo {volume} {3}},\ \bibinfo {pages} {174} (\bibinfo {year} {2019})}\BibitemShut {NoStop}%
\bibitem [{\citenamefont {Lieu}\ \emph {et~al.}(2020)\citenamefont {Lieu}, \citenamefont {Belyansky}, \citenamefont {Young}, \citenamefont {Lundgren}, \citenamefont {Albert},\ and\ \citenamefont {Gorshkov}}]{Lieu2020SB}%
  \BibitemOpen
  \bibfield  {author} {\bibinfo {author} {\bibfnamefont {Simon}\ \bibnamefont {Lieu}}, \bibinfo {author} {\bibfnamefont {Ron}\ \bibnamefont {Belyansky}}, \bibinfo {author} {\bibfnamefont {Jeremy~T.}\ \bibnamefont {Young}}, \bibinfo {author} {\bibfnamefont {Rex}\ \bibnamefont {Lundgren}}, \bibinfo {author} {\bibfnamefont {Victor~V.}\ \bibnamefont {Albert}}, \ and\ \bibinfo {author} {\bibfnamefont {Alexey~V.}\ \bibnamefont {Gorshkov}},\ }\bibfield  {title} {\enquote {\bibinfo {title} {Symmetry breaking and error correction in open quantum systems},}\ }\href {\doibase 10.1103/PhysRevLett.125.240405} {\bibfield  {journal} {\bibinfo  {journal} {Phys. Rev. Lett.}\ }\textbf {\bibinfo {volume} {125}},\ \bibinfo {pages} {240405} (\bibinfo {year} {2020})}\BibitemShut {NoStop}%
\bibitem [{\citenamefont {Wang}\ \emph {et~al.}(2023{\natexlab{a}})\citenamefont {Wang}, \citenamefont {Dai}, \citenamefont {Wang},\ and\ \citenamefont {Wang}}]{wang2023topologically}%
  \BibitemOpen
  \bibfield  {author} {\bibinfo {author} {\bibfnamefont {Zijian}\ \bibnamefont {Wang}}, \bibinfo {author} {\bibfnamefont {Xu-Dong}\ \bibnamefont {Dai}}, \bibinfo {author} {\bibfnamefont {He-Ran}\ \bibnamefont {Wang}}, \ and\ \bibinfo {author} {\bibfnamefont {Zhong}\ \bibnamefont {Wang}},\ }\bibfield  {title} {\enquote {\bibinfo {title} {Topologically ordered steady states in open quantum systems},}\ }\href@noop {} {\bibfield  {journal} {\bibinfo  {journal} {arXiv preprint arXiv:2306.12482}\ } (\bibinfo {year} {2023}{\natexlab{a}})}\BibitemShut {NoStop}%
\bibitem [{\citenamefont {Liu}\ and\ \citenamefont {Lieu}(2024)}]{liu2024dissipative}%
  \BibitemOpen
  \bibfield  {author} {\bibinfo {author} {\bibfnamefont {Yu-Jie}\ \bibnamefont {Liu}}\ and\ \bibinfo {author} {\bibfnamefont {Simon}\ \bibnamefont {Lieu}},\ }\bibfield  {title} {\enquote {\bibinfo {title} {Dissipative phase transitions and passive error correction},}\ }\href {\doibase 10.1103/PhysRevA.109.022422} {\bibfield  {journal} {\bibinfo  {journal} {Phys. Rev. A}\ }\textbf {\bibinfo {volume} {109}},\ \bibinfo {pages} {022422} (\bibinfo {year} {2024})}\BibitemShut {NoStop}%
\bibitem [{\citenamefont {Dai}\ \emph {et~al.}(2023)\citenamefont {Dai}, \citenamefont {Wang}, \citenamefont {Wang},\ and\ \citenamefont {Wang}}]{dai2023steady}%
  \BibitemOpen
  \bibfield  {author} {\bibinfo {author} {\bibfnamefont {Xu-Dong}\ \bibnamefont {Dai}}, \bibinfo {author} {\bibfnamefont {Zijian}\ \bibnamefont {Wang}}, \bibinfo {author} {\bibfnamefont {He-Ran}\ \bibnamefont {Wang}}, \ and\ \bibinfo {author} {\bibfnamefont {Zhong}\ \bibnamefont {Wang}},\ }\bibfield  {title} {\enquote {\bibinfo {title} {Steady-state topological order},}\ }\href@noop {} {\bibfield  {journal} {\bibinfo  {journal} {arXiv preprint arXiv:2310.17612}\ } (\bibinfo {year} {2023})}\BibitemShut {NoStop}%
\bibitem [{\citenamefont {Rakovszky}\ \emph {et~al.}(2023)\citenamefont {Rakovszky}, \citenamefont {Gopalakrishnan},\ and\ \citenamefont {von Keyserlingk}}]{rakovszky2023defining}%
  \BibitemOpen
  \bibfield  {author} {\bibinfo {author} {\bibfnamefont {Tibor}\ \bibnamefont {Rakovszky}}, \bibinfo {author} {\bibfnamefont {Sarang}\ \bibnamefont {Gopalakrishnan}}, \ and\ \bibinfo {author} {\bibfnamefont {Curt}\ \bibnamefont {von Keyserlingk}},\ }\bibfield  {title} {\enquote {\bibinfo {title} {Defining stable phases of open quantum systems},}\ }\href@noop {} {\bibfield  {journal} {\bibinfo  {journal} {arXiv preprint arXiv:2308.15495}\ } (\bibinfo {year} {2023})}\BibitemShut {NoStop}%
\bibitem [{\citenamefont {Lu}\ \emph {et~al.}(2023)\citenamefont {Lu}, \citenamefont {Zhang}, \citenamefont {Vijay},\ and\ \citenamefont {Hsieh}}]{lu2023mixed}%
  \BibitemOpen
  \bibfield  {author} {\bibinfo {author} {\bibfnamefont {Tsung-Cheng}\ \bibnamefont {Lu}}, \bibinfo {author} {\bibfnamefont {Zhehao}\ \bibnamefont {Zhang}}, \bibinfo {author} {\bibfnamefont {Sagar}\ \bibnamefont {Vijay}}, \ and\ \bibinfo {author} {\bibfnamefont {Timothy~H}\ \bibnamefont {Hsieh}},\ }\bibfield  {title} {\enquote {\bibinfo {title} {Mixed-state long-range order and criticality from measurement and feedback},}\ }\href@noop {} {\bibfield  {journal} {\bibinfo  {journal} {arXiv preprint arXiv:2303.15507}\ } (\bibinfo {year} {2023})}\BibitemShut {NoStop}%
\bibitem [{\citenamefont {McGinley}\ and\ \citenamefont {Cooper}(2020)}]{mcginley2020fragility}%
  \BibitemOpen
  \bibfield  {author} {\bibinfo {author} {\bibfnamefont {Max}\ \bibnamefont {McGinley}}\ and\ \bibinfo {author} {\bibfnamefont {Nigel~R.}\ \bibnamefont {Cooper}},\ }\bibfield  {title} {\enquote {\bibinfo {title} {Fragility of time-reversal symmetry protected topological phases},}\ }\href {\doibase 10.1038/s41567-020-0956-z} {\bibfield  {journal} {\bibinfo  {journal} {Nature Physics}\ }\textbf {\bibinfo {volume} {16}},\ \bibinfo {pages} {1181--1183} (\bibinfo {year} {2020})}\BibitemShut {NoStop}%
\bibitem [{\citenamefont {Deng}\ \emph {et~al.}(2021)\citenamefont {Deng}, \citenamefont {Pan}, \citenamefont {Chen},\ and\ \citenamefont {Zhai}}]{deng2021stability}%
  \BibitemOpen
  \bibfield  {author} {\bibinfo {author} {\bibfnamefont {Tian-Shu}\ \bibnamefont {Deng}}, \bibinfo {author} {\bibfnamefont {Lei}\ \bibnamefont {Pan}}, \bibinfo {author} {\bibfnamefont {Yu}~\bibnamefont {Chen}}, \ and\ \bibinfo {author} {\bibfnamefont {Hui}\ \bibnamefont {Zhai}},\ }\bibfield  {title} {\enquote {\bibinfo {title} {Stability of time-reversal symmetry protected topological phases},}\ }\href {\doibase 10.1103/PhysRevLett.127.086801} {\bibfield  {journal} {\bibinfo  {journal} {Phys. Rev. Lett.}\ }\textbf {\bibinfo {volume} {127}},\ \bibinfo {pages} {086801} (\bibinfo {year} {2021})}\BibitemShut {NoStop}%
\bibitem [{\citenamefont {Wang}\ \emph {et~al.}(2021)\citenamefont {Wang}, \citenamefont {Li}, \citenamefont {Li},\ and\ \citenamefont {Cai}}]{wang2021symmetry}%
  \BibitemOpen
  \bibfield  {author} {\bibinfo {author} {\bibfnamefont {Zijian}\ \bibnamefont {Wang}}, \bibinfo {author} {\bibfnamefont {Qiaoyi}\ \bibnamefont {Li}}, \bibinfo {author} {\bibfnamefont {Wei}\ \bibnamefont {Li}}, \ and\ \bibinfo {author} {\bibfnamefont {Zi}~\bibnamefont {Cai}},\ }\bibfield  {title} {\enquote {\bibinfo {title} {Symmetry-protected topological edge modes and emergent partial time-reversal symmetry breaking in open quantum many-body systems},}\ }\href {\doibase 10.1103/PhysRevLett.126.237201} {\bibfield  {journal} {\bibinfo  {journal} {Phys. Rev. Lett.}\ }\textbf {\bibinfo {volume} {126}},\ \bibinfo {pages} {237201} (\bibinfo {year} {2021})}\BibitemShut {NoStop}%
\bibitem [{\citenamefont {de~Groot}\ \emph {et~al.}(2022{\natexlab{a}})\citenamefont {de~Groot}, \citenamefont {Turzillo},\ and\ \citenamefont {Schuch}}]{de2022symmetry}%
  \BibitemOpen
  \bibfield  {author} {\bibinfo {author} {\bibfnamefont {Caroline}\ \bibnamefont {de~Groot}}, \bibinfo {author} {\bibfnamefont {Alex}\ \bibnamefont {Turzillo}}, \ and\ \bibinfo {author} {\bibfnamefont {Norbert}\ \bibnamefont {Schuch}},\ }\bibfield  {title} {\enquote {\bibinfo {title} {Symmetry {P}rotected {T}opological {O}rder in {O}pen {Q}uantum {S}ystems},}\ }\href {\doibase 10.22331/q-2022-11-10-856} {\bibfield  {journal} {\bibinfo  {journal} {{Quantum}}\ }\textbf {\bibinfo {volume} {6}},\ \bibinfo {pages} {856} (\bibinfo {year} {2022}{\natexlab{a}})}\BibitemShut {NoStop}%
\bibitem [{\citenamefont {Ma}\ and\ \citenamefont {Wang}(2023{\natexlab{a}})}]{ma2023average}%
  \BibitemOpen
  \bibfield  {author} {\bibinfo {author} {\bibfnamefont {Ruochen}\ \bibnamefont {Ma}}\ and\ \bibinfo {author} {\bibfnamefont {Chong}\ \bibnamefont {Wang}},\ }\bibfield  {title} {\enquote {\bibinfo {title} {Average symmetry-protected topological phases},}\ }\href@noop {} {\bibfield  {journal} {\bibinfo  {journal} {Physical Review X}\ }\textbf {\bibinfo {volume} {13}},\ \bibinfo {pages} {031016} (\bibinfo {year} {2023}{\natexlab{a}})}\BibitemShut {NoStop}%
\bibitem [{\citenamefont {Ma}\ \emph {et~al.}(2023)\citenamefont {Ma}, \citenamefont {Zhang}, \citenamefont {Bi}, \citenamefont {Cheng},\ and\ \citenamefont {Wang}}]{ma2023topological}%
  \BibitemOpen
  \bibfield  {author} {\bibinfo {author} {\bibfnamefont {Ruochen}\ \bibnamefont {Ma}}, \bibinfo {author} {\bibfnamefont {Jian-Hao}\ \bibnamefont {Zhang}}, \bibinfo {author} {\bibfnamefont {Zhen}\ \bibnamefont {Bi}}, \bibinfo {author} {\bibfnamefont {Meng}\ \bibnamefont {Cheng}}, \ and\ \bibinfo {author} {\bibfnamefont {Chong}\ \bibnamefont {Wang}},\ }\bibfield  {title} {\enquote {\bibinfo {title} {Topological phases with average symmetries: the decohered, the disordered, and the intrinsic},}\ }\href@noop {} {\bibfield  {journal} {\bibinfo  {journal} {arXiv preprint arXiv:2305.16399}\ } (\bibinfo {year} {2023})}\BibitemShut {NoStop}%
\bibitem [{\citenamefont {Lee}\ \emph {et~al.}(2022)\citenamefont {Lee}, \citenamefont {You},\ and\ \citenamefont {Xu}}]{lee2022symmetry}%
  \BibitemOpen
  \bibfield  {author} {\bibinfo {author} {\bibfnamefont {Jong~Yeon}\ \bibnamefont {Lee}}, \bibinfo {author} {\bibfnamefont {Yi-Zhuang}\ \bibnamefont {You}}, \ and\ \bibinfo {author} {\bibfnamefont {Cenke}\ \bibnamefont {Xu}},\ }\bibfield  {title} {\enquote {\bibinfo {title} {Symmetry protected topological phases under decoherence},}\ }\href@noop {} {\bibfield  {journal} {\bibinfo  {journal} {arXiv preprint arXiv:2210.16323}\ } (\bibinfo {year} {2022})}\BibitemShut {NoStop}%
\bibitem [{\citenamefont {Zhang}\ \emph {et~al.}(2022)\citenamefont {Zhang}, \citenamefont {Qi},\ and\ \citenamefont {Bi}}]{zhang2022strange}%
  \BibitemOpen
  \bibfield  {author} {\bibinfo {author} {\bibfnamefont {Jian-Hao}\ \bibnamefont {Zhang}}, \bibinfo {author} {\bibfnamefont {Yang}\ \bibnamefont {Qi}}, \ and\ \bibinfo {author} {\bibfnamefont {Zhen}\ \bibnamefont {Bi}},\ }\bibfield  {title} {\enquote {\bibinfo {title} {Strange correlation function for average symmetry-protected topological phases},}\ }\href@noop {} {\bibfield  {journal} {\bibinfo  {journal} {arXiv preprint arXiv:2210.17485}\ } (\bibinfo {year} {2022})}\BibitemShut {NoStop}%
\bibitem [{\citenamefont {Su}\ \emph {et~al.}(2023)\citenamefont {Su}, \citenamefont {Myerson-Jain}, \citenamefont {Wang}, \citenamefont {Jian},\ and\ \citenamefont {Xu}}]{su2023higher}%
  \BibitemOpen
  \bibfield  {author} {\bibinfo {author} {\bibfnamefont {Kaixiang}\ \bibnamefont {Su}}, \bibinfo {author} {\bibfnamefont {Nayan}\ \bibnamefont {Myerson-Jain}}, \bibinfo {author} {\bibfnamefont {Chong}\ \bibnamefont {Wang}}, \bibinfo {author} {\bibfnamefont {Chao-Ming}\ \bibnamefont {Jian}}, \ and\ \bibinfo {author} {\bibfnamefont {Cenke}\ \bibnamefont {Xu}},\ }\bibfield  {title} {\enquote {\bibinfo {title} {Higher-form symmetries under weak measurement},}\ }\href@noop {} {\bibfield  {journal} {\bibinfo  {journal} {arXiv preprint arXiv:2304.14433}\ } (\bibinfo {year} {2023})}\BibitemShut {NoStop}%
\bibitem [{\citenamefont {Fan}\ \emph {et~al.}(2023)\citenamefont {Fan}, \citenamefont {Bao}, \citenamefont {Altman},\ and\ \citenamefont {Vishwanath}}]{fan2023diagnostics}%
  \BibitemOpen
  \bibfield  {author} {\bibinfo {author} {\bibfnamefont {Ruihua}\ \bibnamefont {Fan}}, \bibinfo {author} {\bibfnamefont {Yimu}\ \bibnamefont {Bao}}, \bibinfo {author} {\bibfnamefont {Ehud}\ \bibnamefont {Altman}}, \ and\ \bibinfo {author} {\bibfnamefont {Ashvin}\ \bibnamefont {Vishwanath}},\ }\bibfield  {title} {\enquote {\bibinfo {title} {Diagnostics of mixed-state topological order and breakdown of quantum memory},}\ }\href@noop {} {\bibfield  {journal} {\bibinfo  {journal} {arXiv preprint arXiv:2301.05689}\ } (\bibinfo {year} {2023})}\BibitemShut {NoStop}%
\bibitem [{\citenamefont {Bao}\ \emph {et~al.}(2023)\citenamefont {Bao}, \citenamefont {Fan}, \citenamefont {Vishwanath},\ and\ \citenamefont {Altman}}]{bao2023mixed}%
  \BibitemOpen
  \bibfield  {author} {\bibinfo {author} {\bibfnamefont {Yimu}\ \bibnamefont {Bao}}, \bibinfo {author} {\bibfnamefont {Ruihua}\ \bibnamefont {Fan}}, \bibinfo {author} {\bibfnamefont {Ashvin}\ \bibnamefont {Vishwanath}}, \ and\ \bibinfo {author} {\bibfnamefont {Ehud}\ \bibnamefont {Altman}},\ }\bibfield  {title} {\enquote {\bibinfo {title} {Mixed-state topological order and the errorfield double formulation of decoherence-induced transitions},}\ }\href@noop {} {\bibfield  {journal} {\bibinfo  {journal} {arXiv preprint arXiv:2301.05687}\ } (\bibinfo {year} {2023})}\BibitemShut {NoStop}%
\bibitem [{\citenamefont {Lee}\ \emph {et~al.}(2023)\citenamefont {Lee}, \citenamefont {Jian},\ and\ \citenamefont {Xu}}]{lee2023quantum}%
  \BibitemOpen
  \bibfield  {author} {\bibinfo {author} {\bibfnamefont {Jong~Yeon}\ \bibnamefont {Lee}}, \bibinfo {author} {\bibfnamefont {Chao-Ming}\ \bibnamefont {Jian}}, \ and\ \bibinfo {author} {\bibfnamefont {Cenke}\ \bibnamefont {Xu}},\ }\bibfield  {title} {\enquote {\bibinfo {title} {Quantum criticality under decoherence or weak measurement},}\ }\href@noop {} {\bibfield  {journal} {\bibinfo  {journal} {arXiv preprint arXiv:2301.05238}\ } (\bibinfo {year} {2023})}\BibitemShut {NoStop}%
\bibitem [{\citenamefont {Wang}\ \emph {et~al.}(2023{\natexlab{b}})\citenamefont {Wang}, \citenamefont {Wu},\ and\ \citenamefont {Wang}}]{wang2023intrinsic}%
  \BibitemOpen
  \bibfield  {author} {\bibinfo {author} {\bibfnamefont {Zijian}\ \bibnamefont {Wang}}, \bibinfo {author} {\bibfnamefont {Zhengzhi}\ \bibnamefont {Wu}}, \ and\ \bibinfo {author} {\bibfnamefont {Zhong}\ \bibnamefont {Wang}},\ }\bibfield  {title} {\enquote {\bibinfo {title} {Intrinsic mixed-state topological order without quantum memory},}\ }\href@noop {} {\bibfield  {journal} {\bibinfo  {journal} {arXiv preprint arXiv:2307.13758}\ } (\bibinfo {year} {2023}{\natexlab{b}})}\BibitemShut {NoStop}%
\bibitem [{\citenamefont {Chen}\ and\ \citenamefont {Grover}(2023{\natexlab{a}})}]{chen2023separability}%
  \BibitemOpen
  \bibfield  {author} {\bibinfo {author} {\bibfnamefont {Yu-Hsueh}\ \bibnamefont {Chen}}\ and\ \bibinfo {author} {\bibfnamefont {Tarun}\ \bibnamefont {Grover}},\ }\bibfield  {title} {\enquote {\bibinfo {title} {Separability transitions in topological states induced by local decoherence},}\ }\href@noop {} {\bibfield  {journal} {\bibinfo  {journal} {arXiv preprint arXiv:2309.11879}\ } (\bibinfo {year} {2023}{\natexlab{a}})}\BibitemShut {NoStop}%
\bibitem [{\citenamefont {Chen}\ and\ \citenamefont {Grover}(2023{\natexlab{b}})}]{chen2023symmetry}%
  \BibitemOpen
  \bibfield  {author} {\bibinfo {author} {\bibfnamefont {Yu-Hsueh}\ \bibnamefont {Chen}}\ and\ \bibinfo {author} {\bibfnamefont {Tarun}\ \bibnamefont {Grover}},\ }\bibfield  {title} {\enquote {\bibinfo {title} {Symmetry-enforced many-body separability transitions},}\ }\href@noop {} {\bibfield  {journal} {\bibinfo  {journal} {arXiv preprint arXiv:2310.07286}\ } (\bibinfo {year} {2023}{\natexlab{b}})}\BibitemShut {NoStop}%
\bibitem [{\citenamefont {Chen}\ and\ \citenamefont {Grover}(2024)}]{chen2024unconventional}%
  \BibitemOpen
  \bibfield  {author} {\bibinfo {author} {\bibfnamefont {Yu-Hsueh}\ \bibnamefont {Chen}}\ and\ \bibinfo {author} {\bibfnamefont {Tarun}\ \bibnamefont {Grover}},\ }\bibfield  {title} {\enquote {\bibinfo {title} {Unconventional topological mixed-state transition and critical phase induced by self-dual coherent errors},}\ }\href@noop {} {\bibfield  {journal} {\bibinfo  {journal} {arXiv preprint arXiv:2403.06553}\ } (\bibinfo {year} {2024})}\BibitemShut {NoStop}%
\bibitem [{\citenamefont {Sang}\ \emph {et~al.}(2023)\citenamefont {Sang}, \citenamefont {Zou},\ and\ \citenamefont {Hsieh}}]{sang2023mixed}%
  \BibitemOpen
  \bibfield  {author} {\bibinfo {author} {\bibfnamefont {Shengqi}\ \bibnamefont {Sang}}, \bibinfo {author} {\bibfnamefont {Yijian}\ \bibnamefont {Zou}}, \ and\ \bibinfo {author} {\bibfnamefont {Timothy~H}\ \bibnamefont {Hsieh}},\ }\bibfield  {title} {\enquote {\bibinfo {title} {Mixed-state quantum phases: Renormalization and quantum error correction},}\ }\href@noop {} {\bibfield  {journal} {\bibinfo  {journal} {arXiv preprint arXiv:2310.08639}\ } (\bibinfo {year} {2023})}\BibitemShut {NoStop}%
\bibitem [{\citenamefont {Su}\ \emph {et~al.}(2024)\citenamefont {Su}, \citenamefont {Yang},\ and\ \citenamefont {Jian}}]{su2024tapestry}%
  \BibitemOpen
  \bibfield  {author} {\bibinfo {author} {\bibfnamefont {Kaixiang}\ \bibnamefont {Su}}, \bibinfo {author} {\bibfnamefont {Zhou}\ \bibnamefont {Yang}}, \ and\ \bibinfo {author} {\bibfnamefont {Chao-Ming}\ \bibnamefont {Jian}},\ }\bibfield  {title} {\enquote {\bibinfo {title} {Tapestry of dualities in decohered quantum error correction codes},}\ }\href@noop {} {\bibfield  {journal} {\bibinfo  {journal} {arXiv preprint arXiv:2401.17359}\ } (\bibinfo {year} {2024})}\BibitemShut {NoStop}%
\bibitem [{\citenamefont {Lyons}(2024)}]{lyons2024understanding}%
  \BibitemOpen
  \bibfield  {author} {\bibinfo {author} {\bibfnamefont {Anasuya}\ \bibnamefont {Lyons}},\ }\bibfield  {title} {\enquote {\bibinfo {title} {Understanding stabilizer codes under local decoherence through a general statistical mechanics mapping},}\ }\href@noop {} {\bibfield  {journal} {\bibinfo  {journal} {arXiv preprint arXiv:2403.03955}\ } (\bibinfo {year} {2024})}\BibitemShut {NoStop}%
\bibitem [{\citenamefont {Li}\ and\ \citenamefont {Mong}(2024)}]{li2024replica}%
  \BibitemOpen
  \bibfield  {author} {\bibinfo {author} {\bibfnamefont {Zhuan}\ \bibnamefont {Li}}\ and\ \bibinfo {author} {\bibfnamefont {Roger~SK}\ \bibnamefont {Mong}},\ }\bibfield  {title} {\enquote {\bibinfo {title} {Replica topological order in quantum mixed states and quantum error correction},}\ }\href@noop {} {\bibfield  {journal} {\bibinfo  {journal} {arXiv preprint arXiv:2402.09516}\ } (\bibinfo {year} {2024})}\BibitemShut {NoStop}%
\bibitem [{\citenamefont {Buca}\ and\ \citenamefont {Prosen}(2012)}]{Buca:2012symmetry}%
  \BibitemOpen
  \bibfield  {author} {\bibinfo {author} {\bibfnamefont {Berislav}\ \bibnamefont {Buca}}\ and\ \bibinfo {author} {\bibfnamefont {Tomaz}\ \bibnamefont {Prosen}},\ }\bibfield  {title} {\enquote {\bibinfo {title} {{A note on symmetry reductions of the Lindblad equation: Transport in constrained open spin chains}},}\ }\href {\doibase 10.1088/1367-2630/14/7/073007} {\bibfield  {journal} {\bibinfo  {journal} {New J. Phys.}\ }\textbf {\bibinfo {volume} {14}},\ \bibinfo {pages} {073007} (\bibinfo {year} {2012})},\ \Eprint {http://arxiv.org/abs/1203.0943} {arXiv:1203.0943 [quant-ph]} \BibitemShut {NoStop}%
\bibitem [{\citenamefont {Lessa}\ \emph {et~al.}(2024)\citenamefont {Lessa}, \citenamefont {Cheng},\ and\ \citenamefont {Wang}}]{lessa2024mixed}%
  \BibitemOpen
  \bibfield  {author} {\bibinfo {author} {\bibfnamefont {Leonardo~A}\ \bibnamefont {Lessa}}, \bibinfo {author} {\bibfnamefont {Meng}\ \bibnamefont {Cheng}}, \ and\ \bibinfo {author} {\bibfnamefont {Chong}\ \bibnamefont {Wang}},\ }\bibfield  {title} {\enquote {\bibinfo {title} {Mixed-state quantum anomaly and multipartite entanglement},}\ }\href@noop {} {\bibfield  {journal} {\bibinfo  {journal} {arXiv preprint arXiv:2401.17357}\ } (\bibinfo {year} {2024})}\BibitemShut {NoStop}%
\bibitem [{\citenamefont {Callan}\ and\ \citenamefont {Harvey}(1985)}]{callan1985anomalies}%
  \BibitemOpen
  \bibfield  {author} {\bibinfo {author} {\bibfnamefont {C.G.}\ \bibnamefont {Callan}}\ and\ \bibinfo {author} {\bibfnamefont {J.A.}\ \bibnamefont {Harvey}},\ }\bibfield  {title} {\enquote {\bibinfo {title} {Anomalies and fermion zero modes on strings and domain walls},}\ }\href {\doibase https://doi.org/10.1016/0550-3213(85)90489-4} {\bibfield  {journal} {\bibinfo  {journal} {Nuclear Physics B}\ }\textbf {\bibinfo {volume} {250}},\ \bibinfo {pages} {427--436} (\bibinfo {year} {1985})}\BibitemShut {NoStop}%
\bibitem [{\citenamefont {Chen}\ \emph {et~al.}(2011{\natexlab{a}})\citenamefont {Chen}, \citenamefont {Gu},\ and\ \citenamefont {Wen}}]{PhysRevB.83.035107}%
  \BibitemOpen
  \bibfield  {author} {\bibinfo {author} {\bibfnamefont {Xie}\ \bibnamefont {Chen}}, \bibinfo {author} {\bibfnamefont {Zheng-Cheng}\ \bibnamefont {Gu}}, \ and\ \bibinfo {author} {\bibfnamefont {Xiao-Gang}\ \bibnamefont {Wen}},\ }\bibfield  {title} {\enquote {\bibinfo {title} {Classification of gapped symmetric phases in one-dimensional spin systems},}\ }\href {\doibase 10.1103/PhysRevB.83.035107} {\bibfield  {journal} {\bibinfo  {journal} {Phys. Rev. B}\ }\textbf {\bibinfo {volume} {83}},\ \bibinfo {pages} {035107} (\bibinfo {year} {2011}{\natexlab{a}})}\BibitemShut {NoStop}%
\bibitem [{\citenamefont {Chen}\ \emph {et~al.}(2011{\natexlab{b}})\citenamefont {Chen}, \citenamefont {Liu},\ and\ \citenamefont {Wen}}]{PhysRevB.84.235141}%
  \BibitemOpen
  \bibfield  {author} {\bibinfo {author} {\bibfnamefont {Xie}\ \bibnamefont {Chen}}, \bibinfo {author} {\bibfnamefont {Zheng-Xin}\ \bibnamefont {Liu}}, \ and\ \bibinfo {author} {\bibfnamefont {Xiao-Gang}\ \bibnamefont {Wen}},\ }\bibfield  {title} {\enquote {\bibinfo {title} {Two-dimensional symmetry-protected topological orders and their protected gapless edge excitations},}\ }\href {\doibase 10.1103/PhysRevB.84.235141} {\bibfield  {journal} {\bibinfo  {journal} {Phys. Rev. B}\ }\textbf {\bibinfo {volume} {84}},\ \bibinfo {pages} {235141} (\bibinfo {year} {2011}{\natexlab{b}})}\BibitemShut {NoStop}%
\bibitem [{\citenamefont {Chen}\ \emph {et~al.}(2012)\citenamefont {Chen}, \citenamefont {Gu}, \citenamefont {Liu},\ and\ \citenamefont {Wen}}]{chen2012symmetry}%
  \BibitemOpen
  \bibfield  {author} {\bibinfo {author} {\bibfnamefont {Xie}\ \bibnamefont {Chen}}, \bibinfo {author} {\bibfnamefont {Zheng-Cheng}\ \bibnamefont {Gu}}, \bibinfo {author} {\bibfnamefont {Zheng-Xin}\ \bibnamefont {Liu}}, \ and\ \bibinfo {author} {\bibfnamefont {Xiao-Gang}\ \bibnamefont {Wen}},\ }\bibfield  {title} {\enquote {\bibinfo {title} {Symmetry-protected topological orders in interacting bosonic systems},}\ }\href@noop {} {\bibfield  {journal} {\bibinfo  {journal} {Science}\ }\textbf {\bibinfo {volume} {338}},\ \bibinfo {pages} {1604--1606} (\bibinfo {year} {2012})}\BibitemShut {NoStop}%
\bibitem [{\citenamefont {Levin}\ and\ \citenamefont {Gu}(2012)}]{PhysRevB.86.115109}%
  \BibitemOpen
  \bibfield  {author} {\bibinfo {author} {\bibfnamefont {Michael}\ \bibnamefont {Levin}}\ and\ \bibinfo {author} {\bibfnamefont {Zheng-Cheng}\ \bibnamefont {Gu}},\ }\bibfield  {title} {\enquote {\bibinfo {title} {Braiding statistics approach to symmetry-protected topological phases},}\ }\href {\doibase 10.1103/PhysRevB.86.115109} {\bibfield  {journal} {\bibinfo  {journal} {Phys. Rev. B}\ }\textbf {\bibinfo {volume} {86}},\ \bibinfo {pages} {115109} (\bibinfo {year} {2012})}\BibitemShut {NoStop}%
\bibitem [{\citenamefont {Chen}\ \emph {et~al.}(2013)\citenamefont {Chen}, \citenamefont {Gu}, \citenamefont {Liu},\ and\ \citenamefont {Wen}}]{PhysRevB.87.155114}%
  \BibitemOpen
  \bibfield  {author} {\bibinfo {author} {\bibfnamefont {Xie}\ \bibnamefont {Chen}}, \bibinfo {author} {\bibfnamefont {Zheng-Cheng}\ \bibnamefont {Gu}}, \bibinfo {author} {\bibfnamefont {Zheng-Xin}\ \bibnamefont {Liu}}, \ and\ \bibinfo {author} {\bibfnamefont {Xiao-Gang}\ \bibnamefont {Wen}},\ }\bibfield  {title} {\enquote {\bibinfo {title} {Symmetry protected topological orders and the group cohomology of their symmetry group},}\ }\href {\doibase 10.1103/PhysRevB.87.155114} {\bibfield  {journal} {\bibinfo  {journal} {Phys. Rev. B}\ }\textbf {\bibinfo {volume} {87}},\ \bibinfo {pages} {155114} (\bibinfo {year} {2013})}\BibitemShut {NoStop}%
\bibitem [{\citenamefont {Freed}(2014)}]{freed2014anomalies}%
  \BibitemOpen
  \bibfield  {author} {\bibinfo {author} {\bibfnamefont {Daniel~S.}\ \bibnamefont {Freed}},\ }\href@noop {} {\enquote {\bibinfo {title} {Anomalies and invertible field theories},}\ } (\bibinfo {year} {2014}),\ \Eprint {http://arxiv.org/abs/1404.7224} {arXiv:1404.7224 [hep-th]} \BibitemShut {NoStop}%
\bibitem [{\citenamefont {Kapustin}\ and\ \citenamefont {Thorngren}(2014{\natexlab{a}})}]{PhysRevLett.112.231602}%
  \BibitemOpen
  \bibfield  {author} {\bibinfo {author} {\bibfnamefont {Anton}\ \bibnamefont {Kapustin}}\ and\ \bibinfo {author} {\bibfnamefont {Ryan}\ \bibnamefont {Thorngren}},\ }\bibfield  {title} {\enquote {\bibinfo {title} {Anomalous discrete symmetries in three dimensions and group cohomology},}\ }\href {\doibase 10.1103/PhysRevLett.112.231602} {\bibfield  {journal} {\bibinfo  {journal} {Phys. Rev. Lett.}\ }\textbf {\bibinfo {volume} {112}},\ \bibinfo {pages} {231602} (\bibinfo {year} {2014}{\natexlab{a}})}\BibitemShut {NoStop}%
\bibitem [{\citenamefont {Kapustin}(2014)}]{Kapustin:2014tfa}%
  \BibitemOpen
  \bibfield  {author} {\bibinfo {author} {\bibfnamefont {Anton}\ \bibnamefont {Kapustin}},\ }\bibfield  {title} {\enquote {\bibinfo {title} {{Symmetry Protected Topological Phases, Anomalies, and Cobordisms: Beyond Group Cohomology}},}\ }\href@noop {} {\  (\bibinfo {year} {2014})},\ \Eprint {http://arxiv.org/abs/1403.1467} {arXiv:1403.1467 [cond-mat.str-el]} \BibitemShut {NoStop}%
\bibitem [{\citenamefont {Kapustin}\ and\ \citenamefont {Thorngren}(2014{\natexlab{b}})}]{kapustin2014anomalies}%
  \BibitemOpen
  \bibfield  {author} {\bibinfo {author} {\bibfnamefont {Anton}\ \bibnamefont {Kapustin}}\ and\ \bibinfo {author} {\bibfnamefont {Ryan}\ \bibnamefont {Thorngren}},\ }\bibfield  {title} {\enquote {\bibinfo {title} {Anomalies of discrete symmetries in various dimensions and group cohomology},}\ }\href@noop {} {\bibfield  {journal} {\bibinfo  {journal} {arXiv preprint arXiv:1404.3230}\ } (\bibinfo {year} {2014}{\natexlab{b}})}\BibitemShut {NoStop}%
\bibitem [{\citenamefont {Senthil}(2015)}]{Senthil_2015}%
  \BibitemOpen
  \bibfield  {author} {\bibinfo {author} {\bibfnamefont {T.}~\bibnamefont {Senthil}},\ }\bibfield  {title} {\enquote {\bibinfo {title} {Symmetry-protected topological phases of quantum matter},}\ }\href {\doibase 10.1146/annurev-conmatphys-031214-014740} {\bibfield  {journal} {\bibinfo  {journal} {Annual Review of Condensed Matter Physics}\ }\textbf {\bibinfo {volume} {6}},\ \bibinfo {pages} {299–324} (\bibinfo {year} {2015})}\BibitemShut {NoStop}%
\bibitem [{\citenamefont {Witten}(2016)}]{witten2016fermion}%
  \BibitemOpen
  \bibfield  {author} {\bibinfo {author} {\bibfnamefont {Edward}\ \bibnamefont {Witten}},\ }\bibfield  {title} {\enquote {\bibinfo {title} {Fermion path integrals and topological phases},}\ }\href@noop {} {\bibfield  {journal} {\bibinfo  {journal} {Reviews of Modern Physics}\ }\textbf {\bibinfo {volume} {88}},\ \bibinfo {pages} {035001} (\bibinfo {year} {2016})}\BibitemShut {NoStop}%
\bibitem [{\citenamefont {Yonekura}(2016)}]{yonekura2016dai}%
  \BibitemOpen
  \bibfield  {author} {\bibinfo {author} {\bibfnamefont {Kazuya}\ \bibnamefont {Yonekura}},\ }\bibfield  {title} {\enquote {\bibinfo {title} {Dai-freed theorem and topological phases of matter},}\ }\href@noop {} {\bibfield  {journal} {\bibinfo  {journal} {Journal of High Energy Physics}\ }\textbf {\bibinfo {volume} {2016}},\ \bibinfo {pages} {1--34} (\bibinfo {year} {2016})}\BibitemShut {NoStop}%
\bibitem [{\citenamefont {de~Groot}\ \emph {et~al.}(2022{\natexlab{b}})\citenamefont {de~Groot}, \citenamefont {Turzillo},\ and\ \citenamefont {Schuch}}]{deGroot2022spt}%
  \BibitemOpen
  \bibfield  {author} {\bibinfo {author} {\bibfnamefont {Caroline}\ \bibnamefont {de~Groot}}, \bibinfo {author} {\bibfnamefont {Alex}\ \bibnamefont {Turzillo}}, \ and\ \bibinfo {author} {\bibfnamefont {Norbert}\ \bibnamefont {Schuch}},\ }\bibfield  {title} {\enquote {\bibinfo {title} {Symmetry {P}rotected {T}opological {O}rder in {O}pen {Q}uantum {S}ystems},}\ }\href {\doibase 10.22331/q-2022-11-10-856} {\bibfield  {journal} {\bibinfo  {journal} {{Quantum}}\ }\textbf {\bibinfo {volume} {6}},\ \bibinfo {pages} {856} (\bibinfo {year} {2022}{\natexlab{b}})}\BibitemShut {NoStop}%
\bibitem [{\citenamefont {Ma}\ and\ \citenamefont {Wang}(2023{\natexlab{b}})}]{ma2023aspt}%
  \BibitemOpen
  \bibfield  {author} {\bibinfo {author} {\bibfnamefont {Ruochen}\ \bibnamefont {Ma}}\ and\ \bibinfo {author} {\bibfnamefont {Chong}\ \bibnamefont {Wang}},\ }\bibfield  {title} {\enquote {\bibinfo {title} {Average symmetry-protected topological phases},}\ }\href {\doibase 10.1103/PhysRevX.13.031016} {\bibfield  {journal} {\bibinfo  {journal} {Phys. Rev. X}\ }\textbf {\bibinfo {volume} {13}},\ \bibinfo {pages} {031016} (\bibinfo {year} {2023}{\natexlab{b}})}\BibitemShut {NoStop}%
\bibitem [{\citenamefont {Ma}\ and\ \citenamefont {Turzillo}(2024)}]{ma2024symmetry}%
  \BibitemOpen
  \bibfield  {author} {\bibinfo {author} {\bibfnamefont {Ruochen}\ \bibnamefont {Ma}}\ and\ \bibinfo {author} {\bibfnamefont {Alex}\ \bibnamefont {Turzillo}},\ }\href@noop {} {\enquote {\bibinfo {title} {Symmetry protected topological phases of mixed states in the doubled space},}\ } (\bibinfo {year} {2024}),\ \Eprint {http://arxiv.org/abs/2403.13280} {arXiv:2403.13280 [quant-ph]} \BibitemShut {NoStop}%
\bibitem [{\citenamefont {Guo}\ \emph {et~al.}(2024)\citenamefont {Guo}, \citenamefont {Zhang}, \citenamefont {Yang},\ and\ \citenamefont {Bi}}]{guo2024locally}%
  \BibitemOpen
  \bibfield  {author} {\bibinfo {author} {\bibfnamefont {Yuchen}\ \bibnamefont {Guo}}, \bibinfo {author} {\bibfnamefont {Jian-Hao}\ \bibnamefont {Zhang}}, \bibinfo {author} {\bibfnamefont {Shuo}\ \bibnamefont {Yang}}, \ and\ \bibinfo {author} {\bibfnamefont {Zhen}\ \bibnamefont {Bi}},\ }\bibfield  {title} {\enquote {\bibinfo {title} {Locally purified density operators for symmetry-protected topological phases in mixed states},}\ }\href@noop {} {\bibfield  {journal} {\bibinfo  {journal} {arXiv preprint arXiv:2403.16978}\ } (\bibinfo {year} {2024})}\BibitemShut {NoStop}%
\bibitem [{\citenamefont {Xue}\ \emph {et~al.}(2024)\citenamefont {Xue}, \citenamefont {Lee},\ and\ \citenamefont {Bao}}]{xue2024tensor}%
  \BibitemOpen
  \bibfield  {author} {\bibinfo {author} {\bibfnamefont {Hanyu}\ \bibnamefont {Xue}}, \bibinfo {author} {\bibfnamefont {Jong~Yeon}\ \bibnamefont {Lee}}, \ and\ \bibinfo {author} {\bibfnamefont {Yimu}\ \bibnamefont {Bao}},\ }\bibfield  {title} {\enquote {\bibinfo {title} {Tensor network formulation of symmetry protected topological phases in mixed states},}\ }\href@noop {} {\bibfield  {journal} {\bibinfo  {journal} {arXiv preprint arXiv:2403.17069}\ } (\bibinfo {year} {2024})}\BibitemShut {NoStop}%
\bibitem [{\citenamefont {Chen}\ \emph {et~al.}(2014)\citenamefont {Chen}, \citenamefont {Lu},\ and\ \citenamefont {Vishwanath}}]{chen2014symmetry}%
  \BibitemOpen
  \bibfield  {author} {\bibinfo {author} {\bibfnamefont {Xie}\ \bibnamefont {Chen}}, \bibinfo {author} {\bibfnamefont {Yuan-Ming}\ \bibnamefont {Lu}}, \ and\ \bibinfo {author} {\bibfnamefont {Ashvin}\ \bibnamefont {Vishwanath}},\ }\bibfield  {title} {\enquote {\bibinfo {title} {Symmetry-protected topological phases from decorated domain walls},}\ }\href@noop {} {\bibfield  {journal} {\bibinfo  {journal} {Nature communications}\ }\textbf {\bibinfo {volume} {5}},\ \bibinfo {pages} {3507} (\bibinfo {year} {2014})}\BibitemShut {NoStop}%
\bibitem [{\citenamefont {Hastings}(2011)}]{hastings2011topological}%
  \BibitemOpen
  \bibfield  {author} {\bibinfo {author} {\bibfnamefont {Matthew~B.}\ \bibnamefont {Hastings}},\ }\bibfield  {title} {\enquote {\bibinfo {title} {Topological order at nonzero temperature},}\ }\href {\doibase 10.1103/PhysRevLett.107.210501} {\bibfield  {journal} {\bibinfo  {journal} {Phys. Rev. Lett.}\ }\textbf {\bibinfo {volume} {107}},\ \bibinfo {pages} {210501} (\bibinfo {year} {2011})}\BibitemShut {NoStop}%
\bibitem [{\citenamefont {Chen}\ \emph {et~al.}(2010)\citenamefont {Chen}, \citenamefont {Gu},\ and\ \citenamefont {Wen}}]{PhysRevB.82.155138}%
  \BibitemOpen
  \bibfield  {author} {\bibinfo {author} {\bibfnamefont {Xie}\ \bibnamefont {Chen}}, \bibinfo {author} {\bibfnamefont {Zheng-Cheng}\ \bibnamefont {Gu}}, \ and\ \bibinfo {author} {\bibfnamefont {Xiao-Gang}\ \bibnamefont {Wen}},\ }\bibfield  {title} {\enquote {\bibinfo {title} {Local unitary transformation, long-range quantum entanglement, wave function renormalization, and topological order},}\ }\href {\doibase 10.1103/PhysRevB.82.155138} {\bibfield  {journal} {\bibinfo  {journal} {Phys. Rev. B}\ }\textbf {\bibinfo {volume} {82}},\ \bibinfo {pages} {155138} (\bibinfo {year} {2010})}\BibitemShut {NoStop}%
\bibitem [{Note1()}]{Note1}%
  \BibitemOpen
  \bibinfo {note} {Nevertheless, in some recent papers anomalies of weak symmetries are discussed from other perspectives \cite {hsin2023anomalies,zang2023detecting}. There the anomaly has very different meanings than our notion.}\BibitemShut {Stop}%
\bibitem [{\citenamefont {Cheng}\ and\ \citenamefont {Seiberg}(2023{\natexlab{b}})}]{10.21468/SciPostPhys.15.2.051}%
  \BibitemOpen
  \bibfield  {author} {\bibinfo {author} {\bibfnamefont {Meng}\ \bibnamefont {Cheng}}\ and\ \bibinfo {author} {\bibfnamefont {Nathan}\ \bibnamefont {Seiberg}},\ }\bibfield  {title} {\enquote {\bibinfo {title} {{Lieb-Schultz-Mattis, Luttinger, and 't Hooft - anomaly matching in lattice systems}},}\ }\href {\doibase 10.21468/SciPostPhys.15.2.051} {\bibfield  {journal} {\bibinfo  {journal} {SciPost Phys.}\ }\textbf {\bibinfo {volume} {15}},\ \bibinfo {pages} {051} (\bibinfo {year} {2023}{\natexlab{b}})}\BibitemShut {NoStop}%
\bibitem [{Note2()}]{Note2}%
  \BibitemOpen
  \bibinfo {note} {The $J_1$ term has a global $U(1)$ symmetry generated by $\DOTSB \sum@ \slimits@ _i(-1)^i\sigma ^z_i\sigma ^z_{i+1}$which contains the $Z_2^{\protect \text {CZ}}$ subgroup. We break the $U(1)$ symmetry down to $Z_2$ by adding the $J_2$ term.}\BibitemShut {Stop}%
\bibitem [{\citenamefont {Wang}\ and\ \citenamefont {Wen}(2023)}]{Wang:2013yta}%
  \BibitemOpen
  \bibfield  {author} {\bibinfo {author} {\bibfnamefont {Juven}\ \bibnamefont {Wang}}\ and\ \bibinfo {author} {\bibfnamefont {Xiao-Gang}\ \bibnamefont {Wen}},\ }\bibfield  {title} {\enquote {\bibinfo {title} {{Nonperturbative regularization of (1+1)-dimensional anomaly-free chiral fermions and bosons: On the equivalence of anomaly matching conditions and boundary gapping rules}},}\ }\href {\doibase 10.1103/PhysRevB.107.014311} {\bibfield  {journal} {\bibinfo  {journal} {Phys. Rev. B}\ }\textbf {\bibinfo {volume} {107}},\ \bibinfo {pages} {014311} (\bibinfo {year} {2023})},\ \Eprint {http://arxiv.org/abs/1307.7480} {arXiv:1307.7480 [hep-lat]} \BibitemShut {NoStop}%
\bibitem [{\citenamefont {Han}\ \emph {et~al.}(2017)\citenamefont {Han}, \citenamefont {Tiwari}, \citenamefont {Hsieh},\ and\ \citenamefont {Ryu}}]{PhysRevB.96.125105}%
  \BibitemOpen
  \bibfield  {author} {\bibinfo {author} {\bibfnamefont {Bo}~\bibnamefont {Han}}, \bibinfo {author} {\bibfnamefont {Apoorv}\ \bibnamefont {Tiwari}}, \bibinfo {author} {\bibfnamefont {Chang-Tse}\ \bibnamefont {Hsieh}}, \ and\ \bibinfo {author} {\bibfnamefont {Shinsei}\ \bibnamefont {Ryu}},\ }\bibfield  {title} {\enquote {\bibinfo {title} {Boundary conformal field theory and symmetry-protected topological phases in $2+1$ dimensions},}\ }\href {\doibase 10.1103/PhysRevB.96.125105} {\bibfield  {journal} {\bibinfo  {journal} {Phys. Rev. B}\ }\textbf {\bibinfo {volume} {96}},\ \bibinfo {pages} {125105} (\bibinfo {year} {2017})}\BibitemShut {NoStop}%
\bibitem [{\citenamefont {Jensen}\ \emph {et~al.}(2018)\citenamefont {Jensen}, \citenamefont {Shaverin},\ and\ \citenamefont {Yarom}}]{Jensen:2017eof}%
  \BibitemOpen
  \bibfield  {author} {\bibinfo {author} {\bibfnamefont {Kristan}\ \bibnamefont {Jensen}}, \bibinfo {author} {\bibfnamefont {Evgeny}\ \bibnamefont {Shaverin}}, \ and\ \bibinfo {author} {\bibfnamefont {Amos}\ \bibnamefont {Yarom}},\ }\bibfield  {title} {\enquote {\bibinfo {title} {{\textquoteright{}t Hooft anomalies and boundaries}},}\ }\href {\doibase 10.1007/JHEP01(2018)085} {\bibfield  {journal} {\bibinfo  {journal} {JHEP}\ }\textbf {\bibinfo {volume} {01}},\ \bibinfo {pages} {085} (\bibinfo {year} {2018})},\ \Eprint {http://arxiv.org/abs/1710.07299} {arXiv:1710.07299 [hep-th]} \BibitemShut {NoStop}%
\bibitem [{\citenamefont {Numasawa}\ and\ \citenamefont {Yamaguchi}(2018)}]{Numasawa:2017crf}%
  \BibitemOpen
  \bibfield  {author} {\bibinfo {author} {\bibfnamefont {Tokiro}\ \bibnamefont {Numasawa}}\ and\ \bibinfo {author} {\bibfnamefont {Satoshi}\ \bibnamefont {Yamaguchi}},\ }\bibfield  {title} {\enquote {\bibinfo {title} {{Mixed global anomalies and boundary conformal field theories}},}\ }\href {\doibase 10.1007/JHEP11(2018)202} {\bibfield  {journal} {\bibinfo  {journal} {JHEP}\ }\textbf {\bibinfo {volume} {11}},\ \bibinfo {pages} {202} (\bibinfo {year} {2018})},\ \Eprint {http://arxiv.org/abs/1712.09361} {arXiv:1712.09361 [hep-th]} \BibitemShut {NoStop}%
\bibitem [{\citenamefont {Li}\ \emph {et~al.}(2022{\natexlab{a}})\citenamefont {Li}, \citenamefont {Hsieh}, \citenamefont {Yao},\ and\ \citenamefont {Oshikawa}}]{Li:2022drc}%
  \BibitemOpen
  \bibfield  {author} {\bibinfo {author} {\bibfnamefont {Linhao}\ \bibnamefont {Li}}, \bibinfo {author} {\bibfnamefont {Chang-Tse}\ \bibnamefont {Hsieh}}, \bibinfo {author} {\bibfnamefont {Yuan}\ \bibnamefont {Yao}}, \ and\ \bibinfo {author} {\bibfnamefont {Masaki}\ \bibnamefont {Oshikawa}},\ }\bibfield  {title} {\enquote {\bibinfo {title} {{Boundary conditions and anomalies of conformal field theories in 1+1 dimensions}},}\ }\href@noop {} {\  (\bibinfo {year} {2022}{\natexlab{a}})},\ \Eprint {http://arxiv.org/abs/2205.11190} {arXiv:2205.11190 [hep-th]} \BibitemShut {NoStop}%
\bibitem [{\citenamefont {Choi}\ \emph {et~al.}(2023)\citenamefont {Choi}, \citenamefont {Rayhaun}, \citenamefont {Sanghavi},\ and\ \citenamefont {Shao}}]{Choi:2023xjw}%
  \BibitemOpen
  \bibfield  {author} {\bibinfo {author} {\bibfnamefont {Yichul}\ \bibnamefont {Choi}}, \bibinfo {author} {\bibfnamefont {Brandon~C.}\ \bibnamefont {Rayhaun}}, \bibinfo {author} {\bibfnamefont {Yaman}\ \bibnamefont {Sanghavi}}, \ and\ \bibinfo {author} {\bibfnamefont {Shu-Heng}\ \bibnamefont {Shao}},\ }\bibfield  {title} {\enquote {\bibinfo {title} {{Remarks on boundaries, anomalies, and noninvertible symmetries}},}\ }\href {\doibase 10.1103/PhysRevD.108.125005} {\bibfield  {journal} {\bibinfo  {journal} {Phys. Rev. D}\ }\textbf {\bibinfo {volume} {108}},\ \bibinfo {pages} {125005} (\bibinfo {year} {2023})},\ \Eprint {http://arxiv.org/abs/2305.09713} {arXiv:2305.09713 [hep-th]} \BibitemShut {NoStop}%
\bibitem [{\citenamefont {Thorngren}\ and\ \citenamefont {Wang}(2021)}]{Thorngren:2020yht}%
  \BibitemOpen
  \bibfield  {author} {\bibinfo {author} {\bibfnamefont {Ryan}\ \bibnamefont {Thorngren}}\ and\ \bibinfo {author} {\bibfnamefont {Yifan}\ \bibnamefont {Wang}},\ }\bibfield  {title} {\enquote {\bibinfo {title} {{Anomalous symmetries end at the boundary}},}\ }\href {\doibase 10.1007/JHEP09(2021)017} {\bibfield  {journal} {\bibinfo  {journal} {JHEP}\ }\textbf {\bibinfo {volume} {09}},\ \bibinfo {pages} {017} (\bibinfo {year} {2021})},\ \Eprint {http://arxiv.org/abs/2012.15861} {arXiv:2012.15861 [hep-th]} \BibitemShut {NoStop}%
\bibitem [{Note3()}]{Note3}%
  \BibitemOpen
  \bibinfo {note} {Although the main goal of this paper is to study novel aspects of anomalies in mixed states, we believe the implication of Theorem 2 on pure states has also not been obtained before, and is interesting in its own right.}\BibitemShut {Stop}%
\bibitem [{Note4()}]{Note4}%
  \BibitemOpen
  \bibinfo {note} {In \cite {ma2023aspt,ma2023topological} generalization of SPT to both disordered and decohered systems are investigated. For the latter case that is more relevant to this paper, exact(average) symmetry has the same meaning as strong(weak) symmetry.}\BibitemShut {Stop}%
\bibitem [{\citenamefont {Yoshida}(2016)}]{PhysRevB.93.155131}%
  \BibitemOpen
  \bibfield  {author} {\bibinfo {author} {\bibfnamefont {Beni}\ \bibnamefont {Yoshida}},\ }\bibfield  {title} {\enquote {\bibinfo {title} {Topological phases with generalized global symmetries},}\ }\href {\doibase 10.1103/PhysRevB.93.155131} {\bibfield  {journal} {\bibinfo  {journal} {Phys. Rev. B}\ }\textbf {\bibinfo {volume} {93}},\ \bibinfo {pages} {155131} (\bibinfo {year} {2016})}\BibitemShut {NoStop}%
\bibitem [{\citenamefont {Li}\ and\ \citenamefont {Yao}(2022)}]{PhysRevB.106.224420}%
  \BibitemOpen
  \bibfield  {author} {\bibinfo {author} {\bibfnamefont {Linhao}\ \bibnamefont {Li}}\ and\ \bibinfo {author} {\bibfnamefont {Yuan}\ \bibnamefont {Yao}},\ }\bibfield  {title} {\enquote {\bibinfo {title} {Duality viewpoint of criticality},}\ }\href {\doibase 10.1103/PhysRevB.106.224420} {\bibfield  {journal} {\bibinfo  {journal} {Phys. Rev. B}\ }\textbf {\bibinfo {volume} {106}},\ \bibinfo {pages} {224420} (\bibinfo {year} {2022})}\BibitemShut {NoStop}%
\bibitem [{\citenamefont {Thorngren}\ \emph {et~al.}(2021)\citenamefont {Thorngren}, \citenamefont {Vishwanath},\ and\ \citenamefont {Verresen}}]{PhysRevB.104.075132}%
  \BibitemOpen
  \bibfield  {author} {\bibinfo {author} {\bibfnamefont {Ryan}\ \bibnamefont {Thorngren}}, \bibinfo {author} {\bibfnamefont {Ashvin}\ \bibnamefont {Vishwanath}}, \ and\ \bibinfo {author} {\bibfnamefont {Ruben}\ \bibnamefont {Verresen}},\ }\bibfield  {title} {\enquote {\bibinfo {title} {Intrinsically gapless topological phases},}\ }\href {\doibase 10.1103/PhysRevB.104.075132} {\bibfield  {journal} {\bibinfo  {journal} {Phys. Rev. B}\ }\textbf {\bibinfo {volume} {104}},\ \bibinfo {pages} {075132} (\bibinfo {year} {2021})}\BibitemShut {NoStop}%
\bibitem [{\citenamefont {Yang}\ \emph {et~al.}(2023)\citenamefont {Yang}, \citenamefont {Li}, \citenamefont {Okunishi},\ and\ \citenamefont {Katsura}}]{PhysRevB.107.125158}%
  \BibitemOpen
  \bibfield  {author} {\bibinfo {author} {\bibfnamefont {Hong}\ \bibnamefont {Yang}}, \bibinfo {author} {\bibfnamefont {Linhao}\ \bibnamefont {Li}}, \bibinfo {author} {\bibfnamefont {Kouichi}\ \bibnamefont {Okunishi}}, \ and\ \bibinfo {author} {\bibfnamefont {Hosho}\ \bibnamefont {Katsura}},\ }\bibfield  {title} {\enquote {\bibinfo {title} {Duality, criticality, anomaly, and topology in quantum spin-1 chains},}\ }\href {\doibase 10.1103/PhysRevB.107.125158} {\bibfield  {journal} {\bibinfo  {journal} {Phys. Rev. B}\ }\textbf {\bibinfo {volume} {107}},\ \bibinfo {pages} {125158} (\bibinfo {year} {2023})}\BibitemShut {NoStop}%
\bibitem [{\citenamefont {Li}\ \emph {et~al.}(2022{\natexlab{b}})\citenamefont {Li}, \citenamefont {Oshikawa},\ and\ \citenamefont {Zheng}}]{li2022symmetry}%
  \BibitemOpen
  \bibfield  {author} {\bibinfo {author} {\bibfnamefont {Linhao}\ \bibnamefont {Li}}, \bibinfo {author} {\bibfnamefont {Masaki}\ \bibnamefont {Oshikawa}}, \ and\ \bibinfo {author} {\bibfnamefont {Yunqin}\ \bibnamefont {Zheng}},\ }\bibfield  {title} {\enquote {\bibinfo {title} {Decorated defect construction of gapless-spt states},}\ }\href@noop {} {\bibfield  {journal} {\bibinfo  {journal} {arXiv preprint arXiv:2204.03131}\ }\textbf {\bibinfo {volume} {46}} (\bibinfo {year} {2022}{\natexlab{b}})}\BibitemShut {NoStop}%
\bibitem [{\citenamefont {Ma}\ \emph {et~al.}(2022)\citenamefont {Ma}, \citenamefont {Zou},\ and\ \citenamefont {Wang}}]{10.21468/SciPostPhys.12.6.196}%
  \BibitemOpen
  \bibfield  {author} {\bibinfo {author} {\bibfnamefont {Ruochen}\ \bibnamefont {Ma}}, \bibinfo {author} {\bibfnamefont {Liujun}\ \bibnamefont {Zou}}, \ and\ \bibinfo {author} {\bibfnamefont {Chong}\ \bibnamefont {Wang}},\ }\bibfield  {title} {\enquote {\bibinfo {title} {{Edge physics at the deconfined transition between a quantum spin Hall insulator and a superconductor}},}\ }\href {\doibase 10.21468/SciPostPhys.12.6.196} {\bibfield  {journal} {\bibinfo  {journal} {SciPost Phys.}\ }\textbf {\bibinfo {volume} {12}},\ \bibinfo {pages} {196} (\bibinfo {year} {2022})}\BibitemShut {NoStop}%
\bibitem [{\citenamefont {Wen}\ and\ \citenamefont {Potter}(2023{\natexlab{a}})}]{PhysRevB.107.245127}%
  \BibitemOpen
  \bibfield  {author} {\bibinfo {author} {\bibfnamefont {Rui}\ \bibnamefont {Wen}}\ and\ \bibinfo {author} {\bibfnamefont {Andrew~C.}\ \bibnamefont {Potter}},\ }\bibfield  {title} {\enquote {\bibinfo {title} {Bulk-boundary correspondence for intrinsically gapless symmetry-protected topological phases from group cohomology},}\ }\href {\doibase 10.1103/PhysRevB.107.245127} {\bibfield  {journal} {\bibinfo  {journal} {Phys. Rev. B}\ }\textbf {\bibinfo {volume} {107}},\ \bibinfo {pages} {245127} (\bibinfo {year} {2023}{\natexlab{a}})}\BibitemShut {NoStop}%
\bibitem [{\citenamefont {Li}\ \emph {et~al.}(2023)\citenamefont {Li}, \citenamefont {Oshikawa},\ and\ \citenamefont {Zheng}}]{li2023intrinsically}%
  \BibitemOpen
  \bibfield  {author} {\bibinfo {author} {\bibfnamefont {Linhao}\ \bibnamefont {Li}}, \bibinfo {author} {\bibfnamefont {Masaki}\ \bibnamefont {Oshikawa}}, \ and\ \bibinfo {author} {\bibfnamefont {Yunqin}\ \bibnamefont {Zheng}},\ }\bibfield  {title} {\enquote {\bibinfo {title} {Intrinsically/purely gapless-spt from non-invertible duality transformations},}\ }\href@noop {} {\bibfield  {journal} {\bibinfo  {journal} {arXiv preprint arXiv:2307.04788}\ } (\bibinfo {year} {2023})}\BibitemShut {NoStop}%
\bibitem [{\citenamefont {Wen}\ and\ \citenamefont {Potter}(2023{\natexlab{b}})}]{Wen:2023otf}%
  \BibitemOpen
  \bibfield  {author} {\bibinfo {author} {\bibfnamefont {Rui}\ \bibnamefont {Wen}}\ and\ \bibinfo {author} {\bibfnamefont {Andrew~C.}\ \bibnamefont {Potter}},\ }\bibfield  {title} {\enquote {\bibinfo {title} {{Classification of 1+1D gapless symmetry protected phases via topological holography}},}\ }\href@noop {} {\  (\bibinfo {year} {2023}{\natexlab{b}})},\ \Eprint {http://arxiv.org/abs/2311.00050} {arXiv:2311.00050 [cond-mat.str-el]} \BibitemShut {NoStop}%
\bibitem [{\citenamefont {Gaiotto}\ \emph {et~al.}(2015)\citenamefont {Gaiotto}, \citenamefont {Kapustin}, \citenamefont {Seiberg},\ and\ \citenamefont {Willett}}]{seiberg2014generalized}%
  \BibitemOpen
  \bibfield  {author} {\bibinfo {author} {\bibfnamefont {Davide}\ \bibnamefont {Gaiotto}}, \bibinfo {author} {\bibfnamefont {Anton}\ \bibnamefont {Kapustin}}, \bibinfo {author} {\bibfnamefont {Nathan}\ \bibnamefont {Seiberg}}, \ and\ \bibinfo {author} {\bibfnamefont {Brian}\ \bibnamefont {Willett}},\ }\bibfield  {title} {\enquote {\bibinfo {title} {{Generalized Global Symmetries}},}\ }\href {\doibase 10.1007/JHEP02(2015)172} {\bibfield  {journal} {\bibinfo  {journal} {JHEP}\ }\textbf {\bibinfo {volume} {02}},\ \bibinfo {pages} {172} (\bibinfo {year} {2015})},\ \Eprint {http://arxiv.org/abs/1412.5148} {arXiv:1412.5148 [hep-th]} \BibitemShut {NoStop}%
\bibitem [{\citenamefont {Wen}(2019)}]{wen2019emergent}%
  \BibitemOpen
  \bibfield  {author} {\bibinfo {author} {\bibfnamefont {Xiao-Gang}\ \bibnamefont {Wen}},\ }\bibfield  {title} {\enquote {\bibinfo {title} {Emergent anomalous higher symmetries from topological order and from dynamical electromagnetic field in condensed matter systems},}\ }\href {\doibase 10.1103/PhysRevB.99.205139} {\bibfield  {journal} {\bibinfo  {journal} {Phys. Rev. B}\ }\textbf {\bibinfo {volume} {99}},\ \bibinfo {pages} {205139} (\bibinfo {year} {2019})}\BibitemShut {NoStop}%
\bibitem [{\citenamefont {Liu}(2023)}]{liu2023efficient}%
  \BibitemOpen
  \bibfield  {author} {\bibinfo {author} {\bibfnamefont {Shang}\ \bibnamefont {Liu}},\ }\bibfield  {title} {\enquote {\bibinfo {title} {Efficient preparation of nonabelian topological orders in the doubled hilbert space},}\ }\href@noop {} {\bibfield  {journal} {\bibinfo  {journal} {arXiv preprint arXiv:2311.18497}\ } (\bibinfo {year} {2023})}\BibitemShut {NoStop}%
\bibitem [{\citenamefont {Sohal}\ and\ \citenamefont {Prem}(2024)}]{sohal2024noisy}%
  \BibitemOpen
  \bibfield  {author} {\bibinfo {author} {\bibfnamefont {Ramanjit}\ \bibnamefont {Sohal}}\ and\ \bibinfo {author} {\bibfnamefont {Abhinav}\ \bibnamefont {Prem}},\ }\bibfield  {title} {\enquote {\bibinfo {title} {A noisy approach to intrinsically mixed-state topological order},}\ }\href@noop {} {\bibfield  {journal} {\bibinfo  {journal} {arXiv preprint arXiv:2403.13879}\ } (\bibinfo {year} {2024})}\BibitemShut {NoStop}%
\bibitem [{\citenamefont {Ellison}\ and\ \citenamefont {Cheng}(2024)}]{cheng2024towards}%
  \BibitemOpen
  \bibfield  {author} {\bibinfo {author} {\bibfnamefont {Tyler}\ \bibnamefont {Ellison}}\ and\ \bibinfo {author} {\bibfnamefont {Meng}\ \bibnamefont {Cheng}},\ }\bibfield  {title} {\enquote {\bibinfo {title} {Towards a classification of mixed-state topological orders in two dimensions},}\ }\href@noop {} {\bibfield  {journal} {\bibinfo  {journal} {arXiv preprint arXiv:2405.02390}\ } (\bibinfo {year} {2024})}\BibitemShut {NoStop}%
\bibitem [{\citenamefont {Lieb}\ \emph {et~al.}(1961)\citenamefont {Lieb}, \citenamefont {Schultz},\ and\ \citenamefont {Mattis}}]{Lieb:1961aa}%
  \BibitemOpen
  \bibfield  {author} {\bibinfo {author} {\bibfnamefont {Elliott}\ \bibnamefont {Lieb}}, \bibinfo {author} {\bibfnamefont {Theodore}\ \bibnamefont {Schultz}}, \ and\ \bibinfo {author} {\bibfnamefont {Daniel}\ \bibnamefont {Mattis}},\ }\bibfield  {title} {\enquote {\bibinfo {title} {Two soluble models of an antiferromagnetic chain},}\ }\href@noop {} {\bibfield  {journal} {\bibinfo  {journal} {Ann. Phys.}\ }\textbf {\bibinfo {volume} {16}},\ \bibinfo {pages} {407--466} (\bibinfo {year} {1961})}\BibitemShut {NoStop}%
\bibitem [{\citenamefont {Oshikawa}(2000)}]{Oshikawa:2000aa}%
  \BibitemOpen
  \bibfield  {author} {\bibinfo {author} {\bibfnamefont {Masaki}\ \bibnamefont {Oshikawa}},\ }\bibfield  {title} {\enquote {\bibinfo {title} {Commensurability, excitation gap, and topology in quantum many-particle systems on a periodic lattice},}\ }\href {https://link.aps.org/doi/10.1103/PhysRevLett.84.1535} {\bibfield  {journal} {\bibinfo  {journal} {Phys. Rev. Lett.}\ }\textbf {\bibinfo {volume} {84}},\ \bibinfo {pages} {1535--1538} (\bibinfo {year} {2000})}\BibitemShut {NoStop}%
\bibitem [{\citenamefont {Hastings}(2004)}]{Hastings:2004ab}%
  \BibitemOpen
  \bibfield  {author} {\bibinfo {author} {\bibfnamefont {M.~B.}\ \bibnamefont {Hastings}},\ }\bibfield  {title} {\enquote {\bibinfo {title} {Lieb-{S}chultz-{M}attis in higher dimensions},}\ }\href {https://link.aps.org/doi/10.1103/PhysRevB.69.104431} {\bibfield  {journal} {\bibinfo  {journal} {Phys. Rev. B}\ }\textbf {\bibinfo {volume} {69}},\ \bibinfo {pages} {104431--} (\bibinfo {year} {2004})}\BibitemShut {NoStop}%
\bibitem [{\citenamefont {Fuji}(2016)}]{Fuji-SymmetryProtection-PRB2016}%
  \BibitemOpen
  \bibfield  {author} {\bibinfo {author} {\bibfnamefont {Yohei}\ \bibnamefont {Fuji}},\ }\bibfield  {title} {\enquote {\bibinfo {title} {Effective field theory for one-dimensional valence-bond-solid phases and their symmetry protection},}\ }\href {\doibase 10.1103/PhysRevB.93.104425} {\bibfield  {journal} {\bibinfo  {journal} {Phys. Rev. B}\ }\textbf {\bibinfo {volume} {93}},\ \bibinfo {pages} {104425} (\bibinfo {year} {2016})}\BibitemShut {NoStop}%
\bibitem [{\citenamefont {Watanabe}\ \emph {et~al.}(2015)\citenamefont {Watanabe}, \citenamefont {Po}, \citenamefont {Vishwanath},\ and\ \citenamefont {Zaletel}}]{Watanabe:2015aa}%
  \BibitemOpen
  \bibfield  {author} {\bibinfo {author} {\bibfnamefont {Haruki}\ \bibnamefont {Watanabe}}, \bibinfo {author} {\bibfnamefont {Hoi~Chun}\ \bibnamefont {Po}}, \bibinfo {author} {\bibfnamefont {Ashvin}\ \bibnamefont {Vishwanath}}, \ and\ \bibinfo {author} {\bibfnamefont {Michael}\ \bibnamefont {Zaletel}},\ }\bibfield  {title} {\enquote {\bibinfo {title} {Filling constraints for spin-orbit coupled insulators in symmorphic and nonsymmorphic crystals},}\ }\href@noop {} {\bibfield  {journal} {\bibinfo  {journal} {Proc. Natl. Acad. Sci. USA}\ }\textbf {\bibinfo {volume} {112}},\ \bibinfo {pages} {14551--14556} (\bibinfo {year} {2015})}\BibitemShut {NoStop}%
\bibitem [{\citenamefont {Huang}\ \emph {et~al.}(2017)\citenamefont {Huang}, \citenamefont {Song}, \citenamefont {Huang},\ and\ \citenamefont {Hermele}}]{PhysRevB.96.205106}%
  \BibitemOpen
  \bibfield  {author} {\bibinfo {author} {\bibfnamefont {Sheng-Jie}\ \bibnamefont {Huang}}, \bibinfo {author} {\bibfnamefont {Hao}\ \bibnamefont {Song}}, \bibinfo {author} {\bibfnamefont {Yi-Ping}\ \bibnamefont {Huang}}, \ and\ \bibinfo {author} {\bibfnamefont {Michael}\ \bibnamefont {Hermele}},\ }\bibfield  {title} {\enquote {\bibinfo {title} {Building crystalline topological phases from lower-dimensional states},}\ }\href {\doibase 10.1103/PhysRevB.96.205106} {\bibfield  {journal} {\bibinfo  {journal} {Phys. Rev. B}\ }\textbf {\bibinfo {volume} {96}},\ \bibinfo {pages} {205106} (\bibinfo {year} {2017})}\BibitemShut {NoStop}%
\bibitem [{\citenamefont {Yao}\ \emph {et~al.}(2019)\citenamefont {Yao}, \citenamefont {Hsieh},\ and\ \citenamefont {Oshikawa}}]{Yao:2019aa}%
  \BibitemOpen
  \bibfield  {author} {\bibinfo {author} {\bibfnamefont {Yuan}\ \bibnamefont {Yao}}, \bibinfo {author} {\bibfnamefont {Chang-Tse}\ \bibnamefont {Hsieh}}, \ and\ \bibinfo {author} {\bibfnamefont {Masaki}\ \bibnamefont {Oshikawa}},\ }\bibfield  {title} {\enquote {\bibinfo {title} {Anomaly matching and symmetry-protected critical phases in {$SU(N)$} spin systems in $1+1$ dimensions},}\ }\href@noop {} {\bibfield  {journal} {\bibinfo  {journal} {Phys. Rev. Lett.}\ }\textbf {\bibinfo {volume} {123}},\ \bibinfo {pages} {180201} (\bibinfo {year} {2019})}\BibitemShut {NoStop}%
\bibitem [{\citenamefont {Thorngren}\ and\ \citenamefont {Else}(2018)}]{PhysRevX.8.011040}%
  \BibitemOpen
  \bibfield  {author} {\bibinfo {author} {\bibfnamefont {Ryan}\ \bibnamefont {Thorngren}}\ and\ \bibinfo {author} {\bibfnamefont {Dominic~V.}\ \bibnamefont {Else}},\ }\bibfield  {title} {\enquote {\bibinfo {title} {Gauging spatial symmetries and the classification of topological crystalline phases},}\ }\href {\doibase 10.1103/PhysRevX.8.011040} {\bibfield  {journal} {\bibinfo  {journal} {Phys. Rev. X}\ }\textbf {\bibinfo {volume} {8}},\ \bibinfo {pages} {011040} (\bibinfo {year} {2018})}\BibitemShut {NoStop}%
\bibitem [{\citenamefont {Yao}\ and\ \citenamefont {Oshikawa}(2021)}]{Yao:2021aa}%
  \BibitemOpen
  \bibfield  {author} {\bibinfo {author} {\bibfnamefont {Yuan}\ \bibnamefont {Yao}}\ and\ \bibinfo {author} {\bibfnamefont {Masaki}\ \bibnamefont {Oshikawa}},\ }\bibfield  {title} {\enquote {\bibinfo {title} {Twisted boundary condition and lieb-schultz-mattis ingappability for discrete symmetries},}\ }\href {\doibase 10.1103/PhysRevLett.126.217201} {\bibfield  {journal} {\bibinfo  {journal} {Phys. Rev. Lett.}\ }\textbf {\bibinfo {volume} {126}},\ \bibinfo {pages} {217201--} (\bibinfo {year} {2021})}\BibitemShut {NoStop}%
\bibitem [{\citenamefont {Yao}\ \emph {et~al.}(2023)\citenamefont {Yao}, \citenamefont {Li}, \citenamefont {Oshikawa},\ and\ \citenamefont {Hsieh}}]{Yao:2023bnj}%
  \BibitemOpen
  \bibfield  {author} {\bibinfo {author} {\bibfnamefont {Yuan}\ \bibnamefont {Yao}}, \bibinfo {author} {\bibfnamefont {Linhao}\ \bibnamefont {Li}}, \bibinfo {author} {\bibfnamefont {Masaki}\ \bibnamefont {Oshikawa}}, \ and\ \bibinfo {author} {\bibfnamefont {Chang-Tse}\ \bibnamefont {Hsieh}},\ }\bibfield  {title} {\enquote {\bibinfo {title} {{Lieb-Schultz-Mattis theorem for 1d quantum magnets with antiunitary translation and inversion symmetries}},}\ }\href@noop {} {\  (\bibinfo {year} {2023})},\ \Eprint {http://arxiv.org/abs/2307.09843} {arXiv:2307.09843 [cond-mat.str-el]} \BibitemShut {NoStop}%
\bibitem [{\citenamefont {Kawabata}\ \emph {et~al.}(2024)\citenamefont {Kawabata}, \citenamefont {Sohal},\ and\ \citenamefont {Ryu}}]{PhysRevLett.132.070402}%
  \BibitemOpen
  \bibfield  {author} {\bibinfo {author} {\bibfnamefont {Kohei}\ \bibnamefont {Kawabata}}, \bibinfo {author} {\bibfnamefont {Ramanjit}\ \bibnamefont {Sohal}}, \ and\ \bibinfo {author} {\bibfnamefont {Shinsei}\ \bibnamefont {Ryu}},\ }\bibfield  {title} {\enquote {\bibinfo {title} {Lieb-schultz-mattis theorem in open quantum systems},}\ }\href {\doibase 10.1103/PhysRevLett.132.070402} {\bibfield  {journal} {\bibinfo  {journal} {Phys. Rev. Lett.}\ }\textbf {\bibinfo {volume} {132}},\ \bibinfo {pages} {070402} (\bibinfo {year} {2024})}\BibitemShut {NoStop}%
\bibitem [{\citenamefont {Zhou}\ \emph {et~al.}(2023)\citenamefont {Zhou}, \citenamefont {Li}, \citenamefont {Zhai}, \citenamefont {Li},\ and\ \citenamefont {Gu}}]{zhou2023reviving}%
  \BibitemOpen
  \bibfield  {author} {\bibinfo {author} {\bibfnamefont {Yi-Neng}\ \bibnamefont {Zhou}}, \bibinfo {author} {\bibfnamefont {Xingyu}\ \bibnamefont {Li}}, \bibinfo {author} {\bibfnamefont {Hui}\ \bibnamefont {Zhai}}, \bibinfo {author} {\bibfnamefont {Chengshu}\ \bibnamefont {Li}}, \ and\ \bibinfo {author} {\bibfnamefont {Yingfei}\ \bibnamefont {Gu}},\ }\bibfield  {title} {\enquote {\bibinfo {title} {Reviving the lieb-schultz-mattis theorem in open quantum systems},}\ }\href@noop {} {\bibfield  {journal} {\bibinfo  {journal} {arXiv preprint arXiv:2310.01475}\ } (\bibinfo {year} {2023})}\BibitemShut {NoStop}%
\bibitem [{\citenamefont {Hsin}\ \emph {et~al.}(2023)\citenamefont {Hsin}, \citenamefont {Luo},\ and\ \citenamefont {Sun}}]{hsin2023anomalies}%
  \BibitemOpen
  \bibfield  {author} {\bibinfo {author} {\bibfnamefont {Po-Shen}\ \bibnamefont {Hsin}}, \bibinfo {author} {\bibfnamefont {Zhu-Xi}\ \bibnamefont {Luo}}, \ and\ \bibinfo {author} {\bibfnamefont {Hao-Yu}\ \bibnamefont {Sun}},\ }\bibfield  {title} {\enquote {\bibinfo {title} {Anomalies of average symmetries: Entanglement and open quantum systems},}\ }\href@noop {} {\bibfield  {journal} {\bibinfo  {journal} {arXiv preprint arXiv:2312.09074}\ } (\bibinfo {year} {2023})}\BibitemShut {NoStop}%
\bibitem [{\citenamefont {Zang}\ \emph {et~al.}(2023)\citenamefont {Zang}, \citenamefont {Gu},\ and\ \citenamefont {Jiang}}]{zang2023detecting}%
  \BibitemOpen
  \bibfield  {author} {\bibinfo {author} {\bibfnamefont {Yunlong}\ \bibnamefont {Zang}}, \bibinfo {author} {\bibfnamefont {Yingfei}\ \bibnamefont {Gu}}, \ and\ \bibinfo {author} {\bibfnamefont {Shenghan}\ \bibnamefont {Jiang}},\ }\bibfield  {title} {\enquote {\bibinfo {title} {Detecting quantum anomalies in open systems},}\ }\href@noop {} {\bibfield  {journal} {\bibinfo  {journal} {arXiv preprint arXiv:2312.11188}\ } (\bibinfo {year} {2023})}\BibitemShut {NoStop}%
\bibitem [{\citenamefont {Seifnashri}(2023)}]{Seifnashri:2023dpa}%
  \BibitemOpen
  \bibfield  {author} {\bibinfo {author} {\bibfnamefont {Sahand}\ \bibnamefont {Seifnashri}},\ }\bibfield  {title} {\enquote {\bibinfo {title} {{Lieb-Schultz-Mattis anomalies as obstructions to gauging (non-on-site) symmetries}},}\ }\href@noop {} {\  (\bibinfo {year} {2023})},\ \Eprint {http://arxiv.org/abs/2308.05151} {arXiv:2308.05151 [cond-mat.str-el]} \BibitemShut {NoStop}%
\end{thebibliography}%

 \clearpage
\appendix 

\section{Boundary correlation enforced by strong-weak mixed anomalies. }\label{app:boundary}
In this appendix, we focus on the $(1+1)$-D open quantum systems with mixed anomalies between strong symmetry $K$ and weak symmetry $G$, and discuss its implication on the boundary correlation. As stated in Section \ref{sec:boundary cor}, in this case it is unclear whether the conclusion in Theorem 2 still holds generally, but at least the anomaly leads to a nontrivial Renyi-2 correlation.

\textbf{Theorem 4.} For a $(1+1)$-D state $\rho$ with anomalous symmetry $\Gamma=K\times G$, under OBC, there must exist long-range boundary correlation for Renyi-2 correlators. $\frac{\Tr( \rho O_lO_r \rho O'_lO'_r)}{\Tr(\rho^2)}-\frac{\Tr (\rho O_l\rho O'_l)}{\Tr(\rho^2)}  \frac{\Tr(\rho O_r\rho O'_r)}{\Tr(\rho^2)}\neq 0$.

Proof. We can repeat the steps \eqref{eq:OBC1},\eqref{eq:OBC2},\eqref{eq:OBC3},\eqref{eq:W_l} in the proof of Theorem 2, replacing $U_{\text{OBC}},W$ with $\mathscr{U}_{\text{OBC}},\mathscr{W}$. Then, similar to the analysis below \eqref{eq:W_l}, we have
\[
\mathscr{W}(\gamma_1,\gamma_2)[\rho]=\lambda(\gamma_1)\lambda(\gamma_2)\lambda^{-1}(\gamma_1,\gamma_2)\rho,
\label{eq:eigenW}
\]
from the condition that $\rho$ is $\Gamma$-symmetric. On the other hand, $\exists \gamma=k
\cdot g, \gamma'=k'\cdot g'$, s.t. 
\[
\mathscr{W}_{l(r)}(\gamma,\gamma')[\rho]\equiv W_{l(r)}(\gamma,\gamma')\rho W_{l(r)}^\dagger(g,g')\neq \beta(\gamma,\gamma')\rho
\label{eq:eigenWl}
\] 
for any U(1) phase factor $\beta$, which follows from the condition that $\Gamma$ is anomalous. Since the eigenvalues of $\mathscr{W}_{l(r)}(\gamma,\gamma')$ are all distributed on the unit circle on the complex plane, 
\[
|\frac{\Tr(\rho \mathscr{W}_{l(r)}(\gamma,\gamma')[\rho])}{\Tr(\rho^2)}|<1.
\]
Therefore,
\[
\begin{split}
&\frac{\Tr[ \rho W_l(\gamma,\gamma')W_r(\gamma,\gamma') \rho W_l(g,g')W_r(g,g'))]}{\Tr(\rho^2)}\\
-&\frac{\Tr [\rho W_l(\gamma,\gamma')\rho W_l(g,g')]}{\Tr(\rho^2)}  \frac{\Tr[\rho W_r(\gamma,\gamma')\rho W_r(g,g')]}{\Tr(\rho^2)}\neq 0.
\end{split}
\label{eq:Renyi2}
\]
$\qed$

Furthermore, for onsite weak symmetry $G$, we can take $W_l(g)=W_r(g)=1$. Then \eqref{eq:eigenWl} leads to  $|\Tr[W_{l(r)}(\gamma,\gamma')\rho]|<1$. Together with \eqref{eq:eigenW}, we get
\[
\Tr[W_l(\gamma,\gamma')W_r(\gamma,\gamma')\rho]-\Tr[W_l(\gamma,\gamma')\rho]\Tr[W_r(\gamma,\gamma')\rho]\neq 0,
\]
which proves Theorem 3 in the main text by construction.

\section{$(1+1)$-D steady-state average SPT }\label{app:1+1d ASPT}
In this appendix, we provide the detail of a (1+1)-D spin chain described by Lindbladians, whose steady state is \eqref{eq:rhocluster}. 
Under PBC, the Hamiltonian here is zero and dissipators are as follows:
\begin{equation}\label{eq:cluster-mixed-state}
\begin{split}
&l^1_{2i}=\sigma^z_{2i}\sigma^z_{2i+2}\frac{1-\tau^z_{2i-1}\sigma^x_{2i}\tau^z_{2i+1}}{2}, \\
&l^2_{2i-1}=\tau^z_{2i-1},\\ 
&l^{3}_{2i-1}=\sigma^z_{2i-2}\tau^x_{2i-1}\sigma^z_{2i},
\end{split}
\end{equation} 
 The effect of dissipators $l^1$ is to decorate $K$-charges $\sigma^x_{2i}=-1$ on $G$-domain walls. More precisely, such a term can only move or annihilate the excitation configuration $\tau^z_{2i-1}\sigma^x_{2i}\tau^z_{2i+1}=-1$. The effect of $l^{2}$ is to dephase the $\tau$ spins and to select diagonal domain wall configurations. The effect of $l^{3}$ is to incoherently proliferate the $G$ domain wall configurations. As mentioned in Sec. \ref{sec:1+1dASPT}, $U(\text{CZ})$ can relate this ASPT model and  a Lindbladians :
 \[
 \begin{split}
 &l^1_{2i}=\sigma^z_{2i}\sigma^z_{2i+2}\frac{1-\sigma^x_{2i}}{2},\\&l^2_{2i-1}=\tau^z_{2i-1},\quad l^3_{2i-1}=\tau^x_{2i-1}.
\end{split}
 \]
 The latter has the trivial mixed state $\rho_{\text{trivial}}=\rho_\sigma^\rightarrow\otimes I_\tau$. Thus the Lindbladian with dissipators \eqref{eq:cluster-mixed-state} has the steady state \eqref{eq:rhocluster}. Such steady state has the string order $\text{Tr}( \tau^z_{2j-1}(\prod^k_{i=j}\sigma^x_{2i})\tau^z_{2k+1}\rho_{\text{cluster}})=1$.



As the nontrivial boundary modes protected by the global symmetry is a signature of pure-state SPT, we find that the same is true for the average SPT in Lindbladians, which also comes from an anomaly on the boundary.  We place the spin system on an open chain with length $L\in 2\Z$. Similarly, we focus on the subspace 
\[
\mathcal{C}=\{\rho_j: \mathcal{L}_{\text{bulk}}[\rho_j]=0\}.
\]
Here $\mathcal{L}_{\text{bulk}}$ is defined by only keeping terms fully supported on the open chain in \eqref{eq:cluster-mixed-state}, so $\mathcal{C}$ can be viewed as the space where edge modes live. By using $U(\text{CZ})$, which is now truncated at the edge, we can identify that $\mathcal{C}$ is a 4-dimensional space satisfying $\tau^z_{2i-1}\sigma^x_{2i}\tau^z_{2i+1}\rho_{j}=\rho_{j}$ and $\sigma^z_{2i}\tau^x_{2i+1}\sigma^z_{2i+2}\rho_{j}\sigma^z_{2i}\tau^x_{2i+1}\sigma^z_{2i+2}=\rho_{j}$ where $1\le i\le L/2-1$. Moreover, the Pauli operators, which preserve $\mathcal{C}$, are given by the following dressed edge operator
\[
\tilde{\tau}^{\alpha}_1=U(\text{CZ})\tau^{\alpha}_1 U(\text{CZ}), \quad \tilde{\sigma}^{\alpha}_L=U(\text{CZ})\sigma^{\alpha}_L U(\text{CZ})
\]
where $\alpha=x,y,z$
Following the same method in sec.\ref{sec:2+1dASPT}, we can show the symmetry operator fractionalizes as $U_{K/G}\tilde{\tau}^{\alpha}_1 U_{K/G}=L_{K/G}\tilde{\tau}^{\alpha}_1L_{K/G}, \quad U_{K/G}\tilde{\sigma}^{\alpha}_L U_{K/G}=R_{K/G}\tilde{\sigma}^{\alpha}_L\mathcal{R}_{K/G}$,
where
\[
\begin{split}
 &L_{K}=\tilde{\tau}^z_1=\tau^z_1, R_{K}=\tilde{\sigma}^x_{L}=\tau^z_{L-1}\sigma^x_L, \\ &L_{G}=\tilde{\tau}^x_1=\tau^x_1\sigma^z_2, R_{G}=\tilde{\sigma}^z_{L}=\sigma^z_L.
\end{split}
\]
Thus the symmetry action on the edge modes ($\rho_j\in\mathcal{C}$) reads:
\[
\begin{split}
\mathscr{U}_{K/G}&=\mathscr{L}_{K/G}\circ\mathscr{R}_{K/G},\\
\mathscr{L}_K[\rho_j]&\equiv L_K\rho_j,\mathscr{R}_K[\rho_j]\equiv R_K\rho_j,\\
\mathscr{L}_G[\rho_j]&\equiv L_G\rho_jL^\dagger_G,\mathscr{R}_G[\rho_j]\equiv R_G\rho_j R^\dagger_G
\end{split}
\]
Thus the effective symmetry forms a projective representation on each edge, $\mathscr{L}_K\circ \mathscr{L}_G=-\mathscr{L}_G\circ\mathscr{L}_K$, and similarly for $\mathscr{R}$. As in pure state, the nontrivial projective representation can be defined as an anomaly in (0+1) dimension. of $\mathcal{L}$ and $\mathcal{R}$ gives rise to at least 2-dimensional edge modes at each boundary.
\section{$(2+1)$-D steady-state average SPT with $K=\Z^B_2\times \Z^C_2$ and $G=\Z^A_2$ }\label{app:2+1d ASPT}
In this appendix, we will study the Lindbladians \eqref{eq:2+1d ASPT-2}, mainly on the SPT feature of steady state and the  anomalous symmetry of edge theory.

At first, the SPT feature can be detected by adding a $\Z^A_2$ domain wall on a loop $ N_{A}$ in the dual lattice of $A$ sublattice, i.e., $B$ and $C$ sublattice. This domain wall is realized by conjugating the steady state by $\prod_{i\in M_A}\sigma^x_i$, where $M_A$ only involves the $A$ sites enclosed by $N_A$. Then we have 
\[
\begin{split}
\prod_{M_A}\sigma^x_i \rho^{ss}_{\text{ASPT}}\prod_{ M_A}\sigma^x_i=U_{N_A}(\text{CZ})\rho^{ss}_{\text{ASPT}}U^{\dagger}_{N_A}(\text{CZ}).
\end{split}
\]
The $U_{N_A}(\text{CZ})$ is known to stack a $(1+1)$-D $\Z_2\times \Z_2$ SPT on $N_A$ in closed systems.
Moreover, we can further add a strong $\Z^B_2$ domain wall on $N_B$, which is realized by applying $\prod_{i\in M_B}\tau^x_i$ on the left of the steady state. By the same proof as Eq.~\eqref{eq:ccz-decorate-1} and Eq.~\eqref{eq:ccz-decorate-2}, we have
\[
\begin{split}
&\prod_{ M_A}\sigma^x_i \prod_{ M_B}\tau^x_j \rho^{ss}_{\text{ASPT}}\prod_{ M_B}\tau^x_j\\=&\prod_{N_A\cap N_B }(\mu^z_i)^{s_i}
 U_{N_A\cup N_B}(\text{CZ})\rho^{ss}_{\text{ASPT}} \prod_{ M_B}\tau^x_j.
\end{split}
\]
Here $s_i$ is the number of 1-links of $C$ vertex $i$ whose endpoints are in $M_B$ and $M_A$. This shows the codimension-2 defect, i.e.,  intersection points of $\Z^A_2$ and $\Z^B_2$ domain walls with odd $s_i$ are decorated with $\Z^C_2$ charge.


Indeed, this model also realizes an ASPT when we only consider strong $\Z^{BC}_2$ symmetry and weak $\Z^A_2$ symmetry. This statement can be seen from the corresponding group cohomology and boundary anomaly as follows. Let us consider a typical cocycle $\omega(\gamma_1,\gamma_2,\gamma_3)=(-1)^{b_1c_2a_3}$ belonging to nontrivial class of $H^1(\Z^A_2,H^2(\Z^B_2\times\Z^C_2,U(1))$, where  $\gamma_i=(a_i,b_i,c_i)\in \Z^A_2\times\Z^B_2\times\Z^C_2$ and $a,b,c=0,1$ represent trivial and nontrivial elements, respectively. 
Then one can directly calculate the gauge invariant combinations:
\[
\begin{split}
&\omega(A,BC,BC)\omega(ABC,A,BC)\omega(ABC,ABC,A)\omega(A,I,A)\\
&=1\times 1\times (-1)\times 1=-1,\\
&\omega(BC,BC,BC)\omega(BC,I,BC)=1\times 1=1,
\end{split}
\]
where $A=(1,0,0), BC=(0,1,1)$. The first combination and second combination classify $H^2(\Z^{A}_2, H^1(\Z^{BC}_2,U(1))$ and $H^3(\Z^{BC}_2,U(1))$. Thus this result implies that the cocycle of $\Z^A_2$ and $\Z^{BC}_2$ symmetry of this ASPT belongs to nontrivial class of $H^2(\Z^{A}_2, H^1(\Z^{BC}_2,U(1))$.

Moreover, we can also study how the global symmetry acts on the edge DOFs on an open lattice in Fig.~\ref{fig: Triangle lattice OBC}. We consider Lindblad $\mathcal{L}=\mathcal{L}_{\text{bulk}}+\mathcal{L}_{\text{edge}}$ and focus on the subspace $\mathcal{C'}$ which involves steady states of $\mathcal{L}_{\text{bulk}}$:
\[
\mathcal{C'}=\{\rho_j: \mathcal{L}_{\text{bulk}}[\rho_j]=0\}.
\] By conjugating $U(\text{CCZ})$, which is truncated at the edge, the bulk and edge DOFs are decoupled. In the bulk, there is only a single steady state $\rho^{ss}_{\tau,\mu}=|\rightarrow \rightarrow\cdots \rightarrow\rangle\langle \rightarrow\rightarrow\cdots\rightarrow|$ with $\prod_{i,j\in \text{bulk}}\tau^x_i\mu^x_j=1$ and $\rho^{ss}_{\sigma}=I_{\sigma}$. It is easy to check that the dressed edge operators \eqref{eq:dressed operators} also preserve $\mathcal{C'}$  and compose a complete set of Pauli operators for edge DOFs in $\mathcal{C'}$. Due to \eqref{eq:edge symmetry},   strong $U_{B}U_C$ and weak $U_A$ symmetry  will restrict to   strong $\prod_{i,j \in \text{edge}} \tilde{\tau}^x_i\tilde{\mu}^x_j$ and weak $\tilde{U}(\text{CZ})$ symmetry on the edge DOFs, which is same as {\it Example 3} in section \ref{sec:lattice model}. Since the  strong $\prod_{i,j \in \text{edge}} \tilde{\tau}^x_i\tilde{\mu}^x_j$ and weak $\tilde{U}(\text{CZ})$ has the mixed anomaly, the bulk model \eqref{eq:2+1d ASPT-2} should realize a $\Z^{BC}_2\times \Z^A_2$ ASPT.


\end{document}